\documentclass[a4paper,11pt]{article}

\usepackage[utf8]{inputenc}
\usepackage{amsmath,amssymb,amsthm}
\usepackage{hyperref}
\usepackage{graphicx}
\usepackage{epstopdf}
\usepackage{mathrsfs}
\usepackage{mathtools}
\usepackage{float}
\usepackage[normalem]{ulem}  %%% to use strikeout  \sout{unwanted text}
\usepackage{cite}
\usepackage{subcaption}
\usepackage{titling}
\usepackage{authblk}
\usepackage{setspace}
\usepackage{lipsum}
\usepackage{slashed}
\usepackage{comment}

\newtheorem{theorem}{Theorem}

\makeatletter\@addtoreset{equation}{section}\makeatother

\renewcommand{\title}[1]{\vbox{\center\LARGE{#1}}\vspace{5mm}}
\renewcommand{\author}[1]{\vbox{\center#1}\vspace{5mm}}

\newcommand{\ran}{\rangle}
\newcommand{\lan}{\langle}
\newcommand{\f}{\frac}
\newcommand{\nn}{\nonumber}

\def\be{\begin{equation}}
\def\ee{\end{equation}}

\def\l{\lambda}

\def\a{\alpha}

\def\l{\lambda}

\def\nn{\nonumber}
\def\f{\frac}

\begin{document}

\begin{titlepage}
\begin{flushright}
IFT-UAM/CSIC-19-127
\end{flushright}
\begin{center}
\vskip 1cm
\title{Melonic Dominance in Subchromatic Sextic Tensor Models}
%\author{Shiroman Prakash, Ritam Sinha}

\bigskip
\vspace{1cm}{
{\large Shiroman Prakash$^a$ and Ritam Sinha$^b$}
\vspace{0.3cm}
} \\[7mm]
{\it {$^a$\, Department of Physics and Computer Science, Dayalbagh Educational Institute,  Agra 282005, India}}\\

{\it {$^b$\, Instituto de Fisica Teorica IFT-UAM/CSIC, Cantoblanco 28049, Madrid. Spain}}
\end{center}
\bigskip

\vspace{.4cm}

\abstract{We study tensor models based on $O(N)^r$ symmetry groups constructed out of rank-$r$ tensors with order-$q$ interaction vertices. We refer to those tensor models for which $r<q-1$ as \textit{subchromatic}. We focus most of our attention on sextic ($q=6$) models with maximally-single-trace interactions. We show that only three subchromatic sextic maximally-single-trace interaction vertices exist: these are the $r=3$ prism, the $r=3$ wheel (or $K_{3,3}$) and the $r=4$ octahedron. For theories based on these interactions we demonstrate that the set of Feynman diagrams that contribute to the free energy in the large $N$ limit are melonic and thus can be explicitly summed. In order to take into account the prism, we generalize the conventional notion of melonic diagrams slightly to include diagrams generated by a new melonic move -- vertex expansion.}

\vfill

\end{titlepage}

\eject \tableofcontents

\section{Introduction and summary}
 In addition to well-known adjoint/matrix model  \cite{'tHooft:1973jz} and vector model large $N$ limits \cite{Moshe:2003xn}, a new large $N$ limit dominated by melonic diagrams has attracted a great deal of attention recently \cite{Gurau:2009tw, Gurau:2011aq, Gurau:2011xq, Bonzom:2011zz, Tanasa:2011ur, Bonzom:2012hw, Carrozza:2015adg, Witten:2016iux, Klebanov:2016xxf}. The melonic limit was first observed in tensor models (see \cite{Gurau:2016cjo, Delporte:2018iyf, Klebanov:2018fzb, Gurau:2019qag} for reviews), but interest in this limit grew due in large part to its appearance in the SYK model \cite{Sachdev:1992fk, KitaevTalk, Maldacena:2016hyu}  which serves as a very educational toy model for quantum gravity \cite{Polchinski:2016xgd,  Maldacena:2016upp, Engelsoy:2016xyb, Jensen:2016pah,  Jevicki:2016bwu, Garcia-Garcia:2016mno, Cotler:2016fpe, Nishinaka:2016nxg, Narayan:2017qtw, Garcia-Garcia:2017pzl, Gross:2017hcz, Das:2017pif, Das:2017wae, Das:2017hrt, Mandal:2017thl, Gao:2016bin, Maldacena:2017axo,Maldacena:2018lmt, Kim:2019upg, Rosenhaus:2019mfr}. 

Here we seek to better understand the entire range of models for which  melonic diagrams dominate. While there are important dynamical differences between the SYK model and tensor models \cite{Choudhury:2017tax, Bulycheva:2017ilt}, tensor models provide a very natural context for understanding the diagrammatics of the melonic large $N$ limit, and its possible generalizations.

Motivated by \cite{Ferrari:2017jgw, Gubser:2018yec, GKPPT, KPP}, we consider the large $N$ limit of tensor models constructed out of rank-$r$ tensors which transform in a representation of $O(N)^r$, with order-$q$, i.e., $\phi^q$, interactions \cite{Klebanov:2018fzb}. Each index transforms in the fundamental representation of its corresponding $O(N)$ symmetry group. 

It is natural to restrict our attention to a class of interaction vertices that are \textit{maximally-single-trace} (MST), first defined in \cite{Ferrari:2017jgw}. (We review the definition of maximally-single-trace, and other basic features of these theories in section \ref{preliminaries}.) In the large $N$ limit, we expect that the maximally-single-trace interactions are the ``most interesting'' interactions, in the same sense that the tetrahedron is more interesting than the pillow and double-trace interactions for $q=4$ theories.\footnote{To provide some justification for this expectation, we observe that all the sextic non-MST interactions in \cite{GKPPT} can be obtained from quartic pillow and double-trace interactions via an auxiliary field.} We also remark that the restriction to maximally-single-trace operators reduces the number of interactions to a much more manageable number. 

\begin{figure}[h]
    \centering
    \includegraphics[height=1.8in]{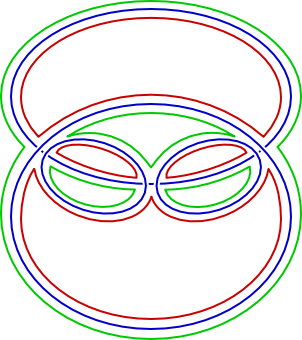}
    \caption{A maximal Feynman diagram in a theory with prism interactions that is not a conventionally melonic Feynman diagram. The diagram is proportional to $g^3 N^{12}=\lambda^3 N^3$.}
    \label{nonmelonicMaximal}
\end{figure}

Let us discuss the large $N$ limit of theories based on such interactions. When $r=2$, these theories define the familiar bifundamental model \cite{Aharony:2008ug,Gurucharan:2014cva},  in the large $N$ limit, in which all planar diagrams survive. When $r=q-1$, these theories are dominated by melonic diagrams, as recently argued in \cite{KPP}. For interaction vertices with intermediate values of $r$, i.e., $2<r<q-1$, we can attempt to determine the set of diagrams which survive in the natural large $N$ limit on a case-by-case basis -- these will certainly include melonic diagrams, but additional diagrams may also contribute. An example of this is the prismatic\footnote{The prismatic limit is solved by introducing an auxiliary field to rewrite the theory as a quartic tensor model with the familiar tetrahedron interaction \cite{Carrozza:2018ewt, GKPPT}. So one can argue that it is effectively still melonic. This limit can also be realized in a theory with random couplings, discussed in \cite{Murugan:2017eto}.} limit of \cite{Azeyanagi:2017mre, GKPPT}, where additional diagrams contribute compared to the $r=5$ sextic tensor model \cite{Ferrari:2017jgw, KPP}, such as the one shown in Figure \ref{nonmelonicMaximal}. Because $r$ determines the number of colours used in a multi-line 't Hooft notation, we may also refer to $r$ as the number of ``colours'' of the model. We refer to models with $r<q-1$ as \textit{subchromatic}\footnote{An alternative name for these models is \textit{subvalent} as each field vertex in the interaction graphs described in the next section have submaximal valence.}.

In this paper, we focus our attention on the case of $q=6$ with $O(N)$ interactions. This case is the smallest non-trivial case to consider. This case is also interesting because, in addition to studying quantum mechanical models in one dimension, one can hope to define quantum field theories with $\phi^6$ interactions in $d\leq 3$ dimensions. 

While one might be concerned that a large number of subchromatic vertices exist for generic $q$, the set of maximally-single-trace subchromatic vertices for $q=6$ turns out to be very small, as we show in section \ref{sexticMSTvertices}. For $r=3$, there are two interaction vertices -- namely the prism and the wheel (which also corresponds to the complete bipartite graph $K_{3,3}$) defined in \cite{GKPPT}; and for $r=4$, there is exactly one such interaction vertex, which corresponds to the graph of a regular octahedron. In section \ref{sexticMSTvertices} we also discuss the discrete symmetries of these interactions.

In section \ref{subchromatic-general} we then discuss basic features of the large $N$ limit of such theories: including the natural 't Hooft coupling and the existence of loops passing through one or two vertices. We also define the notion of a 1-cycles and 2-cycles, which are used in subsequent analysis. These results (which mostly are a special case of a slightly more general analysis in (\cite{Ferrari:2017jgw}) apply to all subchromatic maximally-single-trace interactions. 

In section \ref{melonic-dominance}, we use the results of section \ref{subchromatic-general} to characterize the set of free energy diagrams which survive in the large $N$ limit for theories based on each of the three subchromatic maximally-single-trace interactions we identified in section \ref{sexticMSTvertices}. We also consider theories based on rank-$3$ tensors that contain both prism and wheel interactions. In all cases we find the diagrams can be explicitly summed, and these theories are effectively melonic -- although, the case of the prism \cite{Azeyanagi:2017mre,GKPPT} might not be considered melonic in the conventional sense. 

In section \ref{melonic-dominance}, we also argue that, in the most general ``solvable'' melonic theory, we expect that the set of diagrams that survive in the large $N$ limit can be generated by three types of \textit{melonic moves}: replacing propagators by elementary melons, replacing propagators by elementary snails and vertex expansion. The third-move, vertex-expansion, is not present in most models previously studied; the prismatic theory is the first example of a theory for which all three types of moves are present.

Let us point out\footnote{We thank Igor Klebanov for pointing out the reference \cite{Lionni:2017xvn}.} that related work on sextic $U(N)^3$ and $U(N)^4$ theories appear in \cite{Lionni:2017xvn}, where the melonic dominance of the wheel/$K_{3,3}$ interaction, is discussed following \cite{bonzom2015colored}. We believe our work is complementary to that of \cite{Lionni:2017xvn}, as we consider $O(N)^r$ models, which allow for a larger number of maximally-single-trace interactions. In particular, we prove melonic dominance in the $r=4$ theory based on the octahedron and the $r=3$ theory involving both a prism and wheel interaction. 

In section \ref{SD-section} we briefly discuss the implications for bosonic and fermionic conformal field theories based on these interactions.

In section \ref{conclusion}, we present conclusions and several avenues for future work.

\section{Preliminaries}
\label{preliminaries}

Rank-$3$ tensor models based on fields with $3$ indices\footnote{An alternative class of tensor models is defined using  symmetric traceless or anti-symmetric representations of a single $O(N)$ symmetry group \cite{Klebanov:2017nlk, Benedetti:2017qxl, Carrozza:2018ewt}. We do not consider such theories here.}, $\phi^{a b c}$, that transform under the symmetry group $O(N)^3$ were introduced in \cite{Carrozza:2015adg} and studied in \cite{Klebanov:2016xxf}. Here, $a=1, \ldots, N$, $b=1,\ldots, N$, and $c=1,\ldots, N$ are indices that transform in the fundamental representation of each $O(N)$ symmetry group. Similarly, theories based on rank-$r$ indices are constructed out of fields with $r$ indices that transform under the symmetry group $O(N)^r$. 

The simplest theories one can define are based on a single tensor-field which is either bosonic or fermionic. Theories with quartic interactions of both types, as well as supersymmetric theories, were introduced, and subsequently studied and generalized in, e.g., \cite{ Giombi:2017dtl, Prakash:2017hwq,  BenGeloun:2017jbi, Benedetti:2017fmp, Peng:2016mxj, Peng:2017spg, Mironov:2017aqv, Itoyama:2017xid, Itoyama:2017wjb, Benedetti:2018ghn, Benedetti:2018goh, Chang:2018sve,  Benedetti:2019eyl, Chang:2019yug, Popov:2019nja, Ferrari:2019ogc}. In this paper, we focus primarily on sextic interactions, for which it is natural to consider bosonic theories in $d\leq 3$ dimensions, and fermionic  theories in $d\leq 1$ (i.e., quantum mechanical models). 

A variety of interactions for tensor models exist, which are obtained by contracting the indices in various ways. For example, the wheel interaction \cite{GKPPT} is represented by the following interaction term:
 \begin{equation}
  \mathcal L_{\text{wheel}}=\int d^dx \frac{g_\text{wheel}}{6} \phi^{a_1 b_1 c_1} \phi^{a_2 b_1 c_2} \phi^{a_2 b_2 c_3} \phi^{a_3 b_2 c_1} \phi^{a_3 b_3 c_2} \phi^{a_1 b_3 c_3} .
\end{equation}
We divide the coupling constant by $6$ because that is the size of the automorphism symmetry group of this interaction, as discussed in section \ref{discrete}. 

\begin{figure}
    \centering
    \begin{subfigure}[b]{0.4\textwidth}
      \centering
    \includegraphics[width=.8\textwidth]{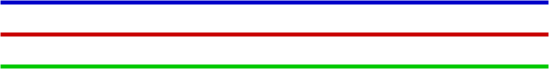}
    \vspace{1.6cm}
    \caption{The propagator.} \label{prop}
    \end{subfigure}
    \begin{subfigure}[b]{0.4\textwidth}
      \centering
    \includegraphics[height=1.70in]{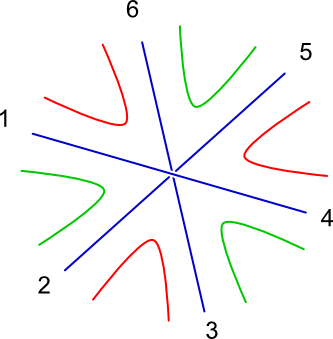}
    \caption{The wheel (or $K_{3,3}$ interaction vertex.}\label{wheel-fat-1}
    \end{subfigure}
    \caption{Feynman diagrams in rank-$3$ tensor models, can be represented by an $3$-line notation, with the propagator and interaction vertex as shown above. Each colour corresponds to a different $O(N)$ symmetry group.} \label{wheel-fat}
\end{figure}

One way of drawing Feynman diagrams for the rank-$3$ tensors is via a triple-line notation that is a straightforward generalization of 't Hooft's double-line notation. Each propagator $\langle \phi^{a_1 b_1 c_1} \phi^{a_2 b_2 c_2} \rangle \sim \delta^{a_1 a_2}\delta^{b_1 b_2} \delta^{c_1 c_2}$ is represented by three coloured lines, with different colours representing the different indices $a$, $b$ and $c$; as shown in Figure \ref{prop}. The wheel interaction vertex is represented by the vertex shown in Figure \ref{wheel-fat-1}. 

\begin{figure}
    \centering
    \includegraphics[width=.4\textwidth]{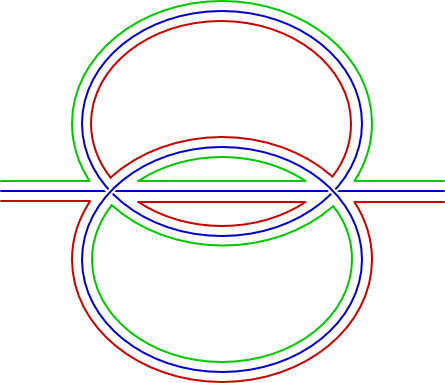}
    \caption{An order $g_\text{wheel}^2$ correction to the propagator in triple-line notation. This diagram is also an \textit{elementary melon} and is proportional to $g_\text{wheel}^2 N^6 = \lambda_\text{wheel}^2$.}
   \label{wheel-melon-fig}
\end{figure}

A two-loop correction to the propagator is shown in Figure \ref{wheel-melon-fig}. As we will show in section \ref{subchromatic-general}, the natural 't Hooft coupling for the wheel is $\lambda_\text{wheel}=g_\text{wheel} N^3$. In the large $N$ limit, with the 't Hooft coupling fixed, this is a leading-order diagram proportional to $\lambda_\text{wheel}^2$.  This diagram is also an elementary melon. In melonic theories, any leading-order diagrams can be obtained by repeatedly replacing propagators by elementary melons. We will discuss this in more detail in section \ref{MelonicSection}.

\begin{figure}
    \centering
    \includegraphics[height=2in]{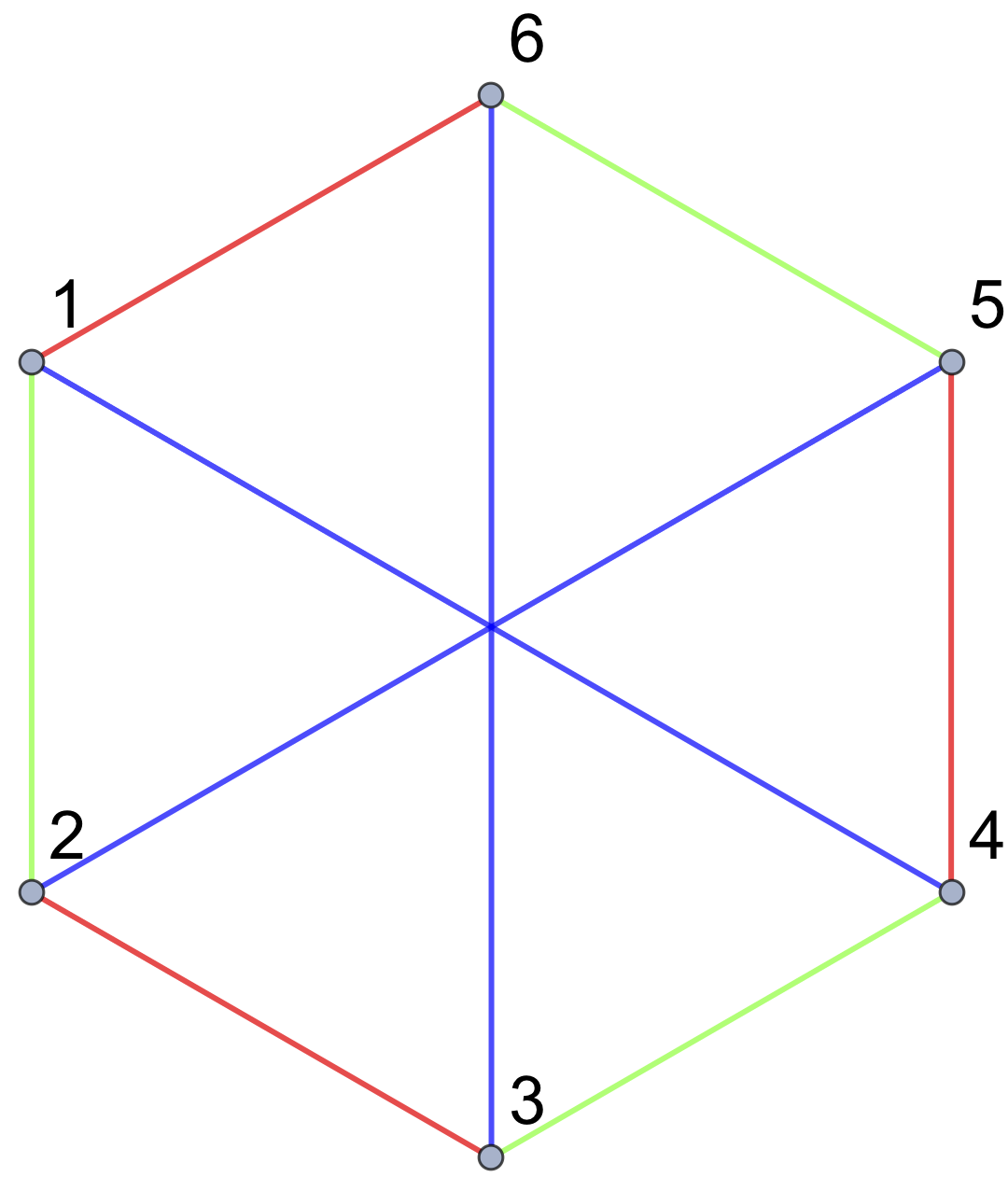}
    \caption{The wheel (or $K_{3,3}$) interaction vertex is represented by the above \textit{labelled interaction graph}.}
    \label{interaction-graph}
\end{figure}

For the purpose of systematically enumerating all possible interactions, it is more convenient to represent interactions by an \textit{interaction graph}, as shown in Figure \ref{interaction-graph}. Each vertex in the interaction graph represents a field. Each symmetry group corresponds to a different colour, and coloured edges denote contractions of the corresponding indices.  We will refer to the vertices of the interaction graph as \textit{field-vertices}, or simply fields, to avoid confusion with ``interaction vertices" in Feynman diagrams.

\noindent
It is convenient to label the fields-vertices of an interaction graph by $i=1$, $2$, \ldots $6$; where $p_i$ represent the momenta of each field, and are ``dummy indices'', as can be seen from the momentum-space representation of the vertex:
\begin{equation}
   \int \f{ d^dp_1 d^d p_2 d^d p_3 d^d p_4 d^d p_5 d^d p_6} {(2\pi)^{6d}} g \phi(p_1) \phi(p_2)\phi(p_3)\phi(p_4) \phi(p_5)\phi(p_6) \delta^d\left(\sum_i p_i\right).
\end{equation}
 Two \textit{labelled interaction graphs} correspond to the same interaction if they can be made identical by a permutation of outgoing-field labels. 

\begin{figure}
    \centering
    \includegraphics[height=1.5in]{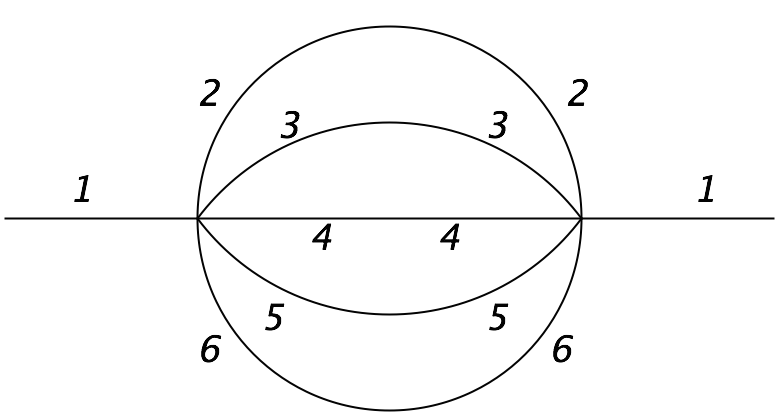}
    \caption{The elementary melon of Figure \ref{wheel-melon-fig} represented as a Feynman diagram in single-line notation using the field-labels of Figure \ref{interaction-graph} to specify the Wick contractions.}
   \label{wheel-melon-fig2}
\end{figure}

Each interaction will be presented with a conventional set of labels for its fields, such as the labels given in Figure \ref{interaction-graph}. One advantage of using these labels is that one can unambiguously represent Feynman diagrams in single-line notation. The elementary melon of Figure \ref{wheel-melon-fig} can also be represented as a single-line Feynman diagram, using field-labels, as shown in Figure \ref{wheel-melon-fig2}.

\begin{figure}
    \centering
    \begin{subfigure}{0.3\textwidth}
     \centering
    \includegraphics[height=2in]{wheel.png}
    \end{subfigure}
    \begin{subfigure}{0.3\textwidth}
      \centering
    \includegraphics[height=2in]{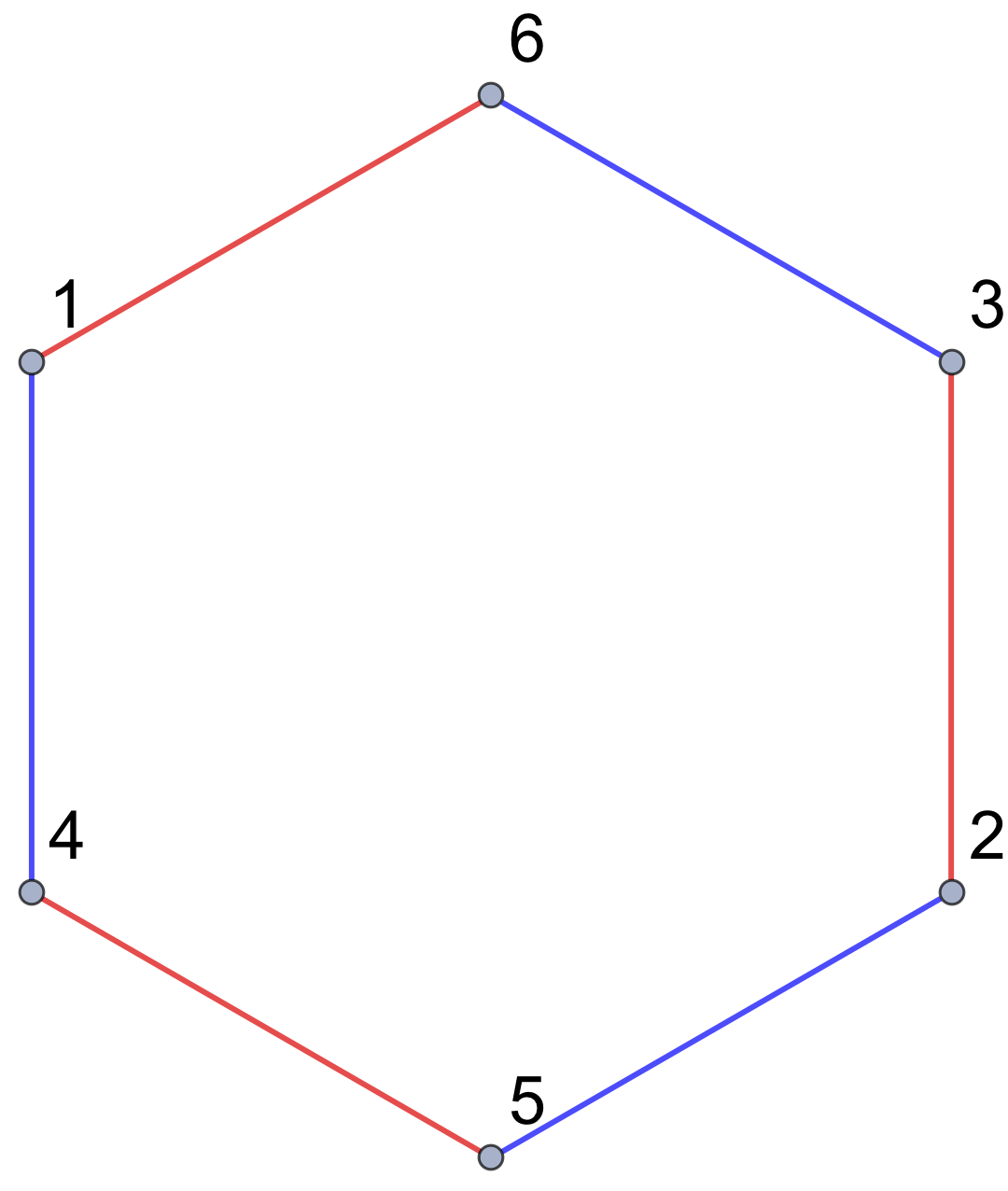}
    \end{subfigure}

    \caption{If we ``forget'' the green edges in the $3$-colour interaction graph on the left, we obtain the $2$-colour interaction (sub)graph on the right.}
    \label{forget}
\end{figure}

An interaction graph representing an interaction of rank-$r$ tensors will involve edges of $r$ different colours. Given a multi-line interaction graph with $r$ colours, one can remove all lines of a given colour, to obtain an interaction graph with $r-1$ colours. For example, in Figure \ref{forget}, if we forget the green edges in the graph on the left, we obtain the $2$-colour interaction graph on the right. We call this process ``forgetting'' a colour. Given an interaction graph, it is convenient to sometimes forget all but $2$ colours. We call the resulting interaction graph a two-colour subgraph. 

\begin{figure}[h]
  \centering
    \begin{subfigure}[b]{0.32\textwidth}
     \centering
    \includegraphics[width=.8\textwidth]{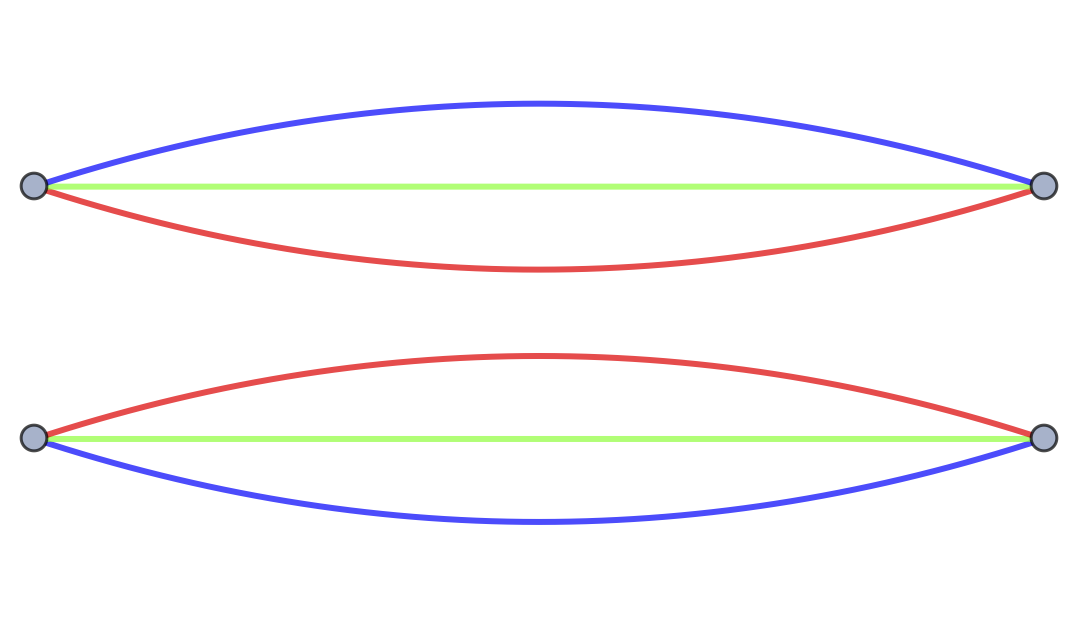}
    \caption{A double-trace interaction.}
    \end{subfigure}
     \begin{subfigure}[b]{0.32\textwidth}
      \centering
    \includegraphics[width=.8\textwidth]{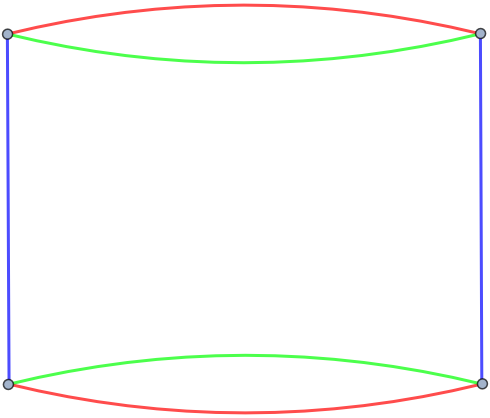}
    \caption{The pillow interaction.}
    \end{subfigure}
     \begin{subfigure}[b]{0.32\textwidth}
      \centering
    \includegraphics[width=.8\textwidth]{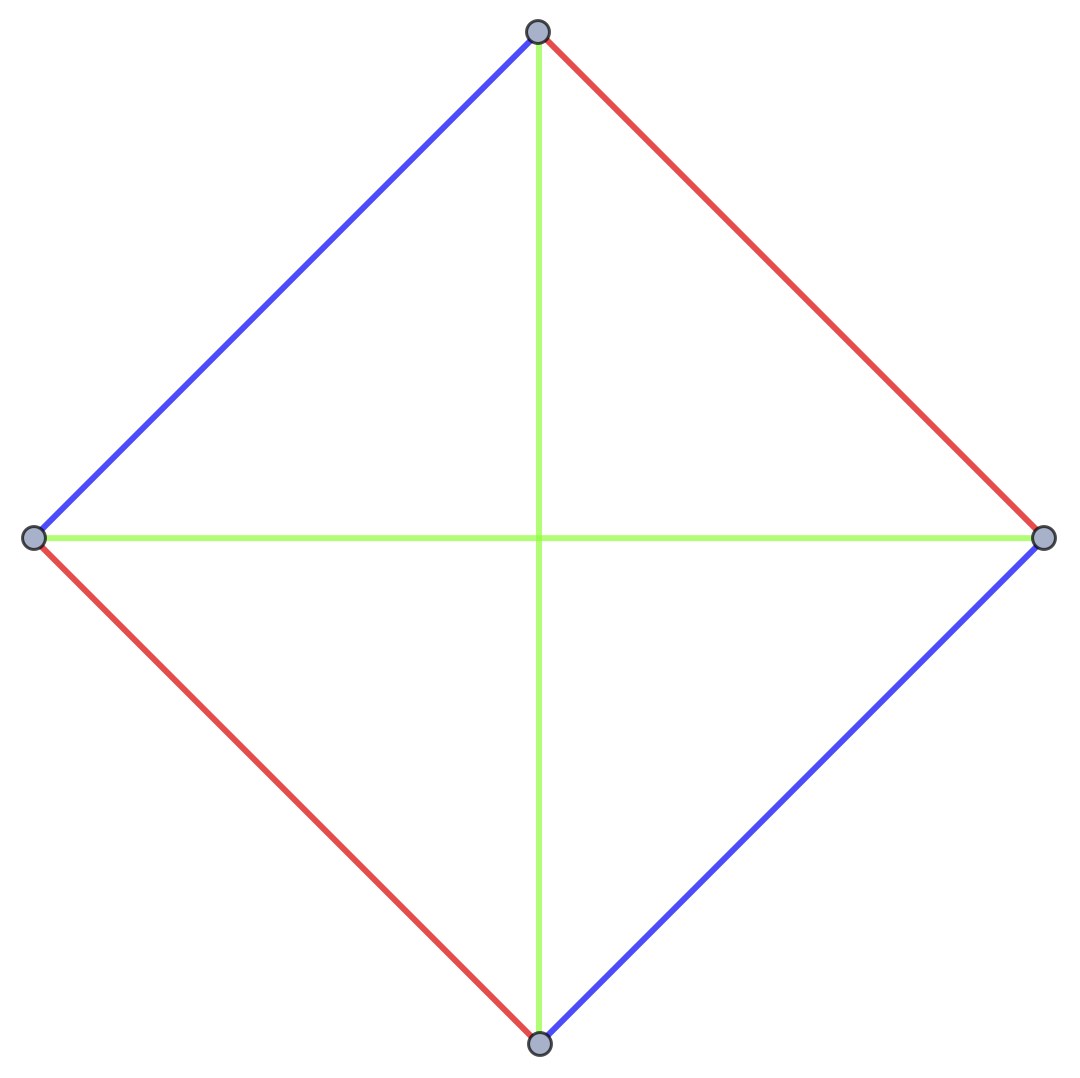}
     \caption{The tetrahedron interaction.}
    \end{subfigure}
    \caption{Representatives of all $r=3$, $q=4$ interaction vertices are pictured above. The first interaction on the left is not single-trace, as it is disconnected. The second interaction, the pillow, is single-trace but not maximally-single-trace, because forgetting the blue edges leaves us with a disconnected interaction graph. The last interaction, the tetrahedron, is maximally-single-trace.  \label{MST-example}}
\end{figure}

We say that an interaction vertex is \textit{single-trace} if it is represented by a connected interaction graph. An interaction is \textit{maximally-single-trace} (MST) if all its two-colour subgraphs are single-trace \cite{Ferrari:2017jgw}. As an example, representatives for all the quartic interaction vertices are pictured in Figure \ref{MST-example}. The tetrahedron vertex is maximally-single-trace; the pillow interaction is single-trace, but not maximally-single-trace; and the double-trace interaction is not single-trace.

\section{Sextic maximally-single-trace interactions}
\label{sexticMSTvertices}

The only quartic maximally-single-trace interaction is the tetrahedron. In this section, we enumerate all the sextic maximally-single-trace interactions and discuss their symmetries. Related discussion for tensor models of maximal rank appears in \cite{Ferrari:2017jgw, KPP}.   

\subsection{Constructing all sextic maximally-single-trace interactions}
\begin{figure}
    \centering
    \includegraphics[width=0.3\textwidth]{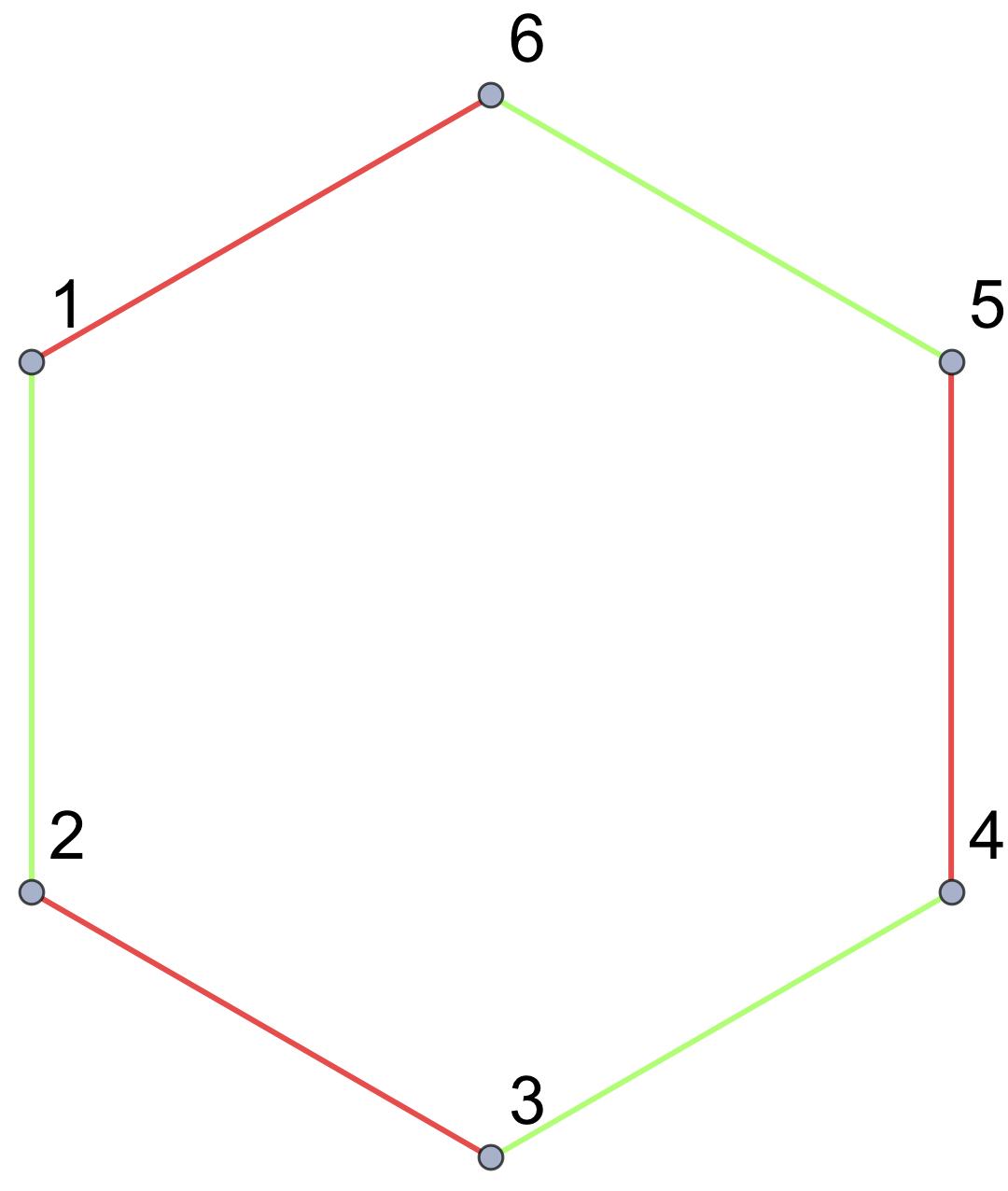}
    \caption{The unique maximally-single-trace interaction vertex for $r=2$ is represented by a cyclic graph. Any two-colour sub-graph of a Maximally-Single-Trace interaction must be a cyclic graph such as this.}
    \label{red-green-cyclic}
\end{figure}

Here we construct all maximally-single-trace sextic vertices for subchromatic tensor models. For $r=2$, there is one MST vertex, the usual single-trace interaction. Note that this must take form of a connected cyclic graph with edges of alternating colours, which we take to be red and green, as shown in Figure \ref{red-green-cyclic}.

\begin{figure}
    \centering
    \begin{subfigure}{0.3\textwidth}
      \centering
    \includegraphics[height=2in]{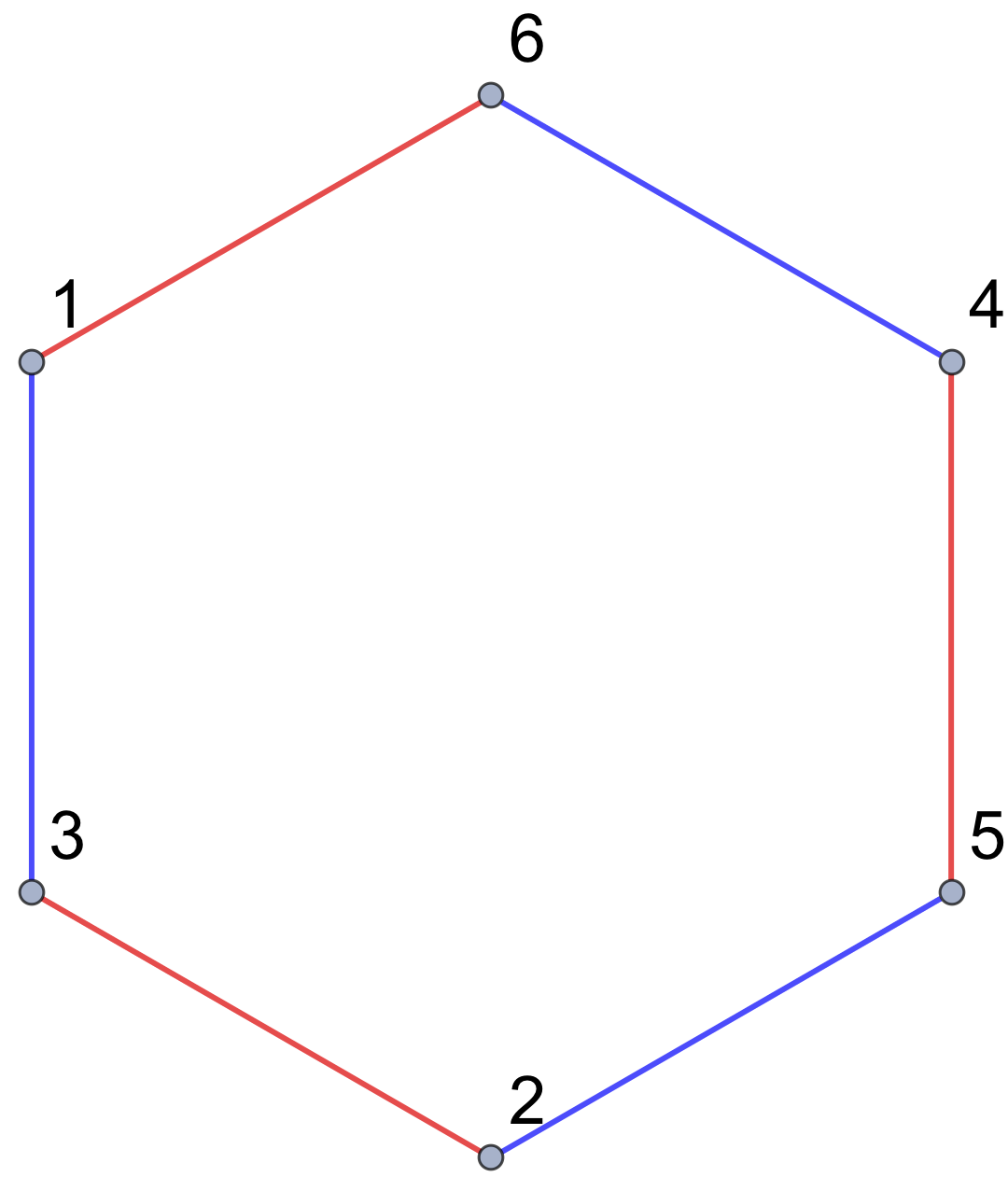}
    \end{subfigure}
    \begin{subfigure}{0.3\textwidth}
     \centering
    \includegraphics[height=2in]{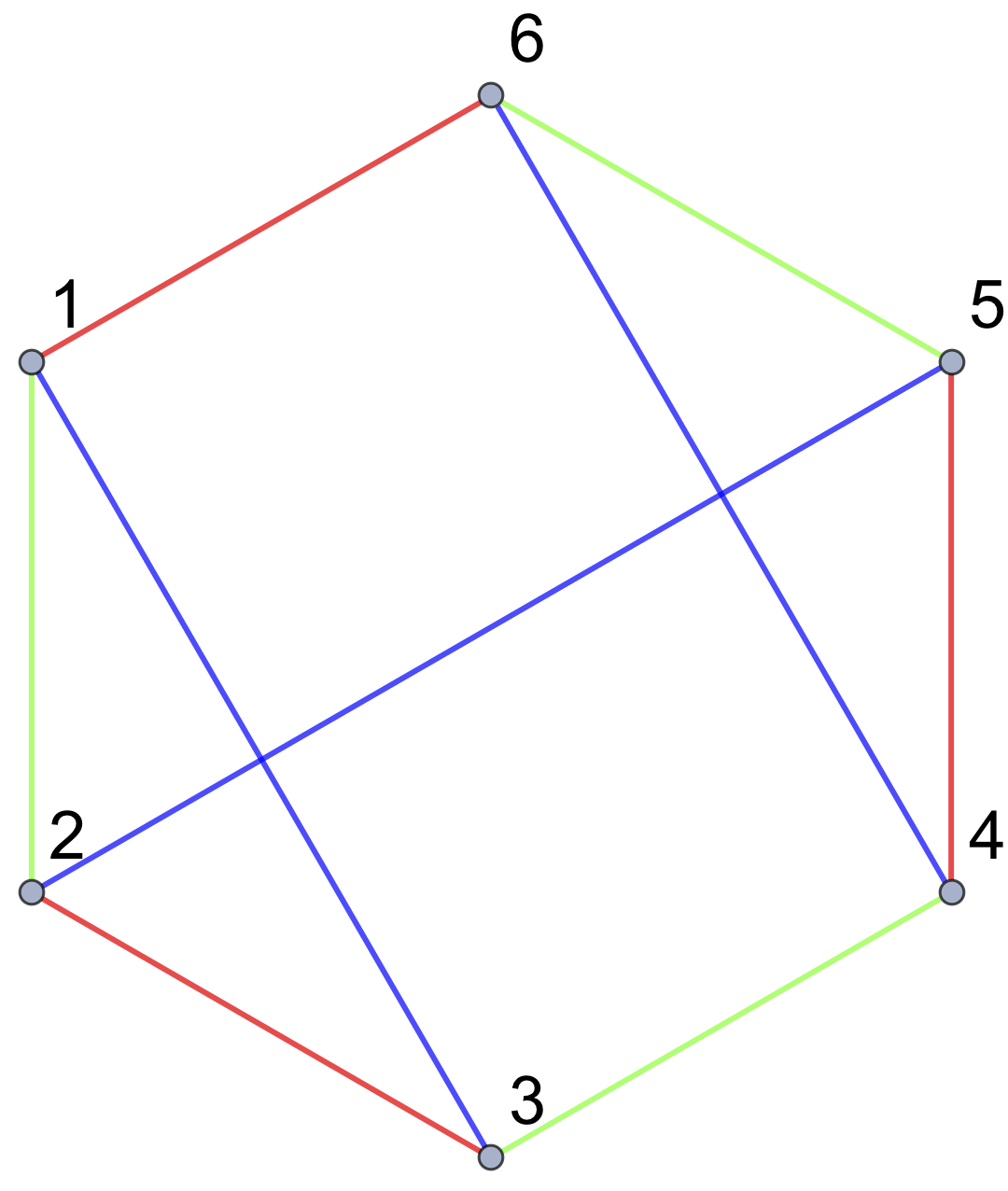}
    \end{subfigure}
    
    \begin{subfigure}{0.4\textwidth}
     \centering
    \includegraphics[height=2in]{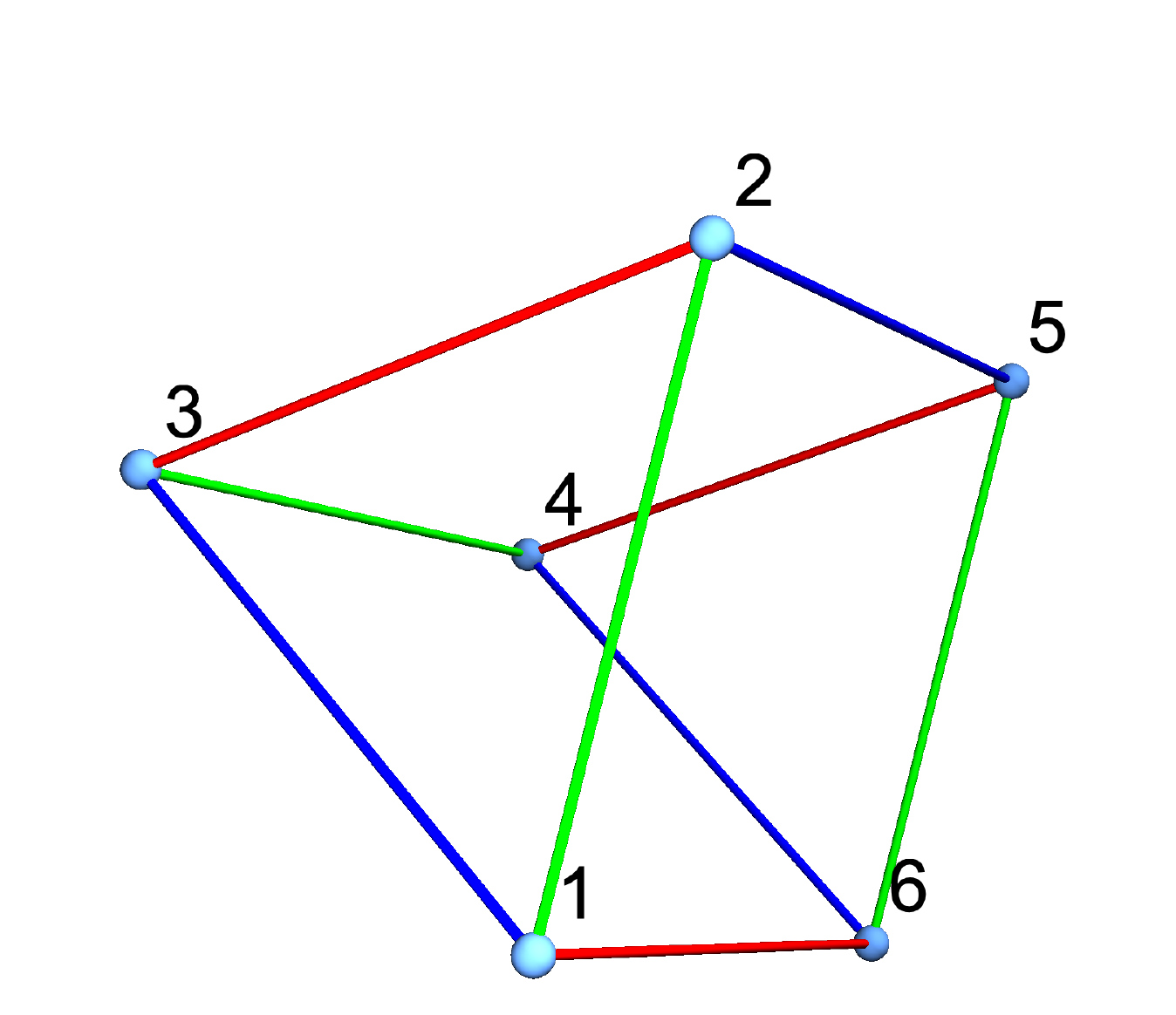}
    \end{subfigure}
    \caption{The \textit{prism} interaction vertex, (above-right) is a maximally-single-trace $r=3$ interaction that can be obtained from combining the red-green cycle of Figure \ref{red-green-cyclic} with the blue-green cycle pictured on the above left. This graph corresponds to the skeleton graph of a triangular prism, as shown below.}
    \label{prism-figure}
\end{figure}

\begin{figure}
    \centering
    \begin{subfigure}{0.3\textwidth}
      \centering
    \includegraphics[height=2in]{red-blue-wheel-cycle.png}
    \end{subfigure}
    \begin{subfigure}{0.3\textwidth}
     \centering
    \includegraphics[height=2in]{wheel.png}
    \end{subfigure}
    
    \begin{subfigure}{0.3\textwidth}
     \centering
    \includegraphics[height=2in]{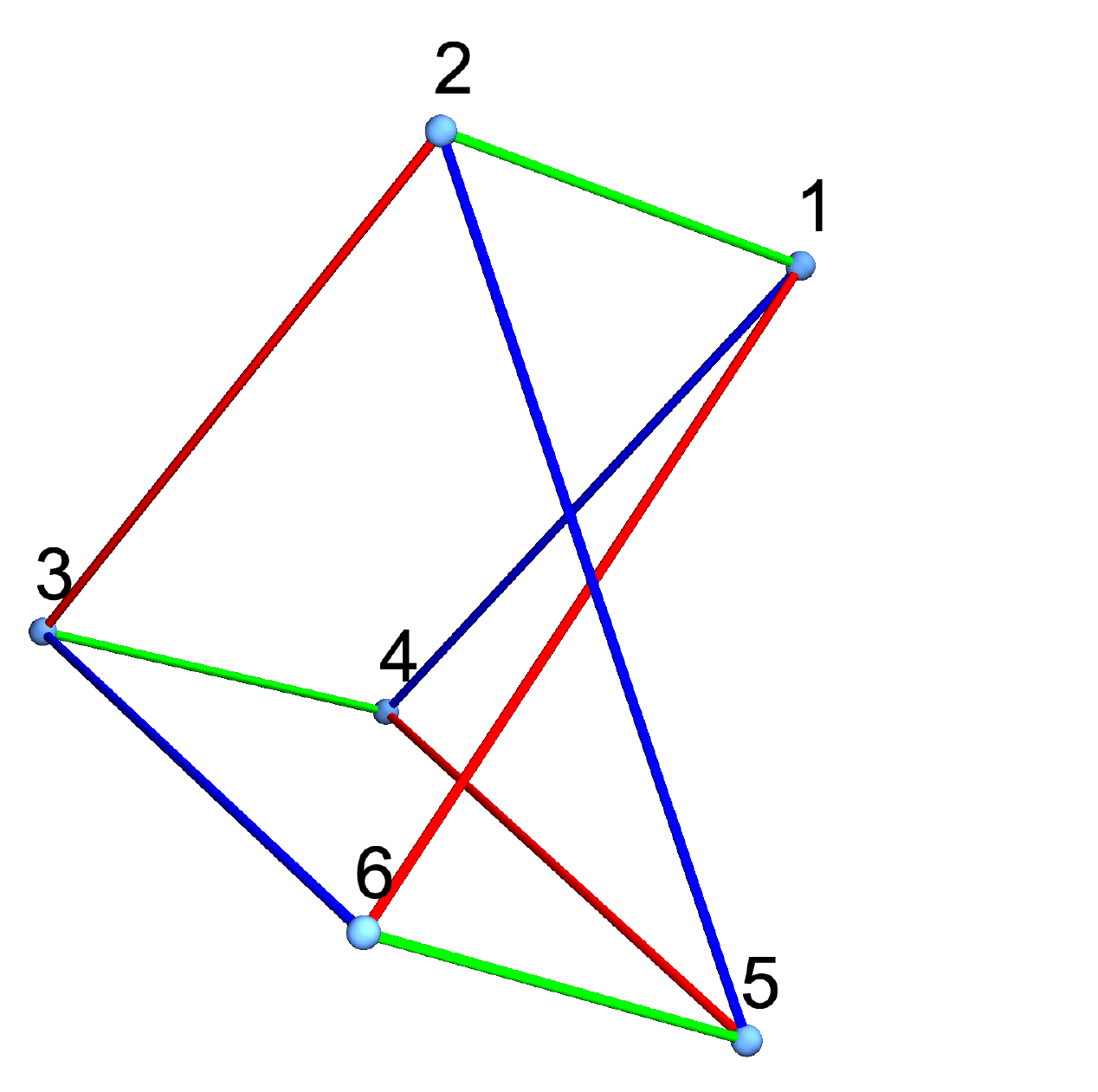}
    \end{subfigure}
    
    \caption{The \textit{wheel} or $K_{3,3}$ interaction vertex, shown on the above right, is a maximally-single-trace $r=3$ interaction that can be obtained from combining the red-green cycle of Figure \ref{red-green-cyclic} with the blue-red cycle pictured on the above left. The graph is the simplest non-planar graph, and can also be drawn as a 3-rung M\"{o}bius ladder, as shown below.}
    \label{wheel}
\end{figure}

Let us now consider the $r=3$ MST interactions. Note that upon forgetting one colour from an $r=3$ MST interaction graph, we are left with the $r=2$ cyclic graph. To construct an $r=3$ MST interaction, we need to add three blue edges to the red-green cyclic graph of Figure \ref{red-green-cyclic}, such that the red-blue and green-blue subgraphs are also cyclic graphs. Note that, if we use a blue edge to connect two vertices that were already connected by a green edge, then the blue-green subgraph will consist of two or more disconnected components -- hence there are not very many possibilities to consider for the locations of the blue edges. One can explicitly check all possibilities to see that there are exactly two ways to add blue edges which result in an MST interaction -- these correspond to the \textit{prism} and the \textit{wheel} of \cite{GKPPT} shown in Figures \ref{prism-figure} and \ref{wheel}. 

The wheel interaction graph is also known as $K_{3,3}$ (the complete bipartite graph consisting of two sets of 3 vertices) in the graph theory literature, which is one of the two simplest non-planar graphs. 

\noindent
We will refer to the three colours of the $r=3$ interactions as $(r,g,b)$.

\begin{figure}
    \centering
    \begin{subfigure}{0.3\textwidth}
      \centering
    \includegraphics[height=2in]{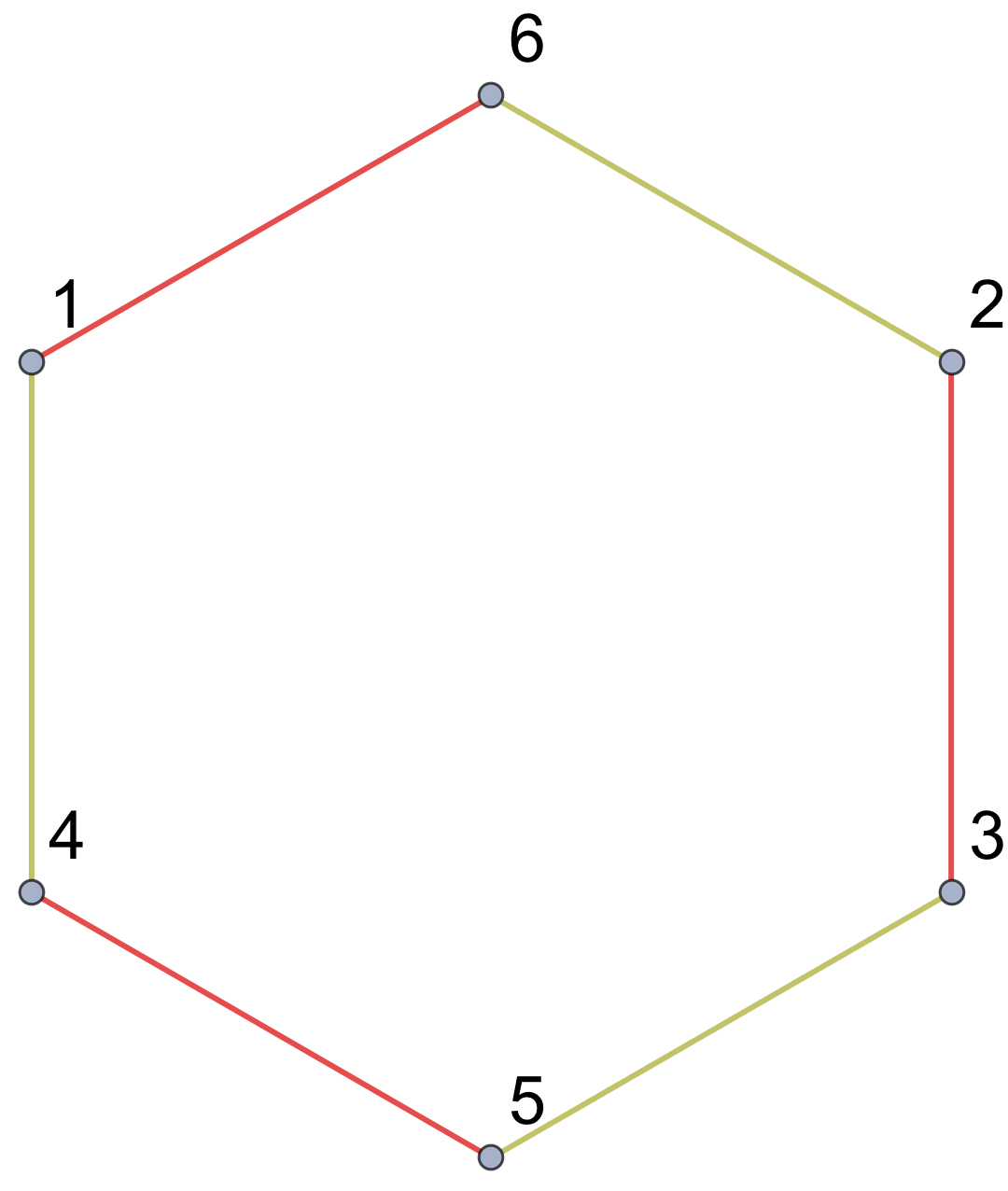}
    \end{subfigure}
    \begin{subfigure}{0.3\textwidth}
     \centering
    \includegraphics[height=2in]{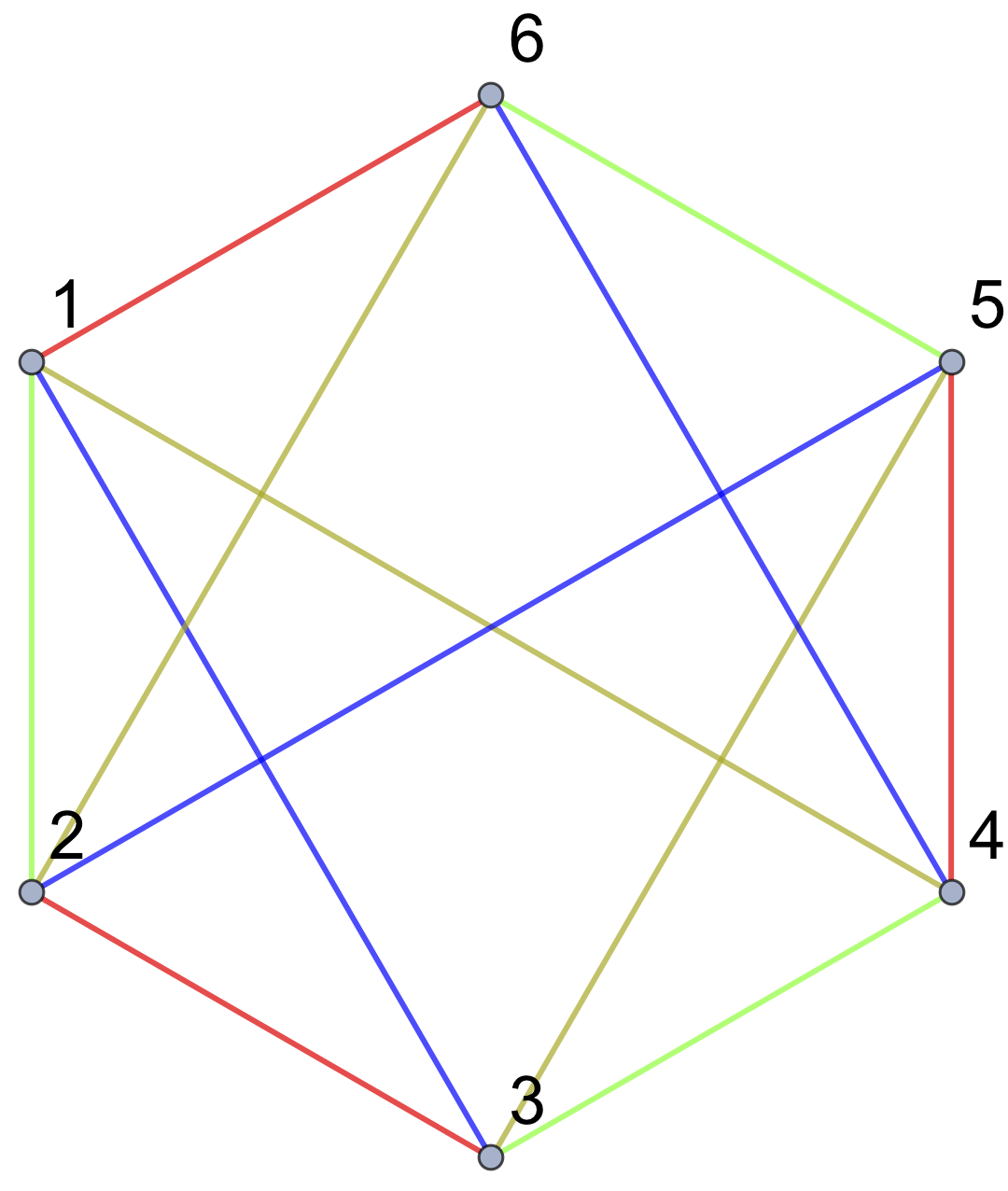}
    \end{subfigure}
    
    \begin{subfigure}{0.4\textwidth}
     \centering
    \includegraphics[height=2in]{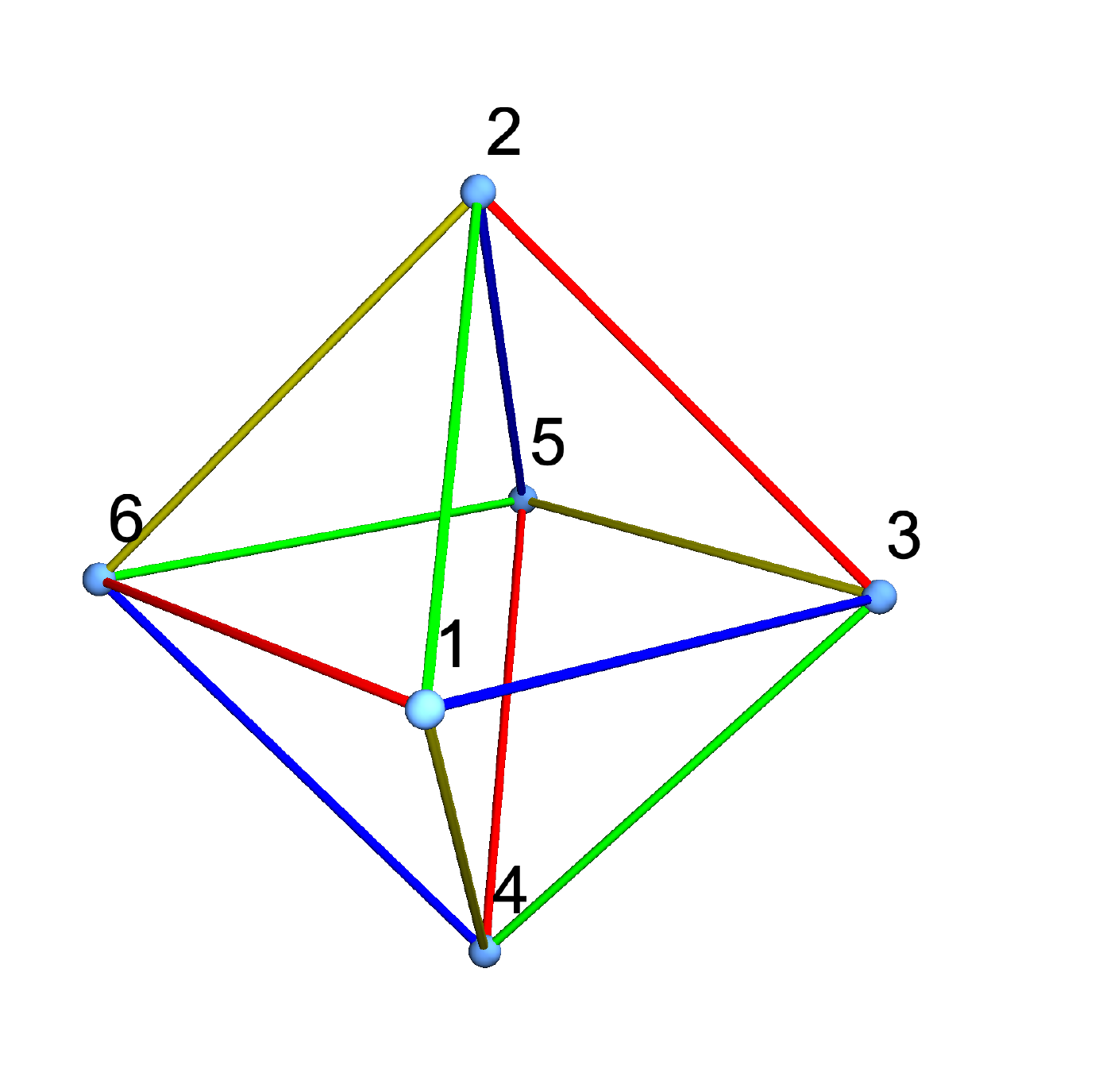}
    \end{subfigure}
    \caption{The \textit{octahedron} interaction vertex (above-right) is the unique maximally-single-trace $r=4$ interaction. It can be obtained from combining the $r=3$ prism interaction of Figure \ref{prism-figure} with the yellow-red cycle pictured on the above-left. The graph can be redrawn as the vertices of a regular octahedron shown below.}
    \label{doublePrism}
\end{figure}

Let us now consider the case of $r=4$. When any one colour is forgotten, the $r=4$ MST interaction must reduce to a prism or wheel. As before, we consider all ways of adding (three) yellow edges to the prism or the wheel, such that the resulting interaction is MST. We find that there is no way of adding yellow edges to the wheel interaction while preserving the MST property; and there is exactly one way to add yellow edges to the prism interaction that preserves MST. Hence there is a unique $r=4$ MST vertex, depicted in Figure \ref{doublePrism}. We will refer to the four colours of the $r=4$ interaction as $(r,g,b,y)$. If one redraws this graph in three-dimensions, one can see that it corresponds to a regular octahedron.\footnote{We thank Igor Klebanov and Martin Ro\v{c}ek for pointing this out to us. Another name  we used for this interaction is the \textit{double-prism}, as both the $rgb$ and $rgy$ subgraphs of this interaction are prism interactions.} Hence we refer to this as the \textit{octahedron} interaction.  

It turns out that one can add a colour to the $r=4$ MST vertex while preserving the MST property. This gives rise to the unique $r=5$ MST vertex. This interaction contains the maximal number of colours for a sextic vertex, and gives rise to a traditionally melonic large $N$ limit, as discussed in \cite{Ferrari:2017jgw, KPP}. 

This recursive procedure of adding colours to subchromatic graphs can be repeated to enumerate all MST interactions for larger values of $q$ as well.

\subsection{Automorphism and colour permutation symmetry}
\label{discrete}
All of the three MST interactions identified above have a discrete symmetry -- they are symmetric under permutation of the $O(N)$ symmetry groups. In terms of the graphical representation of interactions, this is a discrete symmetry under permutation of the various colours in an interaction graph. The interactions may also posses a non-trivial automorphism symmetry group.

We represent symmetry operations by permutations of the field-vertices in the labelled interaction graph. If a field permutation $\hat{\sigma}$ gives rise to a labelled interaction graph isomorphic to the original, $\hat{\sigma}$ is an element of the \textit{automorphism group} of the interaction vertex.  If a field permutation $\hat{\sigma}$ gives rise to a labelled interaction graph isomorphic to the original graph, up to a permutation of colours, then $\hat{\sigma}$ is an element of the \textit{colour permutation symmetry group}. 

In order to show that an interaction is symmetric under all colour permutations, we require that, for each permutation of colours (i.e., relabelling of edges), there is a permutation of field-labels that leaves the labelled interaction graph unchanged. For rank-three interactions, we need to check that all colour permutations which are generated by the two generators $\sigma_{rg}=(r,g)$ and $\sigma_{gb}=(g,b)$, correspond to field-label permutations. For the rank-four interaction, the colour permutation group has 3 generators: $\sigma_{rg}$, $\sigma_{gb}$, and $\sigma_{by}$, and we need to verify that each of these generators corresponds to a permutation of field-labels.

\begin{figure}
  \centering
    \begin{subfigure}{0.3\textwidth}
     \centering
    \includegraphics[height=2in]{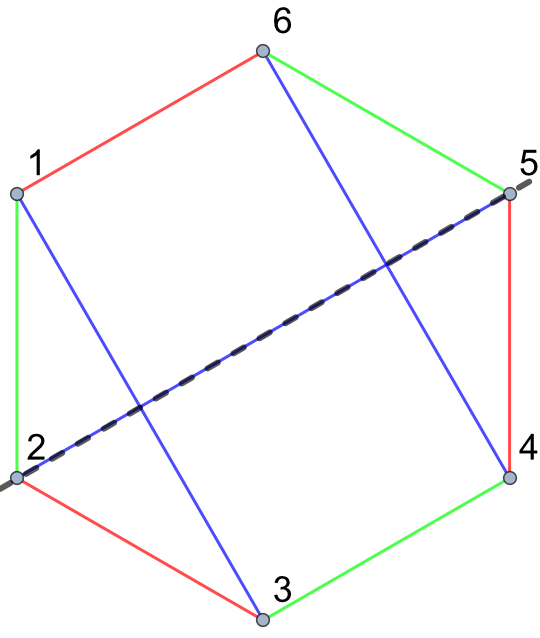}
    \end{subfigure}
     \begin{subfigure}{0.3\textwidth}
      \centering
    \includegraphics[height=2in]{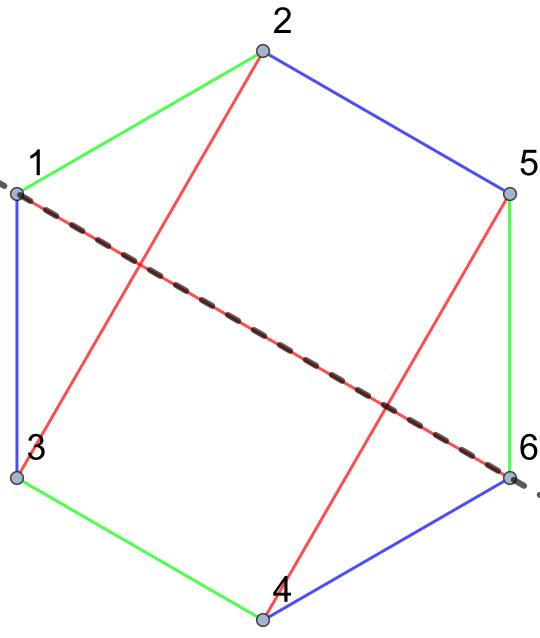}
    \end{subfigure}
    \caption{Two ways of drawing the prism interaction graph that make its colour permutation symmetry manifest. Reflection across the dashed line interchanges two colours in each Figure. \label{prism1and2}}
\end{figure}

Let us first consider the prism. We draw the prism interaction in two different ways in Figure \ref{prism1and2}, from which we can see that $\sigma_{rg}$ is equivalent to the field-vertex permutation $(6,4)(1,3)$, and $\sigma_{gb}$ is equivalent to the field-vertex permutation $(2,3)(5,4)$. We also see that the field-vertex permutation $(1,6)(2,5)(3,4)$  leaves all colours the same. Hence, the prism interaction should conventionally come with a factor of $1/2$ in the Lagrangian.

\begin{figure}
  \centering
    \begin{subfigure}{0.3\textwidth}
     \centering
    \includegraphics[height=2in]{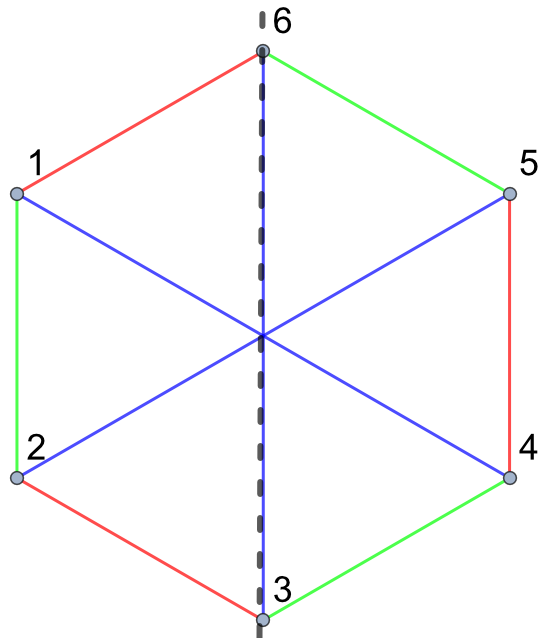}
    \end{subfigure}
     \begin{subfigure}{0.3\textwidth}
      \centering
    \includegraphics[height=2in]{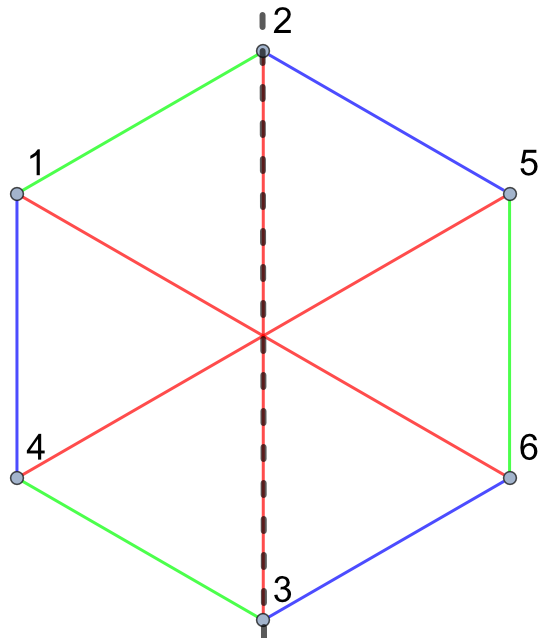}
    \end{subfigure}
    \caption{Two ways of drawing the wheel (or $K_{33}$)  interaction graph. Reflection through the dashed line corresponds to an exchange of two colours.\label{wheel1and2}}
\end{figure}

We next consider the wheel (or $K_{33}$) interaction. We draw the wheel in two different ways in Figure \ref{wheel1and2}, from which we can see that $\sigma_{rg}$ is equivalent to the field-vertex permutation $(1,5)(2,4)$, and $\sigma_{gb}$ is equivalent to the field-vertex permutation $(1,5)(4,6)$. We see that the field-vertex permutations $(1,6)(2,5)(3,4)$ and $(1,2)(4,5)(3,6)$  leave all colours the same. Hence, the wheel (or $K_{33}$)  interaction should conventionally come with a factor of $1/6$ in the Lagrangian.

\begin{figure}
  \centering
    \begin{subfigure}{0.3\textwidth}
     \centering
    \includegraphics[height=2in]{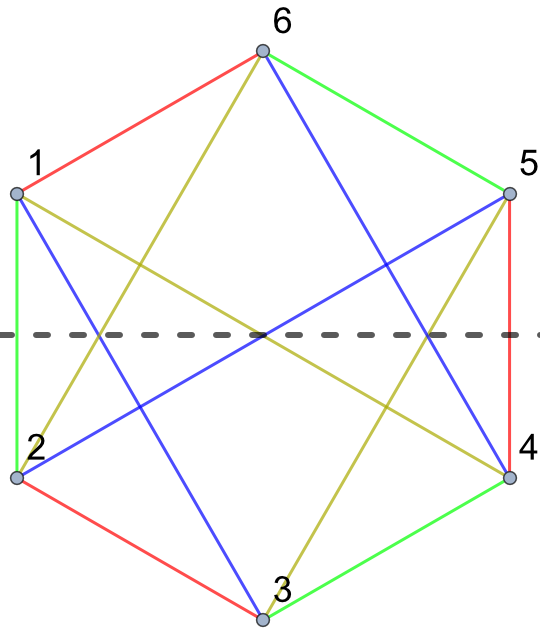}
    \end{subfigure}
     \begin{subfigure}{0.3\textwidth}
      \centering
    \includegraphics[height=2in]{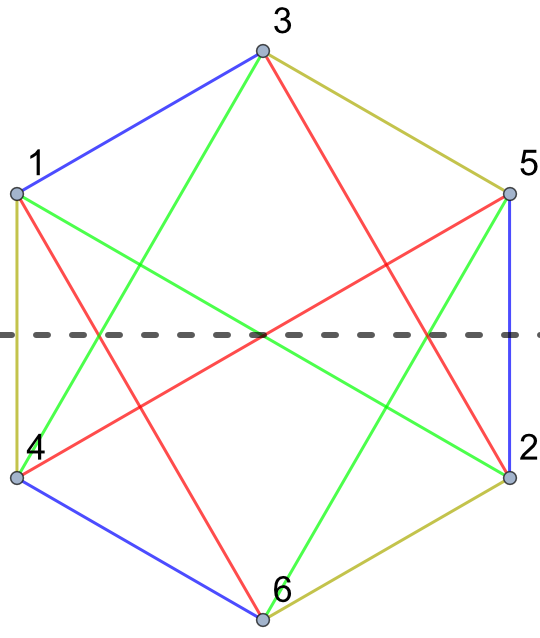}
    \end{subfigure}
     \begin{subfigure}{0.3\textwidth}
      \centering
    \includegraphics[height=2in]{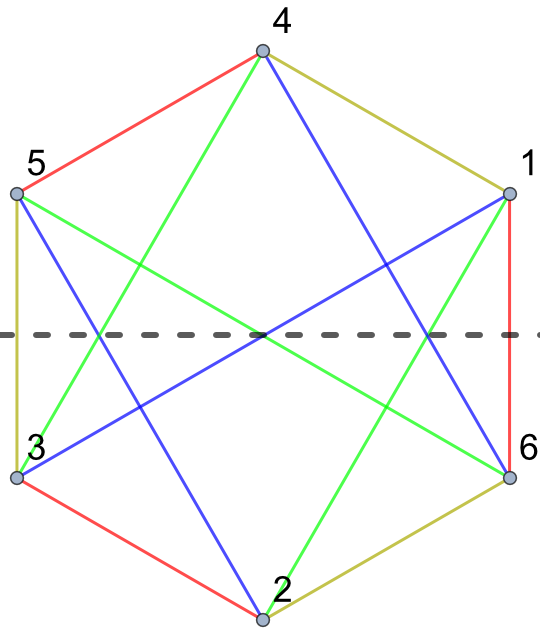}
    \end{subfigure}
    \caption{Three ways of drawing the interaction graph for the $r=4$ octahedron interaction. Reflection through the dashed line corresponds to an exchange of two colours.\label{doublePrism12and3}}
\end{figure}

Finally, we consider the octahedron. We draw the octahedron in three different ways in Figure \ref{doublePrism12and3}, from which we can see that $\sigma_{rg}$ is equivalent to the field-vertex permutation $(1,4)(2,5)(3,6)$, $\sigma_{gb}=(1,6)(2,4)(3,5)$, and $\sigma_{by}=(1,2)(3,6)(4,5)$. There is no field-vertex permutation that leaves all colours the same, so the interaction comes with a factor of $1$ in the Lagrangian.

Let us conclude this section with the observation that, if the automorphism symmetry group includes a field-vertex permutation that is odd, then, in a theory of Majorana fermions, the 1-dimensional fermionic interaction term based on that interaction vanishes  due to anti-commutation of Grassmann variables.

In particular, as observed in \cite{Bulycheva:2017ilt}, for the wheel:
\begin{eqnarray}
L_\text{wheel} & = & \frac{g_{\text{wheel}} }{6}\int dt ~\psi^{a_1 b_1 c_1} \psi^{a_1 b_2 c_2}\psi^{a_2 b_2 c_3} \psi^{a_2 b_3 c_1} \psi^{a_3 b_3 c_2}\psi^{a_3 b_1 c_3} \\
& = &  -\frac{g_{\text{wheel}} }{6} \int dt~ \psi^{a_3 b_1 c_3} \psi^{a_3 b_3 c_2} \psi^{a_2 b_3 c_1} \psi^{a_2 b_2 c_3}\psi^{a_1 b_2 c_2}\psi^{a_1 b_1 c_1} \\
& = &  -L_\text{wheel}.
\end{eqnarray}
In the second line, we applied the field permutation $(1,6)(2,5)(3,4)$. In the third line, because this is an automorphism, we are able to relabel the dummy indices $a_i$, $b_i$ and $c_i$ (via: $a_3 \leftrightarrow a_1$, $b_2 \leftrightarrow b_3$) to undo this permutation. Similarly, for the prism:
\begin{eqnarray}
L_\text{prism} & = & \frac{g_{\text{prism}} }{2}\int dt ~\psi^{a_1 b_1 c_1} \psi^{a_1 b_2 c_2}\psi^{a_2 b_2 c_1} \psi^{a_2 b_3 c_3} \psi^{a_3 b_3 c_2}\psi^{a_3 b_1 c_3} \\
& = &  -\frac{g_{\text{prism}} }{2} \int dt~ \psi^{a_3 b_1 c_3} \psi^{a_3 b_3 c_2} \psi^{a_2 b_3 c_3} \psi^{a_2 b_2 c_1}\psi^{a_1 b_2 c_2}\psi^{a_1 b_1 c_1} \\
& = &  -L_\text{prism}.
\end{eqnarray}
These arguments do not apply to complex fermions.\footnote{We thank Igor Klebanov for discussions on this point.}

Let us also remark that, if we would like to define an theory with complex fields, and promote all the symmetry groups to $U(N)$, then we require the interaction graph to be bipartite. The wheel is bipartite, but the prism, the octahedron and the unique $r=5$ sextic MST interaction of \cite{Ferrari:2017jgw, KPP} are not bipartite. If we wish to promote some, but not all, of the symmetry groups to $U(N)$, this restriction does not apply.

In all cases, real bosonic versions of these theories can be defined, and can be thought of as special cases of the general sextic bosonic theory studied perturbatively in \cite{Osborn:2017ucf}. 

\section{Large $N$ limit of subchromatic maximally-single-trace interactions}
\label{subchromatic-general}

We now consider the large $N$ limit of the maximally-single-trace interaction vertices defined in the previous section. In this section we present results for an arbitrary maximally-single-trace interaction of any order $q$. We follow the approach outlined in \cite{Klebanov:2016xxf}. The results of this section are a special case of a more general analysis of the large $N$ limit of maximally-single-trace interactions given in \cite{Ferrari:2017jgw}. 

\subsection{Large $N$ scaling of coupling constants}
Let us first determine the natural scaling of the coupling constant with $N$, in the large $N$ limit, for an order-$q$ maximally-single-trace interaction with $r$ indices. 

Consider a connected Feynman diagram with no external edges, i.e., one that contributes to the free energy of our theory. As shown in the previous sections, Feynman diagrams for fields based on rank-$r$ tensors can be drawn in a multi-line notation, using $r$ different colours. If we take any such diagram, and erase all but two of the colours, we obtain a \textit{two-colour fat graph}, or simply \textit{fat graph}. Since our multi-line propagator contains $r$ colours, we can generate $r(r-1)/2$ fat graphs -- one for each pair of colours. We shall use the index $\alpha$, where $\alpha=1,\dots, r(r-1)/2$, to label each of these fat graphs. We denote the number of loops for each fat graph by $f_\alpha$. Then, summing over all such pairs of indices, we obtain,
\begin{equation}
    \sum_{\a=1}^{r(r-1)/2}f_\alpha = (r-1)f_{total},
\end{equation}
where $f_{total}$ is the total number of loops in the multi-line Feynman diagram and determines the power of $N$.

Because our interaction vertices are maximally-single-trace, each fat graph will consist of a single connected component. The Euler equation for each $\a$ can be written as,
\begin{align}
&f_\a + v - e = \chi_\a = 2-2g_\a 
\end{align}
where $v$ and $e$ refer to the number of vertices and edges in the graph (which are the same for all fat graphs) and $g_\a$ is the genus of the fat graph labelled by $\a$. We define \textit{maximal} Feynman diagrams to be those with the largest $f_{total}$ for a given $v$. From the above formula, we see that maximal Feynman diagrams satisfy $g_\a=0$, i.e., maximal Feynman diagrams are those diagrams that give rise to only planar fat graphs.

Since our vertices are order-$q$, we have $2e=qv$. Placing this into the above equation, we obtain,
\begin{align}
    &f_\a + v\bigg(1-\f{q}2\bigg)=2-2g_\a\nn\\
    &f_\a + v\bigg(1-\f{q}2\bigg)\leq 2
\end{align}
Now, summing over all $\a$, we get
\begin{align}
    \sum_{\a=1}^{r(r-1)/2}f_\a + \sum_{\a=1}^{r(r-1)/2}v\bigg(1-\f{q}2\bigg) & \leq 2\sum_{\a=1}^{r(r-1)/2} \\
    (r-1)f_{tot} + \f{r(r-1)}2 v\bigg(1-\f{q}2\bigg) & \leq r(r-1)\nn
\end{align}
which imply,
\begin{equation}
      f_{tot}  \leq \f{r}4(q-2)v+r.
\end{equation}
\noindent
Maximal Feynman diagrams saturate the above bound, and must satisfy: 
\begin{equation}
    f_{tot}= \f{r}4(q-2)v+r.
    \label{ftot}
\end{equation}

This relation tells us that we should define the large $N$ limit while keeping the 't Hooft coupling 
\begin{equation}
    \l=g N^{r(q-2)/4}, \label{tHooft}
\end{equation}
fixed. (This is essentially the same as Equation (3.37) of \cite{Ferrari:2017jgw}.) Then, the free energy scales with $N$ as,
\begin{equation}
    N^r f(\l).\hspace{.2cm}
\end{equation}

\subsection{Existence of a loop passing through one or two vertices}

We now show that any connected maximal diagram contributing to the free energy contains a loop passing through exactly one vertex or a loop passing through exactly two vertices.

Let us define $\mathcal F_s$ to be the number of loops passing through $s$ vertices \cite{Klebanov:2016xxf}. Clearly,
\begin{equation}
    \sum_{s=0}^\infty \mathcal F_s = f_{total}
    \label{sum1}
\end{equation}
Also, by considering the total number of coloured lines passing through a vertex in the multi-line notation, one can see that:
\begin{equation}
    \sum_{s=1}^\infty s \mathcal F_s = r q \f{v}2
    \label{sum2}.
\end{equation}
Combining \eqref{sum1} and \eqref{sum2}, and eliminating $f_{total}$ using \eqref{ftot}, we obtain
\begin{equation}
    \sum_s\bigg(\f{2q}{(q-2)} -s \bigg)\mathcal F{}_s =\f{2q}{(q-2)}r.
\end{equation}
When $q\geq 6$, one can separate the terms $\mathcal F_1$ and $\mathcal F_2$ from the sum, to obtain the following inequality:
\begin{equation}
    (q+2)\mathcal F_1 + 4 \mathcal F_2 \geq 2q r.
    \label{f1-f2}
\end{equation}
The above analysis assumed that the number of interaction vertices was greater than or equal to $1$. The trivial free-energy diagram which contains no interaction vertices, is also maximal. Thus, we have the following result:
\begin{theorem}
Any maximal, non-trivial, multi-line Feynman diagram contributing to the free energy contains at least one loop passing through one vertex, or one loop passing through two  vertices.\label{loop-theorem}
\end{theorem}

\subsection{1-cycles and 2-cycles}
In our subsequent analysis, it will be convenient to distinguish between single-line Feynman diagrams and multi-line Feynman diagrams. Each single-line Feynman diagram (constructed from labelled interaction vertices) corresponds to a multi-line Feynman diagram and vice-versa. (There may be multiple ways of labeling the interaction vertices in the single-line diagram corresponding to a given multi-line diagram, due to the automorphism symmetry of each interaction, but this does not affect our discussion.)

A single line Feynman diagram contributing to the free energy is a \textit{graph}, i.e. a collection of vertices and edges. A \textit{walk} of length $s$ in a graph is defined to be an ordered sequence of $s$ edges $(e_1,~e_2,\ldots,~e_s)$, that joins a sequence of vertices $(v_0,~v_1,~v_2,\ldots,~v_s)$. A \textit{closed walk} is a walk that starts and ends on the same vertex, i.e., $v_0=v_s$. Let us define an \textit{$s$-cycle} as a closed walk of length $s$, such that no edge or vertex is repeated, other than the starting vertex which is the same as the end vertex. This is the usual notion of a cycle in graph theory. 

In the previous subsection, we defined $\mathcal F_s$ as the number of loops passing through $s$ vertices in a multi-line Feynman diagram. Any loop passing through $s$ vertices in a multi-line Feynman diagram induces a closed walk of length $s$, with distinct edges, in the corresponding single-line Feynman diagram.  (Of course, the converse is not be true -- a closed walk in the single line Feynman diagram need not correspond to a loop in the multi-line Feynman diagram.) Note that a closed walk of length $1$ is necessarily a $1$-cycle. Note also that a closed walk of length $2$ with distinct edges is either a 2-cycle, or passes through the same vertex twice, in which case it is the union of two (possibly identical) 1-cycles.
 
Theorem \ref{loop-theorem} of the previous subsection therefore implies the following theorem.
\begin{theorem}Any maximal, non-trivial, single-line Feynman diagram contributing to the free energy must contain at least one 1-cycle, or at least one 2-cycle. \label{cycle-theorem}
\end{theorem}
Theorem \ref{cycle-theorem} is slightly weaker than Theorem \ref{loop-theorem} of the previous subsection, but it will be more convenient to apply Theorem \ref{cycle-theorem} in the subsequent analysis.

While Feynman diagrams containing 1-cycles may be unphysical, we include them in the analysis below, in order to obtain purely combinatorial constraints on the diagrams that contribute in the large $N$ limit.

\section{Melonic dominance of subchromatic interaction vertices}
\label{melonic-dominance}
Here, we focus our attention on theories based on the sextic MST interactions obtained in section \ref{sexticMSTvertices}. We can define a theory with $r=3$, to contain a wheel interaction, a prism interaction, or both. We can also define a theory with $r=4$ based on the octahedron interaction. The theory for $r=5$ was discussed in \cite{Ferrari:2017jgw, KPP} so we do not discuss it here. 

\subsection{General strategy}
We wish to explicitly characterize and generate all the maximal Feynman diagrams that contribute to the free energy in any of the above theories. Our arguments are inspired by \cite{Ferrari:2017jgw, Bonzom:2018jfo, KPP}. %melonic proofs citations

\begin{figure}
    \centering

    \begin{subfigure}[b]{0.25 \textwidth}
    \centering
    \includegraphics[width=.5\textwidth]{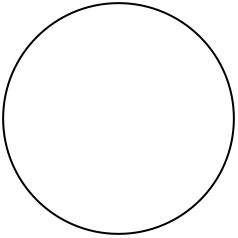}
    \vspace{1cm}
    \caption{\centering The diagram with no vertices.}
    \label{0-cycle}
    \end{subfigure}
    \begin{subfigure}[b]{0.35 \textwidth}
    \centering
    \includegraphics[width=.7\textwidth]{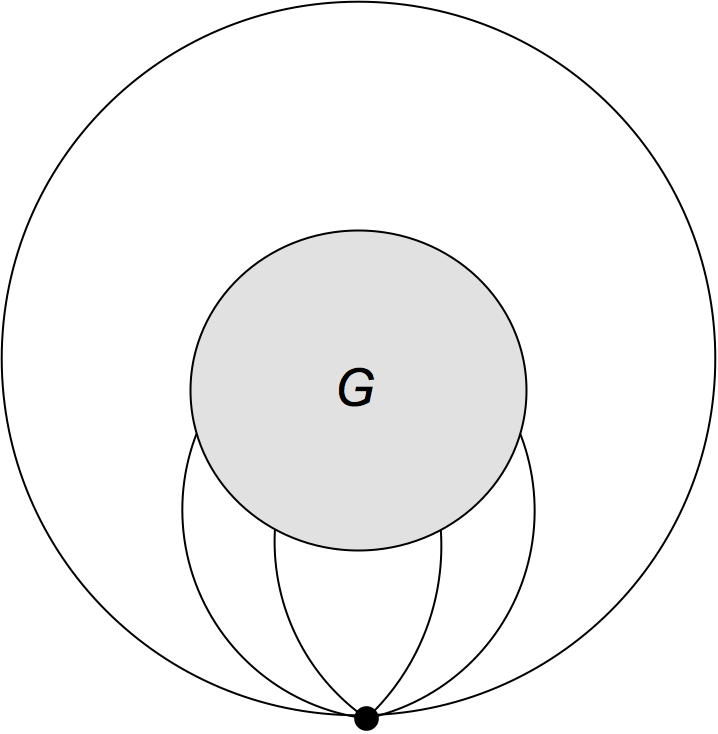}
    \caption{\centering A diagram containing a 1-cycle.}
    \label{1-cycle}
    \end{subfigure}
    \begin{subfigure}[b]{0.35 \textwidth}
    \centering
    \includegraphics[width=.8\textwidth]{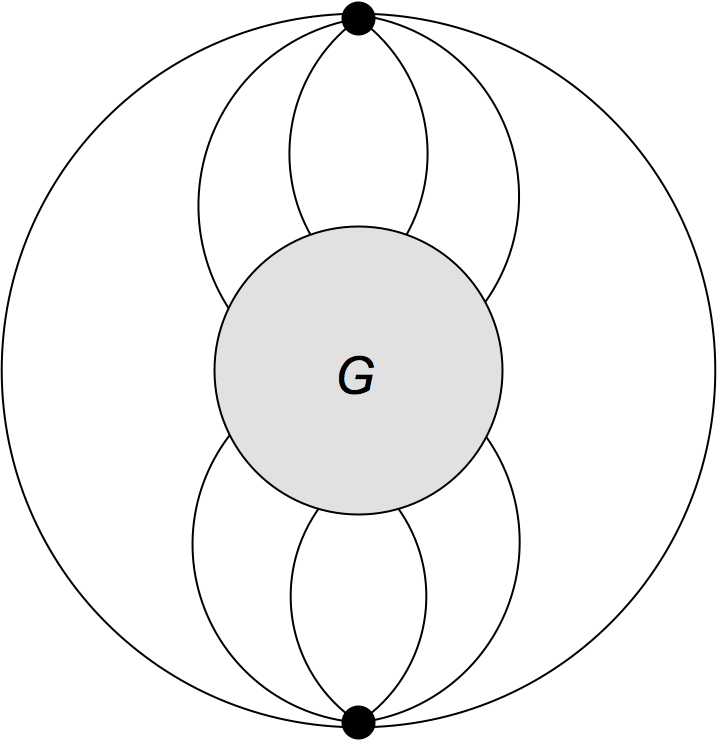}
    \caption{\centering A diagram containing a 2-cycle. }
    \label{2-cycle}
    \end{subfigure}
    \caption{Any maximal Feynman diagram contributing to the free energy must be of one of the three types specified above. To obtain a recursive enumeration of all Feynman diagrams, we must place some constraints on the subgraph $G$.} \label{cases}
\end{figure}

Our strategy is as follows. In the previous section, we demonstrated that any Feynman diagram, drawn in single-line notation, contributing to the free energy that survives in the large $N$ limit must either:
\begin{enumerate}
    \item Contain no vertices, in which case it is just the zeroth-order diagram.
    \item Contain at least one 1-cycle.
    \item Contain at least one 2-cycle.
\end{enumerate}
These cases are illustrated in Figure \ref{cases}. Note also that a 2-cycle cannot pass through the same vertex twice by definition, so we do not need to treat this case separately. (A loop passing through the same vertex twice in the multi-line Feynman diagram would give rise to a 1-cycle in the single-line Feynman diagram.)

To enumerate all the maximal Feynman diagrams contributing to the free energy, we need to enumerate all the inequivalent free energy diagrams containing a 1-cycle or a 2-cycle. (In a slight abuse of terminology, we refer to these as inequivalent 1-cycles or 2-cycles.) For each inequivalent 1-cycle and 2-cycle, we draw each of its $r(r-1)/2$ fat graphs, and impose the restriction that each fat graph is planar. 

%A 1-cycle is specified by the labels of two field-vertices which are contracted to each other $\langle X, Y \rangle$. 

%To specify a 2-cycle, we first introduce two interaction vertices, which we call $L$ and $R$. We then specify two pairs of field vertices that are contracted to each other. Each pair should consist of one field vertex from $L$ and one field vertex from $R$. We can draw a general graph containing a given 1-cycle or 2-cycle as in Figure \ref{1-cycle} and \ref{2-cycle}. 

When we draw a fat graph corresponding to a particular pair of colours, we must arrange the outgoing lines from each interaction vertex in a particular cyclic order, for the fat graph to be manifestly planar. For example, consider the wheel interaction, shown in Figure \ref{wheel-v}. If we draw a blue red fat-graph, we must arrange the outgoing lines of each interaction vertex in the cyclic order $163254$ (or $145236$). If we instead draw a red-green fat-graph, we must arrange the outgoing lines of each interaction vertex in the cyclic order $123456$ (or $654321$).

\begin{figure}
    \centering
    \includegraphics[width=0.3\textwidth]{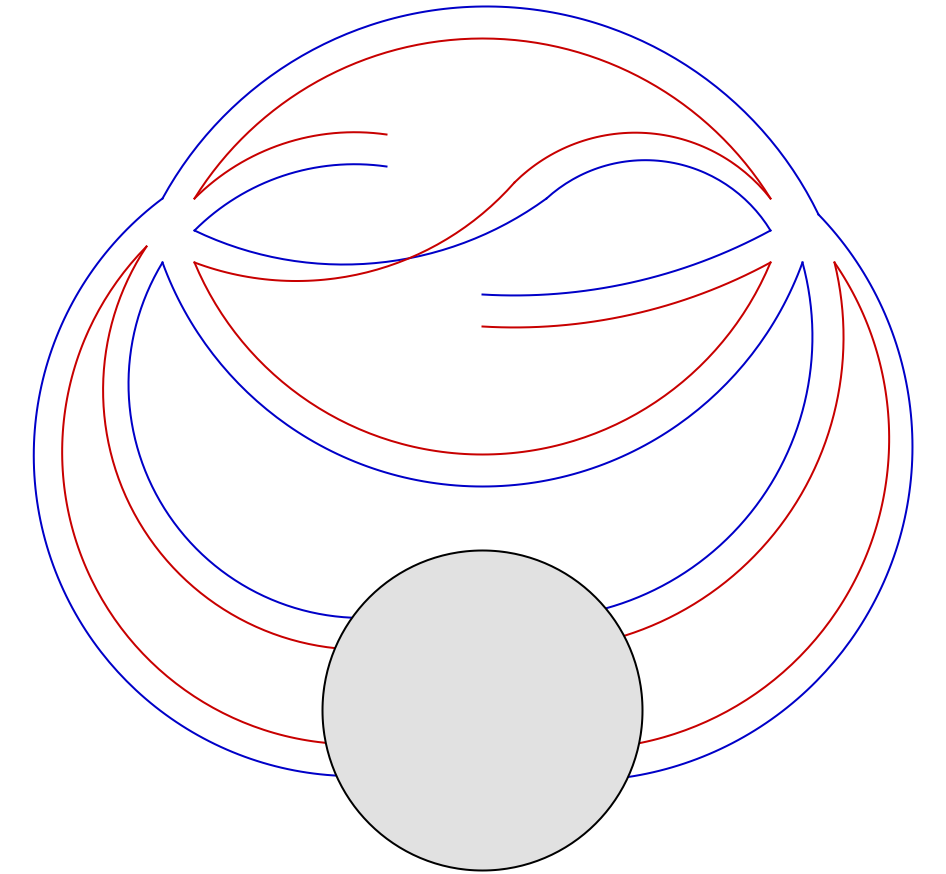}
    \caption{A 2-cycle which contains an odd number of twists and is therefore non-planar.}
    \label{non-planar-2-cycle}
\end{figure}

\begin{figure}
    \centering
    \includegraphics[width=.3\textwidth]{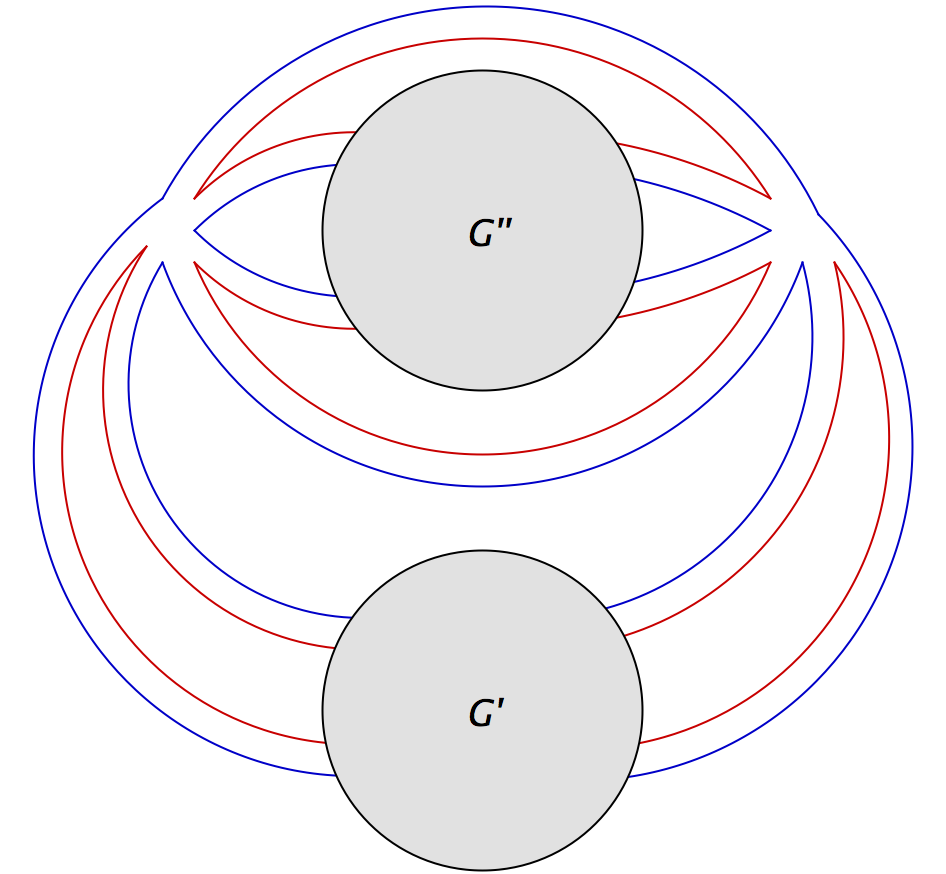}
    \caption{We require each two-colour fat graph to be planar. For a given choice of two-colours, such as red and blue, planarity forces us to arrange the outgoing edges from each interaction vertex in a particular cyclic order. This means that the 2-cycle may split the subgraph $G$ from Figure \ref{2-cycle} into two disconnected pieces, as shown above. If we consider all two-colour fat graphs, we hope to split the graph $G$ into four disconnected pieces as shown in Figure \ref{2-cycle-split}. }
    \label{splitgraph}
\end{figure}

For a given pair of colours, the requirement of fat-graph planarity might rule out a particular 2-cycle entirely, as is the case for the 2-cycle shown in Figure \ref{non-planar-2-cycle} which contains an odd number of twists. Alternatively, the requirement of fat-graph planarity may ``divide'' the subgraph $G$ into two smaller disconnected components $G'$ and $G''$ as shown in Figure \ref{splitgraph}.

\begin{figure}
    \centering
    \begin{subfigure}{.45\textwidth}
    \centering
    \includegraphics[width=.6\textwidth]{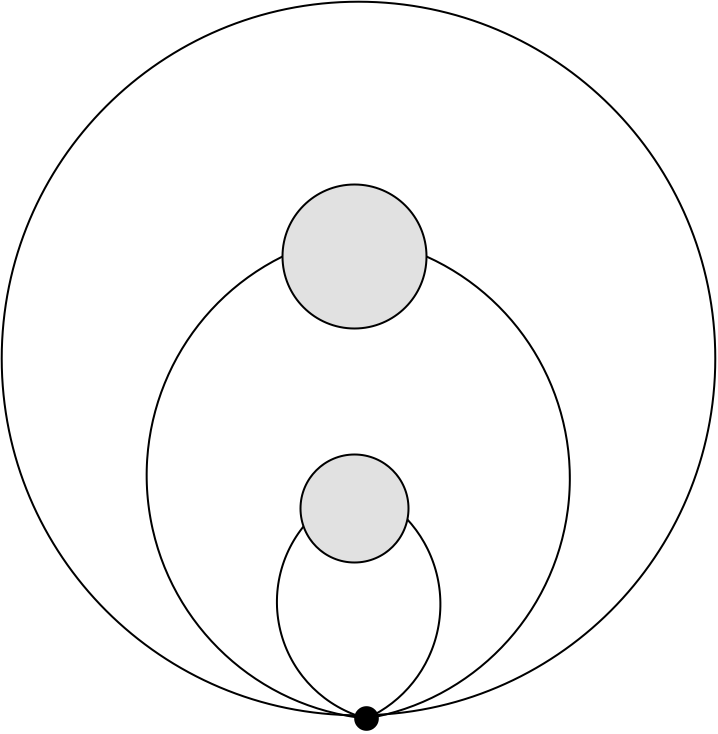}
    \end{subfigure}
    \begin{subfigure}{.45\textwidth}
    \centering
    \includegraphics[width=.6\textwidth]{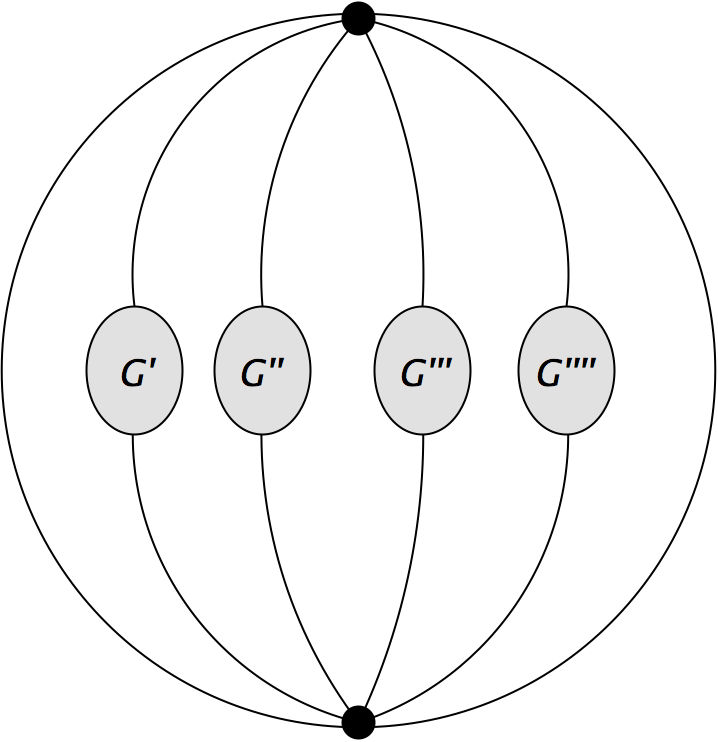}
    \end{subfigure}
    \caption{If, for all possible 1-cycles and 2-cycles, the requirement that all two-colour fat graphs are planar splits up the subgraph $G$ from Figure \ref{1-cycle} and \ref{2-cycle} into disconnected components as shown above, then we can argue the theory is melonic the conventional sense, via the cutting and sewing argument in Figure \ref{cutandsew}.}
    \label{2-cycle-split}
\end{figure}

\begin{figure}
    \centering
    \begin{subfigure}{.8\textwidth}
    \centering
    \includegraphics[width=.8\textwidth]{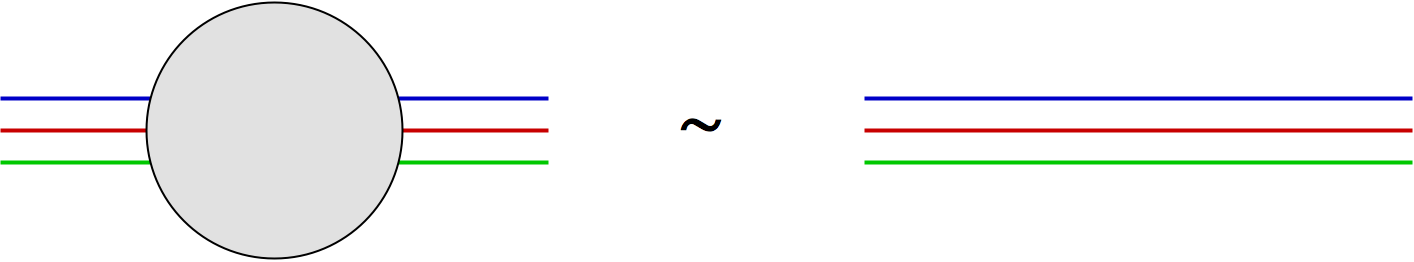}
    \caption{The ``connectivity structure'' of a connected graph with two external edges must be the same as that of a free propagator.}
    \end{subfigure}
     \vspace{1cm}

    \begin{subfigure}{.8\textwidth}
    \centering
    \includegraphics[height=2in]{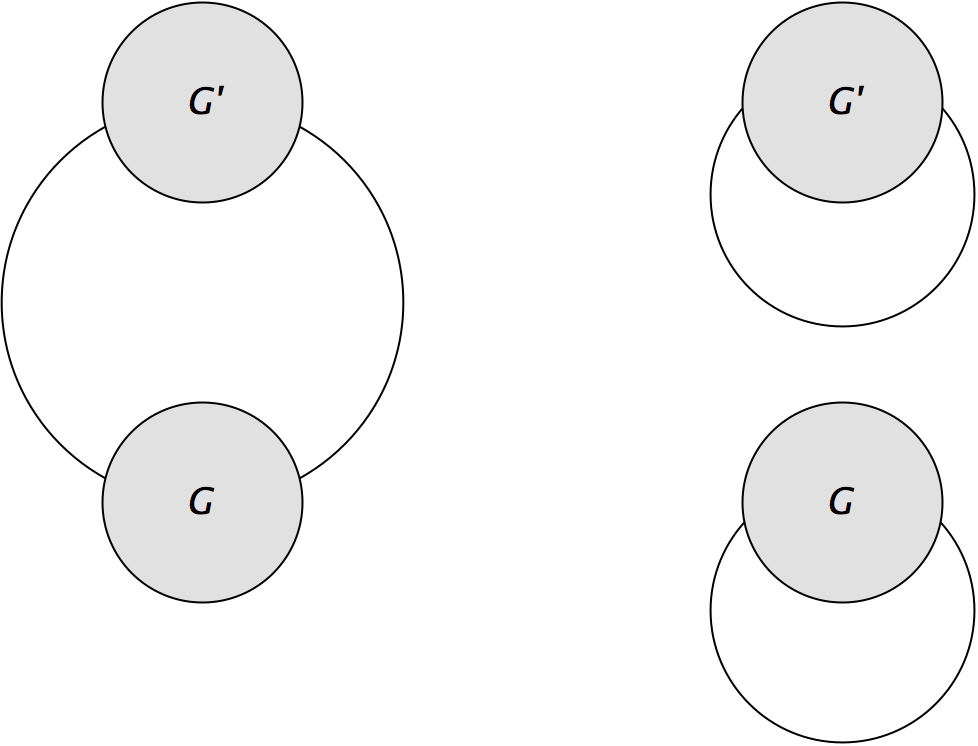}
    \caption{We can argue that the combined graph of $G$ and $G'$ on the left is maximal if and only if both separated components on the right are maximal, since each subgraph graph can be replaced by a free propagator without affecting the $N$ counting.}
    \end{subfigure}
    \caption{The above two Figures illustrate a cutting and sewing argument that can be used to separate the graphs $G'$, $G''$, $G'''$ and $G''''$ from Figure \ref{2-cycle-split}. Either these graphs contain no vertices, or they contain a 1-cycle or 2-cycle. One thus obtains a recursive characterization of all the leading order graphs.}
    \label{cutandsew}
\end{figure}

Suppose, for each inequivalent 1-cycle and 2-cycle consistent with fat-graph planarity, the requirement of fat-graph-planarity splits the subgraphs $G$ of Figure \ref{1-cycle} and \ref{2-cycle} into disconnected components, as shown in Figure \ref{2-cycle-split}. Crucially, each of these sub-graphs in Figure \ref{2-cycle-split} contains only two external edges. This fact allows us to use a cutting and sewing argument to isolate the graphs $G^{(i)}$, as illustrated in Figure \ref{cutandsew}. Each of the cut-and-sewn subgraphs $G^{(i)}$ defines a maximal Feynman diagram contributing to the free energy, with fewer interaction vertices than the original diagram. The arguments of this section apply to the cut-and-sewn subgraph as well, so we obtain a recursive enumeration of all the Feynman diagrams contributing to the free energy.

\subsection{Melonic diagrams}
\label{MelonicSection}

We now discuss the recursive enumeration of the diagrams that contribute in the large $N$ limit, following \cite{Ferrari:2017jgw} with some generalizations.

\begin{figure}[h]
    \centering
    \begin{subfigure}[b]{0.8\textwidth}
    \centering
        \includegraphics[width=0.6\textwidth]{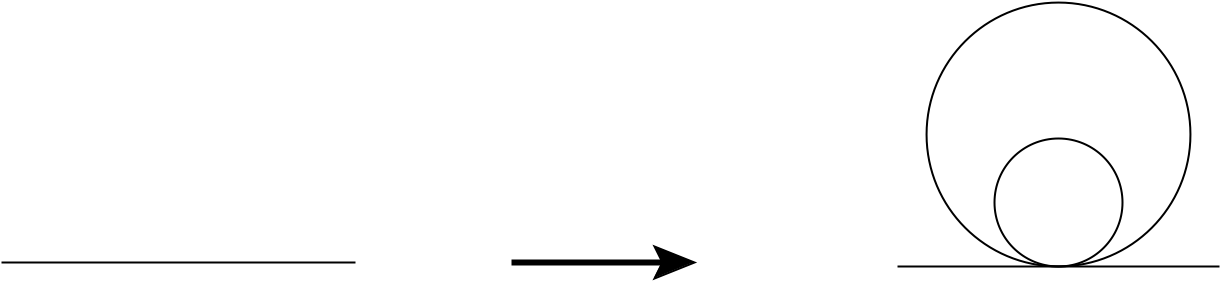}
    \end{subfigure}
    \begin{subfigure}[b]{0.8\textwidth}
    \centering \hspace{1.6cm}
    \includegraphics[width=0.8\textwidth]{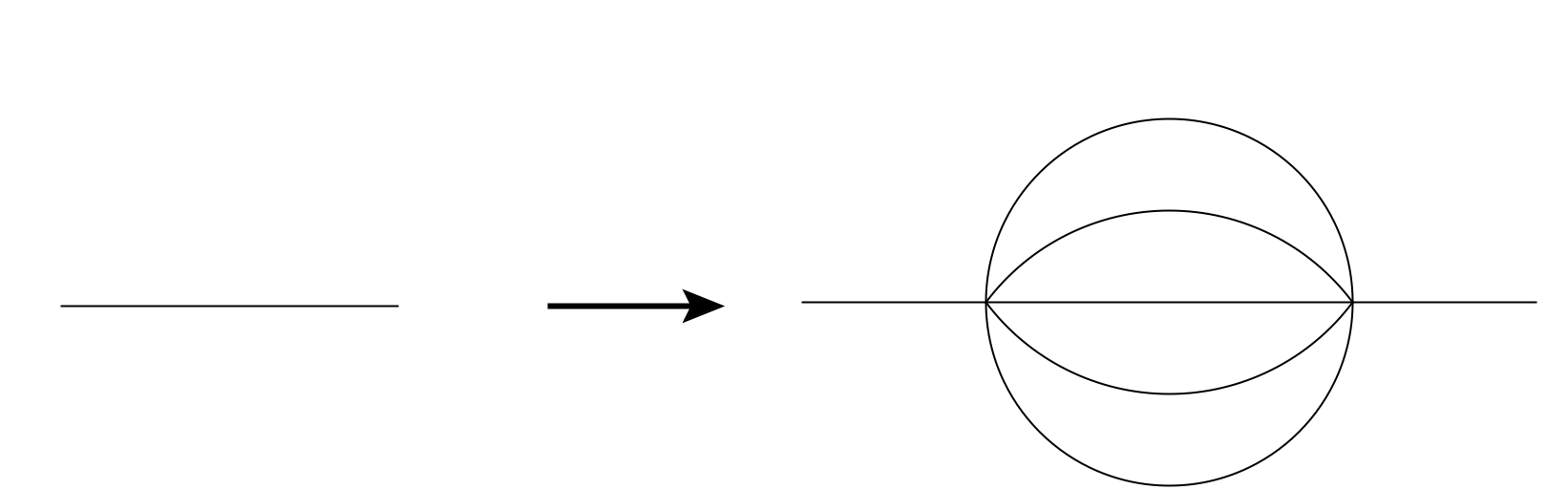}
    \end{subfigure}
    \caption{The diagrams recursively enumerated via the argument in the text can also be generated by repeatedly replacing propagators with an elementary snail (above) or an elementary melon (below). The snail need not be present in all models.}
    \label{melonic-move}
\end{figure}

The set of maximally free energy diagrams enumerated by this recursive procedure above can typically also be obtained starting from the $0$-cycle by repeatedly replacing propagators by \textit{elementary snails} or \textit{elementary melons} as shown in Figure \ref{melonic-move}.  
 
 The set of all elementary melons can be obtained from the list of maximal free-energy graphs containing a $2$-cycle: we first replace all subgraphs (shaded blobs) by free propagators, and then cut any one of the edges open to obtain an elementary melon such as the one shown in Figure \ref{melonic-move}. The set of all elementary snails can be obtained from the list of maximal free energy graphs containing a $1$-cycle in a similar manner. 

The act of replacing a propagator by an elementary snail or elementary melon is called a \textit{melonic move}. More generally, we may consider a theory with additional melonic moves. Given a set of melonic moves, we define the set of \textit{melonic diagrams} to be the set of diagrams that can be generated by making an arbitrary number of melonic moves on the $0$-cycle.

After identifying the complete set of melonic moves, it is straightforward to check that \textit{all melonic diagrams are maximal}. For instance, using the elementary melon defined for the wheel interaction, it is easy to check that if the act of replacing a propagator by an elementary melon causes $v\rightarrow v+2$ and $f_{tot}\rightarrow f_{tot} + \f{r}2(q-2)$. We also have a similar result for the elementary snail. Hence, if one acts on a maximal Feynman diagram with either of the two above elementary melonic moves, the relation \eqref{ftot} is preserved. 

However, we would like to know if the converse is true for a given model, i.e., \textit{are all maximal diagrams in a given theory melonic?} This requires us to carry out the analysis outlined in the previous subsection, which also enumerates the complete set of melonic moves for a given model. We summarize this argument here:

\begin{enumerate}
\item We argued that any free energy diagram must take one of the three forms shown in Figure \ref{cases}. For each of the interaction vertices shown in Figure \ref{cases}, we must consider all possible ways of labeling the outgoing lines. There are a large but finite number of ways to do this. For the interactions we study in this paper, many labellings are equivalent to each other via the colour permutation and automorphisms described in section \ref{discrete}, which reduces the number of cases we need to check. We refer to this stage of the argument as \textit{``enumerating all inequivalent 1-cycles and 2-cycles"}.
\item For each inequivalent 1-cycle or 2-cycle, we then impose the constraint of planarity on each of the two-colour fat graphs, as illustrated in Figures \ref{non-planar-2-cycle} and \ref{splitgraph}. These constraints translate into restrictions on the subgraphs shown in Figure \ref{cases}. We refer to this stage of the argument as \textit{``imposing fat-graph planarity''}.
\item By tracing the index contractions of lines flowing out of each subgraph, we attempt to replace the original Feynman diagram $F$ containing the subgraph with a simpler Feynman diagram $F'$. We must be able to argue that the original Feynman diagram $F$ is maximal if and only if the simpler Feynman diagram $F'$ is maximal. We refer to this stage of the argument, depicted in Figures \ref{2-cycle-split} and \ref{cutandsew}, as \textit{``cutting and sewing''}. (Here, we say $F'$ is ``simpler'' than $F$ if it contains fewer interaction vertices.)
\item The inverse of the ``cutting and sewing'' argument, i.e., replacing $F'$ with $F$, defines a melonic move. If we can find a ``cutting and sewing'' argument for each inequivalent 1-cycle and 2-cycle, then we can enumerate the entire list of melonic moves that apply to our model. We can also conclude that all maximal diagrams can be obtained from the action of melonic moves on simpler diagrams. Therefore, the model is, in principle, solvable in the large $N$ limit. We refer to this stage of the argument as \textit{``collecting all melonic moves"}.
\end{enumerate}
For any given model, this 4-stage procedure may lead to three possible outcomes:
\begin{enumerate} 
\item If the set of melonic moves so obtained involve only replacing propagators by elementary melons or elementary snails, as shown in Figure \ref{melonic-move}, the model is melonic in the conventional sense. 
\item If the set of melonic moves includes any additional move, (such as vertex-expansion defined below for the prism, shown in Figure \ref{prism-cut-and-sew} and \ref{post-melonic}), the model is still solvable in the large $N$ limit, but it is not melonic in the conventional sense.
\item If the cutting and sewing argument fails for any 1-cycle or 2-cycle, (i.e., we cannot find a simpler $F'$ to replace $F$ for some labelling of the vertices) then the theory may not be solvable in the large $N$ limit. (This is the case, for instance, when $r=2$.)
\end{enumerate}
We propose that any theory for which case 1 or 2 applies, should be considered as a generalized melonic theory, even if it is not conventionally melonic. 

We can ask what are the most general set of melonic moves possible? Note that, $F$ contains 1 or 2 interaction vertices (excluding interaction vertices in subgraphs). $F'$ contains at most 1 interaction vertex (excluding interaction vertices in subgraphs). Therefore there are three types of melonic moves possible in our generalized notion of a melonic theory,
\begin{enumerate}
	\item Replacing a diagram containing 0 interaction vertices (i.e., a propagator) with a diagram containing 1 interaction vertex, (e.g., an \textit{elementary snail}).
	\item Replacing a diagram containing 0 interaction vertices with a diagram containing 2 interaction vertices, (e.g., an \textit{elementary melon}).
	\item Replacing a diagram containing 1 interaction vertex with a diagram containing 2 interaction vertices, (i.e., \textit{vertex expansion}).
\end{enumerate}
Most well-known melonic theories only involve moves of the second type, or possibly of the first two types. However, we will see below that the prismatic model is an example of a theory that contains a move of the third type.

\begin{figure}
    \centering
    \includegraphics[width=0.7\textwidth]{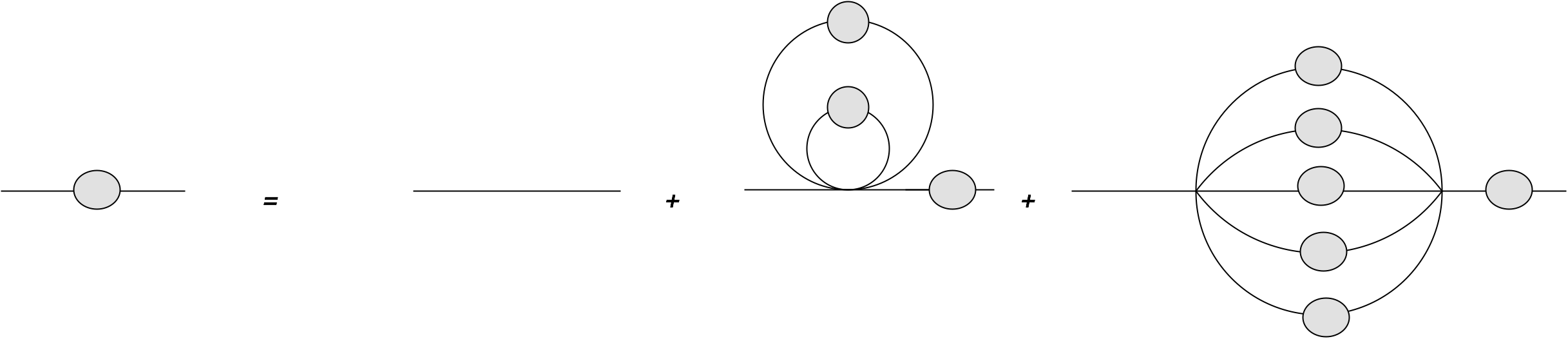}
    \caption{The melonic moves give rise to the above schematic equation for the exact propagator. If one were to connect the two external lines together, the first term on the RHS is a 0-cycle, the second term is a 1-cycle, and the third term is a 2-cycle.}
    \label{exact-propagator}
\end{figure}

For conventionally melonic theories, the melonic moves directly translate into the equation for the exact propagator shown in Figure \ref{exact-propagator}. The snails may or may not be present depending on the interaction. As mentioned earlier, the snails are tadpoles, and would formally vanish in a quantum field theory due to dimensional regularization. 

Let us conclude this section with the caveat that, while it is straightforward to obtain the Schwinger-Dyson equation for the exact propagator, if one wants to actually evaluate the free energy, one must also include the symmetry factors for each of these diagrams. As one can see from the diagrammatic expansion for the free energy in a vector model, e.g., \cite{Giombi:2011kc}, these factors are generally non-trivial to obtain. 

\subsection{Wheel (or $K_{3,3}$) interaction}

\begin{figure}[h!]
    \centering
    \begin{subfigure}[t]{0.5\textwidth}
        \centering
    \includegraphics[height=1.5in]{wheel-labels.png}
    \caption{The wheel interaction is shown as a triple-line fat vertex. }
    \end{subfigure}
    
\vspace{1cm}

    \begin{subfigure}[b]{0.32 \textwidth}\centering
    \includegraphics[height=1.5in]{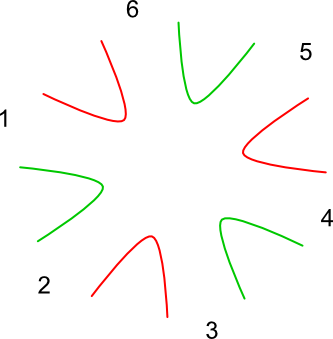}
    \caption{The red-green fat vertex.}
    \end{subfigure}
     \begin{subfigure}[b]{0.32 \textwidth}\centering
    \includegraphics[height=1.5in]{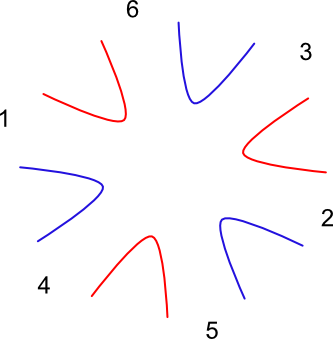}
    \caption{The red-blue fat vertex.}
    \end{subfigure}
     \begin{subfigure}[b]{0.32 \textwidth}\centering
    \includegraphics[height=1.5in]{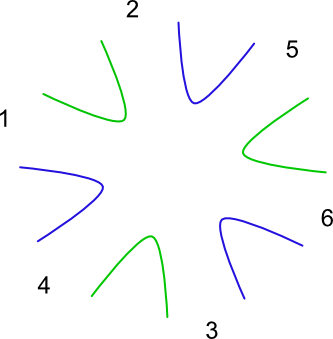}
        \caption{The green-blue fat vertex.}

    \end{subfigure}
    \caption{The wheel interaction and its three two-colour fat-vertices are shown above. For a given choice of two-colours, we must arrange the labelled-fields in one of the two particular cyclic orders in order to maintain manifest planarity.}     \label{wheel-v}
\end{figure}

In this section we demonstrate the melonic dominance of the wheel vertex, according to the general recipe above. The wheel vertex is shown in Figure \ref{wheel-v}, along with its three two-colour fat vertices.

\subsubsection*{Diagrams containing a 2-cycle}
A 2-cycle is specified by two Wick contractions. We label the two interaction vertices $L$ and $R$; each Wick contraction must include one field from each interaction vertex. We thus specify the two Wick contractions that define the 2-cycle in the following notation: $( \langle X_L, Y_R \rangle, \langle Z_L, W_R \rangle)$, where $X$, $Y$, $Z$, and $W$ are integers between $1$ and $6$ corresponding to the field labels given in the labelled interaction graph of Figure \ref{wheel}. The notation means that field from the left interaction vertex labelled by the number $X$ in the labelled interaction graph is contracted with the field labelled by the number $Y$ from the right interaction vertex \cite{KPP}.

\textit{A priori} there are a large number of different 2-cycles that are possible. However, using the automorphism and colour permutation symmetries of the wheel interaction, we can reduce the total number of inequivalent 2-cycles to a very manageable number. The idea behind this is that if a particular 2-cycle $( \langle X_L, Y_R \rangle, \langle Z_L, W_R \rangle)$ induces only planar fat graphs, then the same must be true for any other 2-cycle $( \langle X_L', Y_R' \rangle, \langle Z_L', W_R' \rangle)$ obtained by a permutation of colours. Hence we do not need to check all different 2-cycles: we only need to check the subset of 2-cycles whose orbit under the colour permutation (and automorphism) symmetry group covers all 2-cycles. This is an elementary exercise in group theory, which, for the sake of clarity and completeness we spell out explicitly in the appendix. 

%First use the automorphism symmetry (which can be applied separately to each vertex) to choose $X_L=1_L$ and $Y_R=1_R$. One can then check that the group of colour-permutation symmetries and automorphisms contain the field-permutations $(2,4,6)$ and $(3,5)(4,6)$ which both leave $1$ fixed. Thus, if we were to choose $Z=4_L$ or $6_L$, we could permute the colours to make $Z=2_L$; similarly, if we choose $Z_L=5_L$ , one can permute the colours to obtain an equivalent diagram with $Z_L=3_L$. The inequivalent 2-cycles are thus:
As shown in the appendix, the inequivalent 2-cycles are
\begin{enumerate}
\item $(\langle 1_L, 1_R \rangle, \langle 2_L, W_R \rangle )$, 
\item $(\langle 1_L, 1_R \rangle, \langle 3_L, W_R' \rangle )$,
\end{enumerate}
where $W_R=2,~3$ or $4$, and $W'_R=2,~3$ or $5$. 

Not all these 2-cycles give rise to planar fat-graphs. If $W_R$ is odd, or if $W_R'$ is even, one can check that the red-green fat graph arising from this 2-cycle contains an odd number of twists, and hence is non-planar. Hence there are only four different 2-cycles to consider. 

\begin{figure}[h!]
    \centering
     \begin{subfigure}[b]{0.32 \textwidth}\centering
    \includegraphics[height=1.56in]{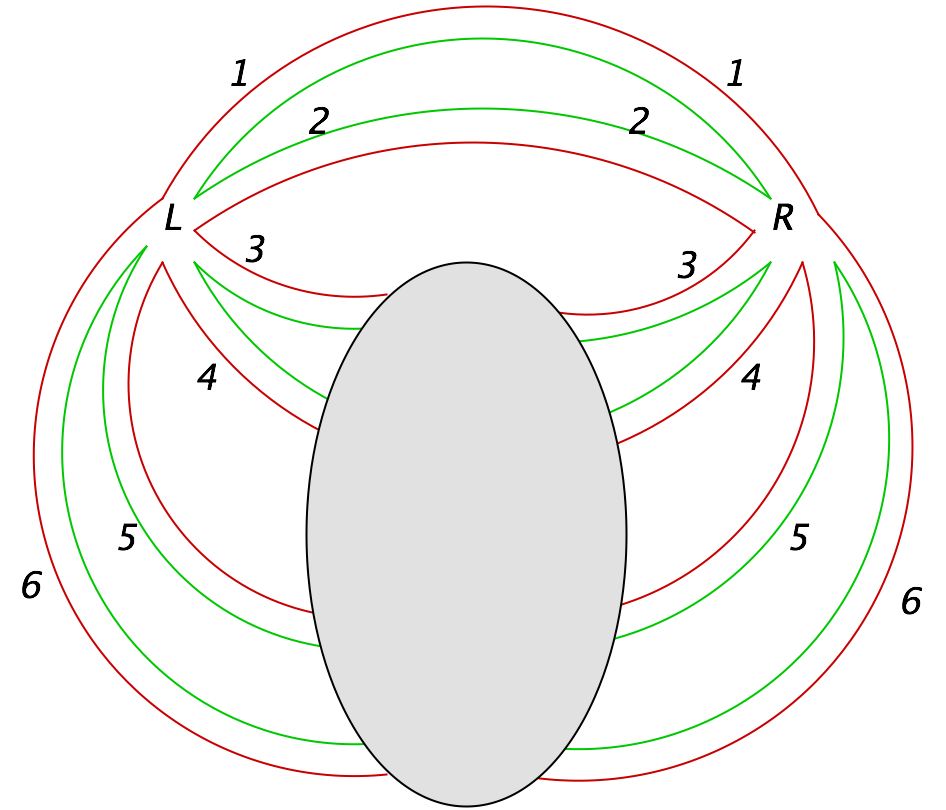}
    \end{subfigure}
     \begin{subfigure}[b]{0.32 \textwidth}\centering
    \includegraphics[height=1.56in]{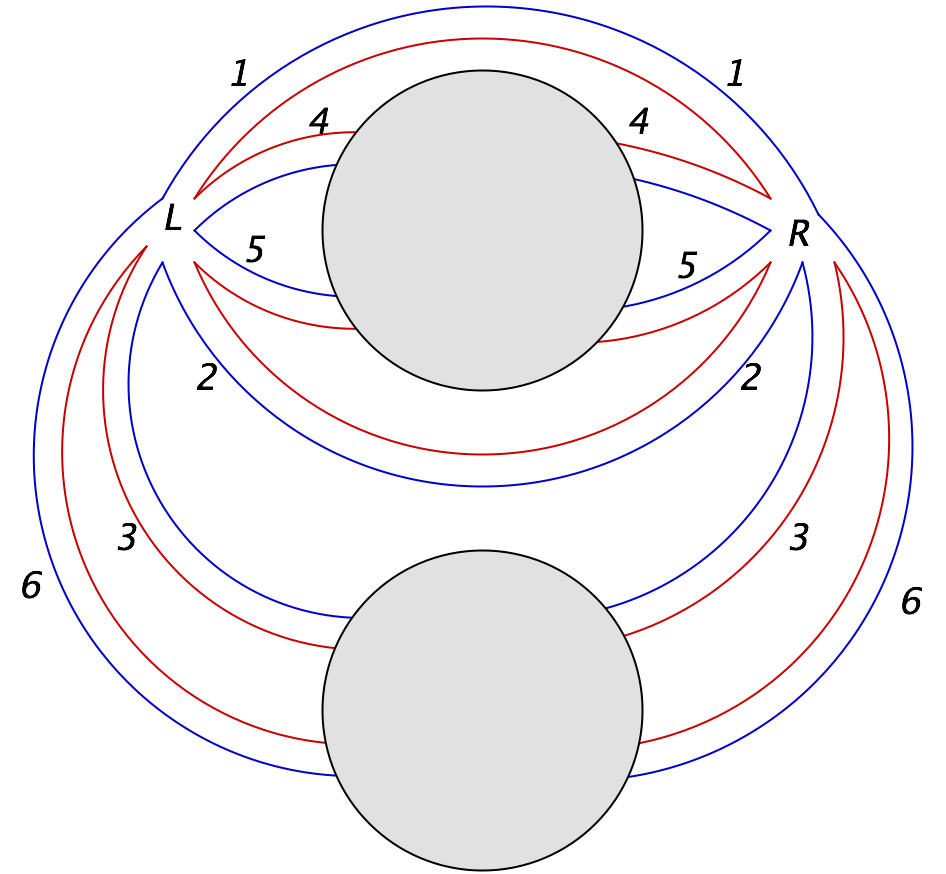}
    \end{subfigure}
     \begin{subfigure}[b]{0.32 \textwidth}\centering
    \includegraphics[height=1.56in]{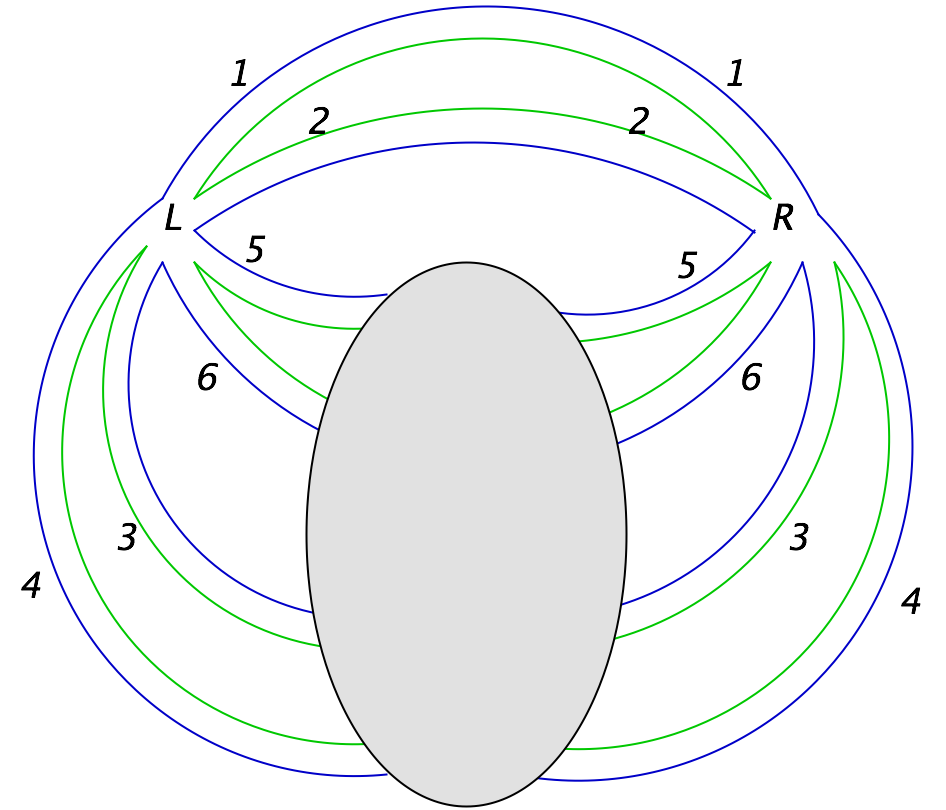}
    \end{subfigure}
    \caption{The fat graphs for the 2-cycle $(\langle 1_L, 1_R \rangle, \langle 2_L, 2_R \rangle )$ involving two wheel interaction vertices. From the blue-red fat-graph, we see that $4_L$ can be connected (via a subgraph) to one of $4_R$, $5_R$ or $5_L$. If $4_L$ is connected to $4_R$ or $5_R$, then we see from the red-green fat graph that $3_L$ must be connected to $3_R$, and $6_L$ must be connected to $6_R$. Then using the blue-green fat-graph, we see $4_L$ must be connected to $4_R$ and $5_L$ must be connected to $5_R$. This corresponds to the first graph on the left in Figure \ref{wheel-1122}. If, instead, $4_L$ is connected to $5_L$, one can similarly work out that the connections must be as depicted in the second graph in Figure \ref{wheel-1122}. }
    \label{1122fatgraphs}
\end{figure}

\begin{figure}[h!]
    \centering
    \begin{subfigure}[b]{0.4 \textwidth}
    \includegraphics[height=1.2in]{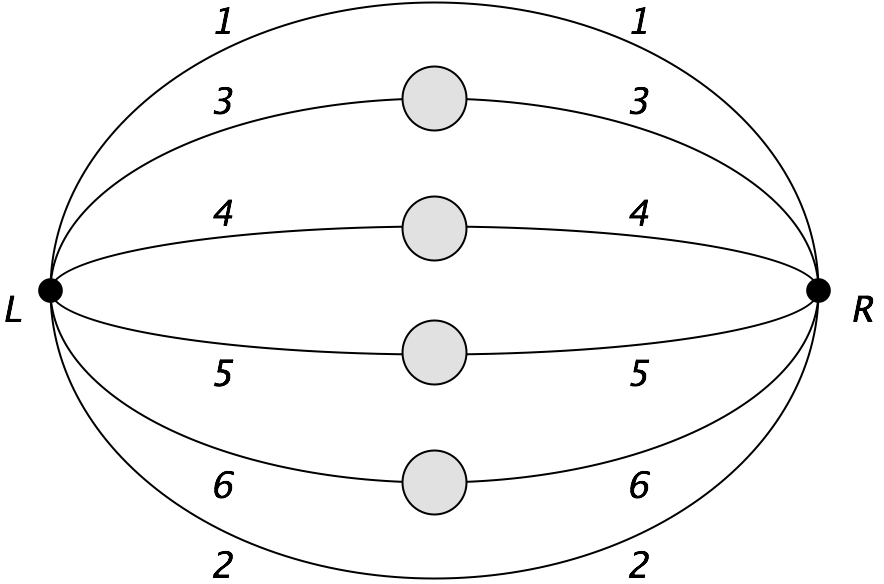}
    \end{subfigure}
    \begin{subfigure}[b]{0.4 \textwidth}
    \centering
    \includegraphics[height=.9in]{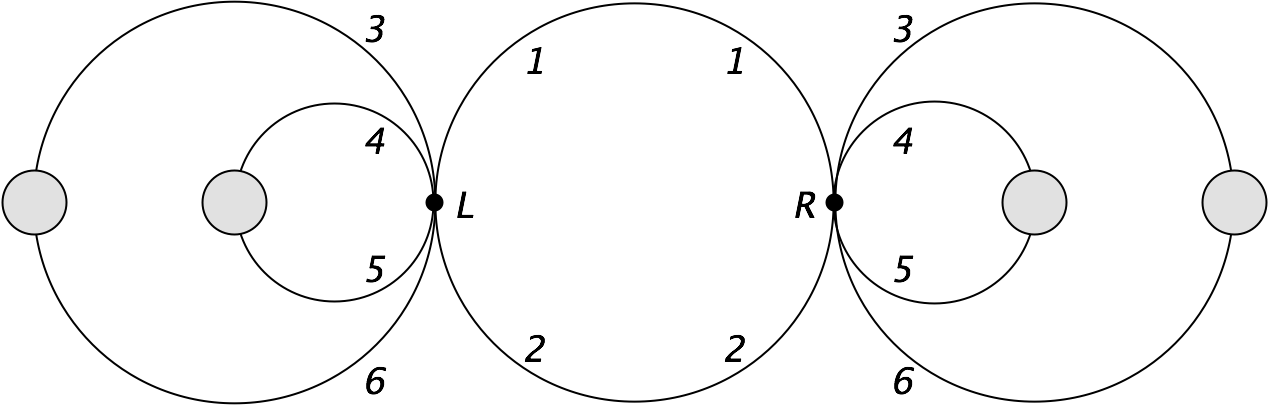}
    \end{subfigure}
    \caption{For the 2-cycle $(\langle 1_L, 1_R \rangle, \langle 2_L, 2_R \rangle )$, the constraint that all fat graphs are planar means the interaction vertices must be connected in one of the two above ways. The second possibility is a ``double-snail" that originates from the insertion two elementary snails.}
    \label{wheel-1122}
\end{figure}

Let us first consider the 2-cycle $(\langle 1_L, 1_R \rangle, \langle 2_L, 2_R \rangle )$. The three fat graphs for this 2-cycle are shown in Figure \ref{1122fatgraphs}. Using these constraints we have two possibilities for the form of a free-energy diagram containing this 2-cycle, shown in Figure \ref{wheel-1122}.


Let us next consider the 2-cycle $(\langle 1_L, 1_R \rangle, \langle 3_L, 3_R \rangle )$. The three fat graphs for this 2-cycle are shown in Figure \ref{wheel-1133}. Using these constraints we have one possibility for the form of a free-energy diagram containing this 2-cycle, also depicted in Figure \ref{wheel-1133}.

\begin{figure}[h!]
\centering
    \includegraphics[height=.9in]{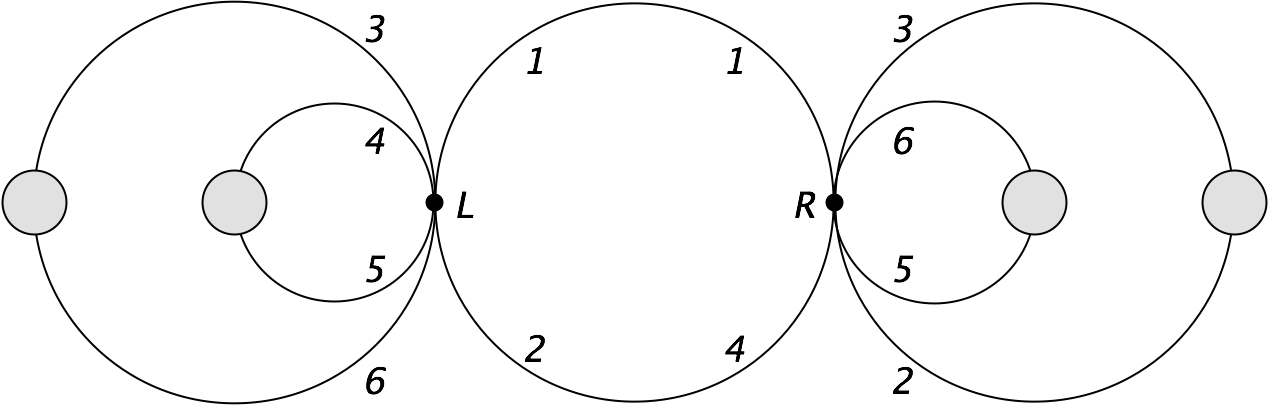}
 
    \caption{For the 2-cycle $(\langle 1_L, 1_R \rangle, \langle 2_L, 4_R \rangle )$ connecting two wheel interaction vertices, the constraint that all fat graphs are planar means the interaction vertices must be connected as shown above.}
    \label{wheel-1124}
\end{figure}

\begin{figure}[h!]
    \centering
         \begin{subfigure}[b]{0.32 \textwidth}\centering
    \includegraphics[height=1.56in]{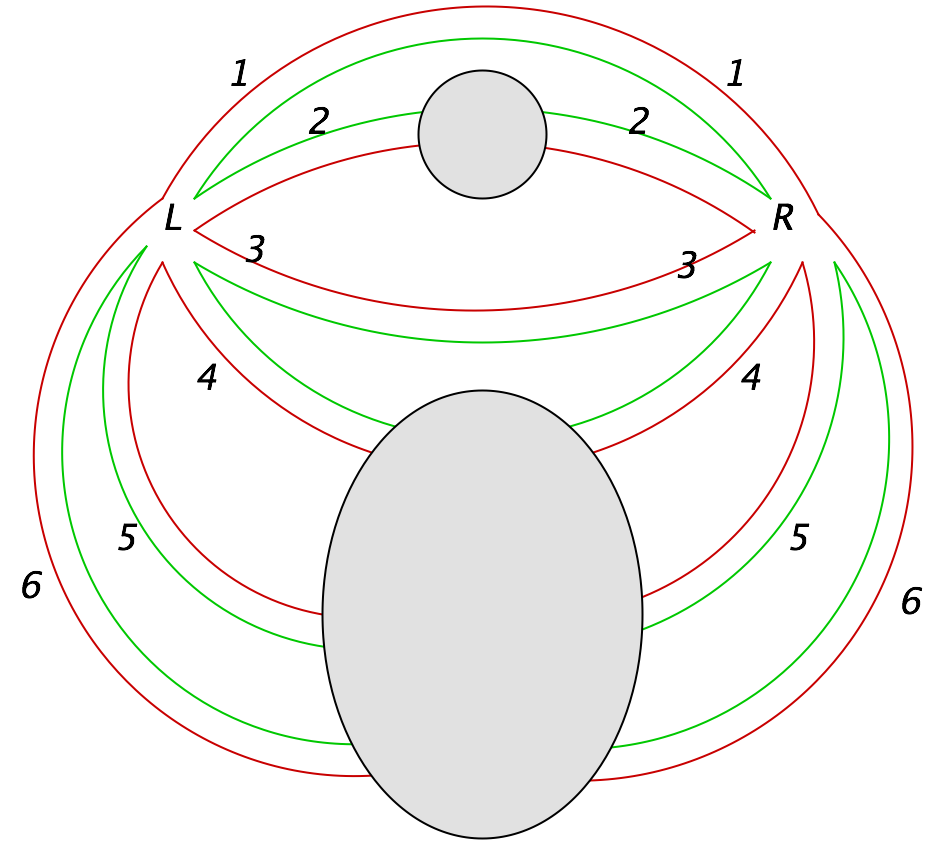}
    \end{subfigure}
     \begin{subfigure}[b]{0.32 \textwidth}\centering
    \includegraphics[height=1.56in]{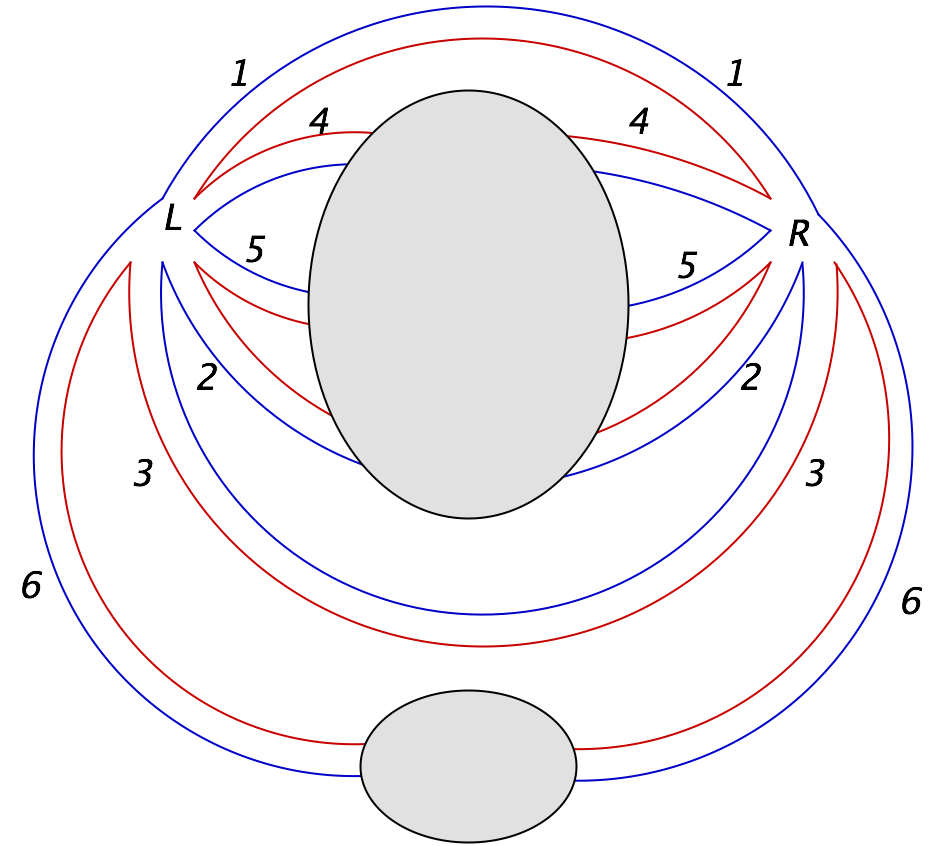}
    \end{subfigure}
     \begin{subfigure}[b]{0.32 \textwidth}\centering
    \includegraphics[height=1.56in]{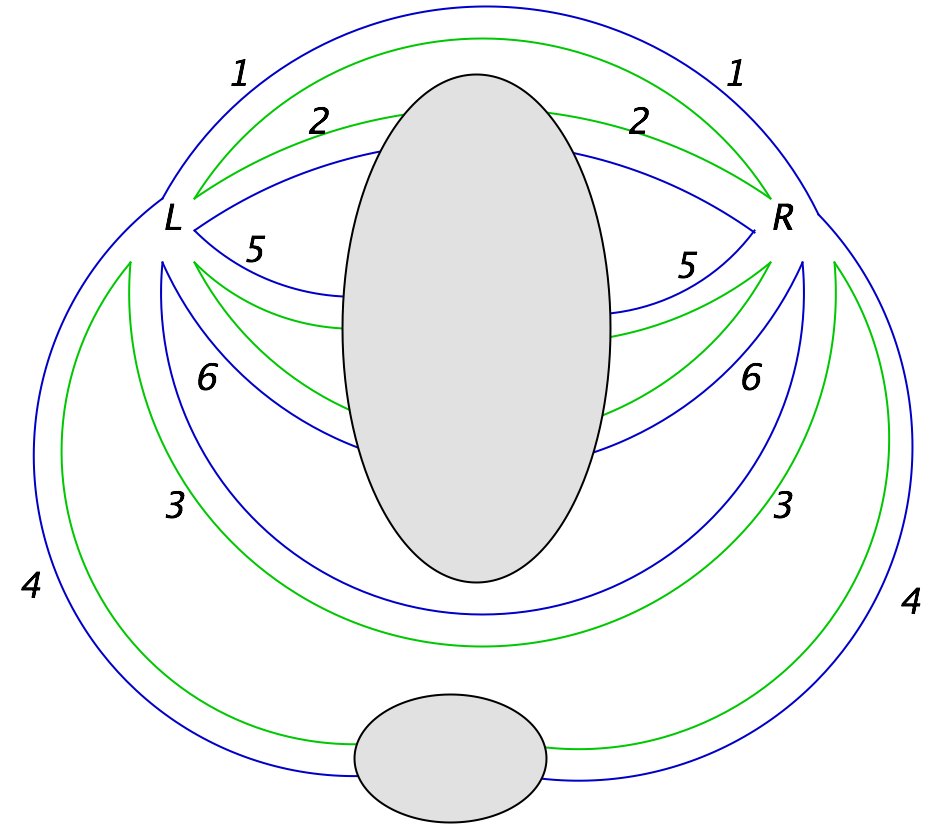}
    \end{subfigure}
    \vspace{1cm}

    \begin{subfigure}[b]{0.4 \textwidth}
    \centering
    \includegraphics[height=1.2in]{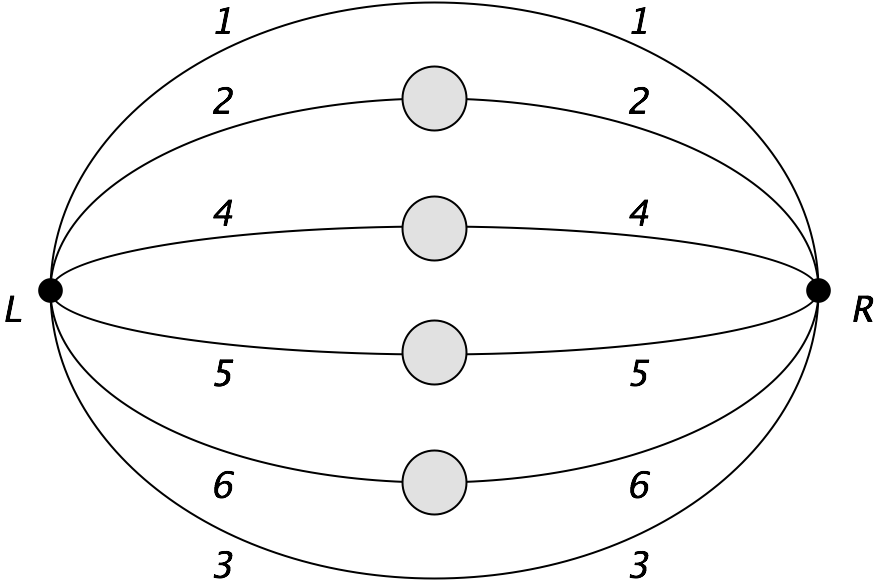}
    \end{subfigure}
    \caption{For the 2-cycle $(\langle 1_L, 1_R \rangle, \langle 3_L, 3_R \rangle )$ connecting two wheel interaction vertices, the constraint that all fat graphs (shown above) are planar means the interaction vertices must be connected as shown below.}
    \label{wheel-1133}
\end{figure}

We carry out a similar analysis for the 2-cycle $(\langle 1_L, 1_R \rangle, \langle 2_L, 4_R \rangle )$; the form of any free-energy diagram consistent with planarity of fat-graphs is shown in Figure \ref{wheel-1124}.

For the 2-cycle $(\langle 1_L, 1_R \rangle, \langle 3_L, 5_R \rangle )$, we find there is no way to satisfy the constraints that all fat-graphs be planar. 

\subsubsection*{Diagrams containing a 1-cycle}
A 1-cycle is formed by contracting two fields from the same interaction vertex, which we denote as $\langle X, Y \rangle$. Using the automorphism symmetry of the wheel interaction, we can always choose the first field $X=1$. Via colour permutation symmetry, the second field $Y$ can be chosen to be $2$ or $3$. There are thus two inequivalent 1-cycles: $\langle 1,2 \rangle$ and $\langle 1, 3 \rangle$. 

\begin{figure}
    \centering
    \includegraphics[height=1.5in]{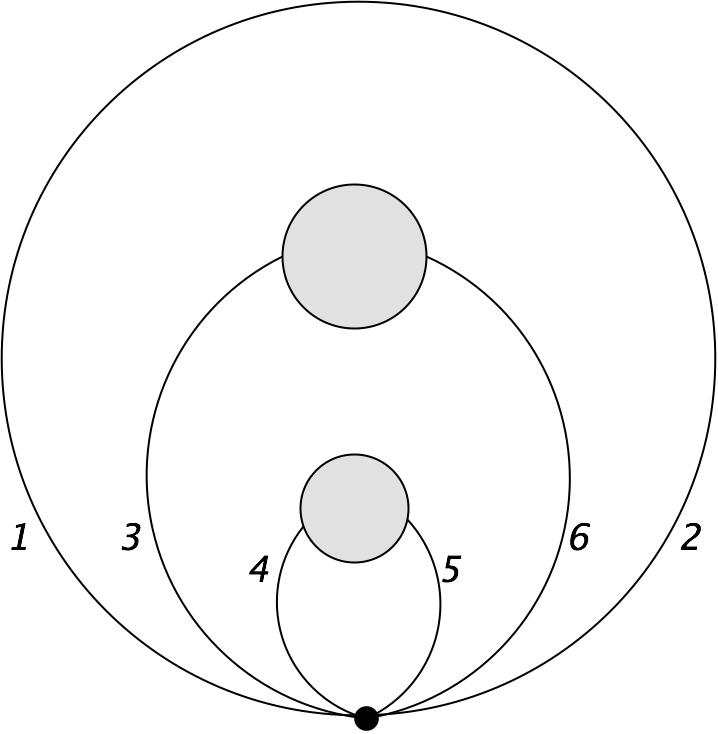}
    \caption{Any Feynman diagram containing a 1-cycle involving one wheel vertex must be of this form, or related to this by permutation of colours.}
    \label{121-cycle-wheel}
\end{figure}

The 1-cycle $\langle 1, 3 \rangle$ results in a non-planar red-green fat graph, so is non-maximal. The $\langle 1, 2 \rangle$ 1-cycle is constrained by planarity of the red-blue fat graph to be of the form in Figure \ref{121-cycle-wheel}.

\subsubsection*{Melonic Moves}
\begin{figure}
    \centering
    \begin{subfigure}[b]{0.4\textwidth}
    \centering
        \includegraphics[height=1.0in]{elementary-melon-wheel.png}
    \end{subfigure}
    \begin{subfigure}[b]{0.4\textwidth}
    \centering
    \includegraphics[height=1.2in]{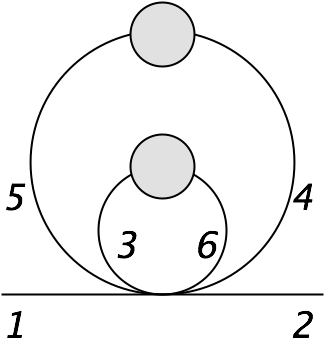}
    \end{subfigure}
    \caption{The maximal diagrams arising from the wheel interaction can also be generated by replacing propagators with the above elementary snail and elementary melon (or their colour permutations).}
    \label{elementary-melons}
\end{figure}

From the inequivalent $1$-cycles and $2$-cycles above, we can extract the elementary melon and elementary snail shown in Figure \ref{elementary-melons}.

\subsection{Octahedron interaction}

\begin{figure}[h!]
    \centering
    \begin{subfigure}[t]{0.5\textwidth}
    \centering
    \includegraphics[height=1.5in]{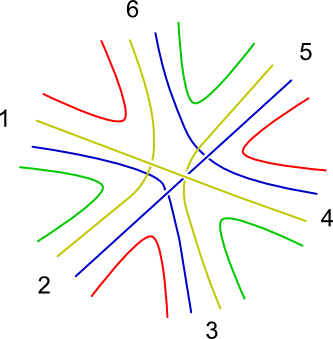}
    \caption{The octahedron interaction as a multi-line fat vertex. }
    \end{subfigure}
    
\vspace{1cm}

    \begin{subfigure}[b]{0.32 \textwidth}\centering
    \includegraphics[height=1.5in]{labelled-vertex-red-green.png}
    \caption{The red-green fat vertex.}
    \end{subfigure}
     \begin{subfigure}[b]{0.32 \textwidth}\centering
    \includegraphics[height=1.5in]{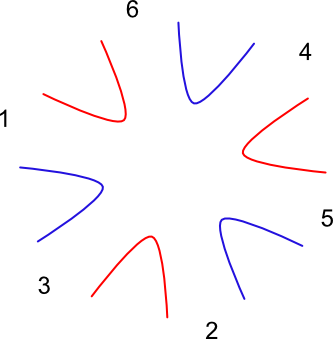}
    \caption{The red-blue fat vertex.}
    \end{subfigure}
     \begin{subfigure}[b]{0.32 \textwidth}\centering
    \includegraphics[height=1.5in]{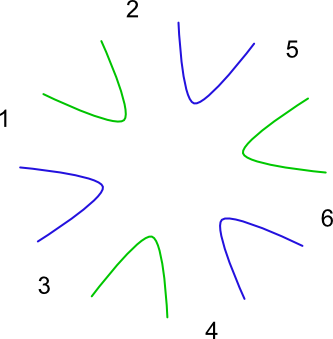}
        \caption{The blue-green fat vertex.}
    \end{subfigure}
\vspace{.5cm}

    \begin{subfigure}[b]{0.32 \textwidth}
        \centering
    \includegraphics[height=1.5in]{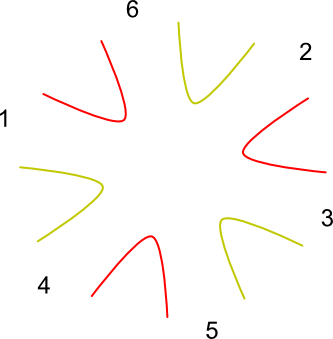}
    \caption{The yellow-red fat vertex.}
    \end{subfigure}
     \begin{subfigure}[b]{0.32 \textwidth}
        \centering
    \includegraphics[height=1.5in]{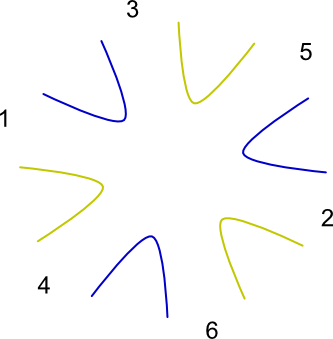}
    \caption{The blue-yellow fat vertex.}
    \end{subfigure}
     \begin{subfigure}[b]{0.32 \textwidth}
     \centering
    \includegraphics[height=1.5in]{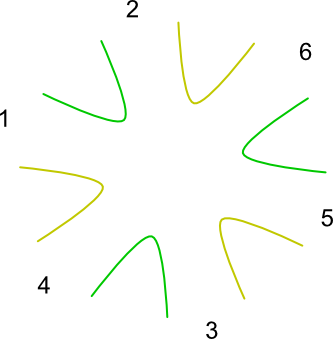}
        \caption{The yellow-green fat vertex.}
    \end{subfigure}
    \caption{The octahedron interaction and its three two-colour fat-vertices are shown above. Note that, for a given choice of two-colours, we must arrange the fields in a particular cyclic order in order to maintain manifest planarity}
    \label{double-prismV}
\end{figure}

Let us now consider the octahedron. 
(We present this case before the prism, because it also gives rise to a conventional melonic limit.) Its fat vertices are pictured in Figure \ref{double-prismV}.

\subsubsection*{Diagrams containing a 1-cycle}
One can check that, for the octahedron, there is no way of forming a 1-cycle such that all six fat-graphs are planar. Hence elementary snails are ruled out on purely combinatorial grounds.

\subsubsection*{Diagrams containing a 2-cycle}

Let us enumerate all inequivalent 2-cycles: $(\lan X_L, Y_R\ran, \lan Z_L, W_R\ran)$ passing through two octahedron interaction vertices.

The octahedron has no automorphism symmetry. However, we can use the colour permutation symmetry to reduce the number of cases one has to check to ensure all fat-graphs are planar. 

By permuting the colours in the Feynman diagram, any 2-cycle can be mapped to one in which $X=1$. This procedure does not use all the colour permutation symmetry, as the permutation $\sigma_{(gbyr)}$ corresponds to the field-vertex permutation $(2,3,4,6)$, which leaves the choice of $X=1$ invariant. In case $Y$ was chosen to be $3,~4$ or $6$, one could use the colour permutation $\sigma_{(gbyr)}$ to map it to an equivalent diagram where $Y=2$. We can thus set $Y=1,~5$ or $2$. If $Y=1,~5$, then the colour permutation symmetry still remains at our disposal to set $Z=1,~5$ or $2$. Finally, if $Z=1,~5$, the unused colour permutation symmetry can be used to set $W=1,~5$ or $2$.

The inequivalent 2-cycles can thus be taken as: 
\begin{enumerate}
    \item $(\lan 1_L, 2_R\ran, \lan Z_L, W_R\ran)$
    \item $(\lan 1_L, 5_R\ran, \lan 2_L, W_R'\ran)$
    \item $(\lan 1_L, 5_R\ran, \lan 5_L, 2_R\ran)$
    \item $(\lan 1_L, 5_R\ran, \lan 5_L, 1_R\ran)$
    \item $(\lan 1_L, 1_R\ran, \lan 2_L, W_R''\ran)$
    \item $(\lan 1_L, 1_R\ran, \lan 5_L, 5_R\ran)$
\end{enumerate}

Unless $W''=2$, $Z=2$, $W=1$, one of the two-colour fat-graphs will contain an odd number of twists. Cases 2 and 3 also give rise to fat graphs with an odd number of twists, so these cases are also ruled out. We finally have only four possibly-planar inequivalent 2-cycles, which are:

\begin{itemize}
    \item $(\lan 1_L, 2_R\ran, \lan 2_L, 1_R\ran)$
    \item $(\lan 1_L, 5_R\ran, \lan 5_L, 1_R\ran)$
    \item $(\lan 1_L, 1_R\ran, \lan 2_L, 2_R\ran)$
    \item $(\lan 1_L, 1_R\ran, \lan 5_L, 5_R\ran)$
\end{itemize}

\noindent

Drawing all fat-graphs for each 2-cycle as we did for the wheel, we obtain the following results. We find that any free energy diagram containing the 2-cycle $(\lan 1_L, 2_R\ran, \lan 2_L, 1_R\ran)$, or $(\lan 1_L, 5_R\ran, \lan 5_L, 1_R\ran)$, gives rise to at least one non-planar fat graphs. Hence any free-energy diagram containing these cycles is non-maximal.

\begin{figure}[h!]
\centering
    \centering
    \includegraphics[height=1.7in]{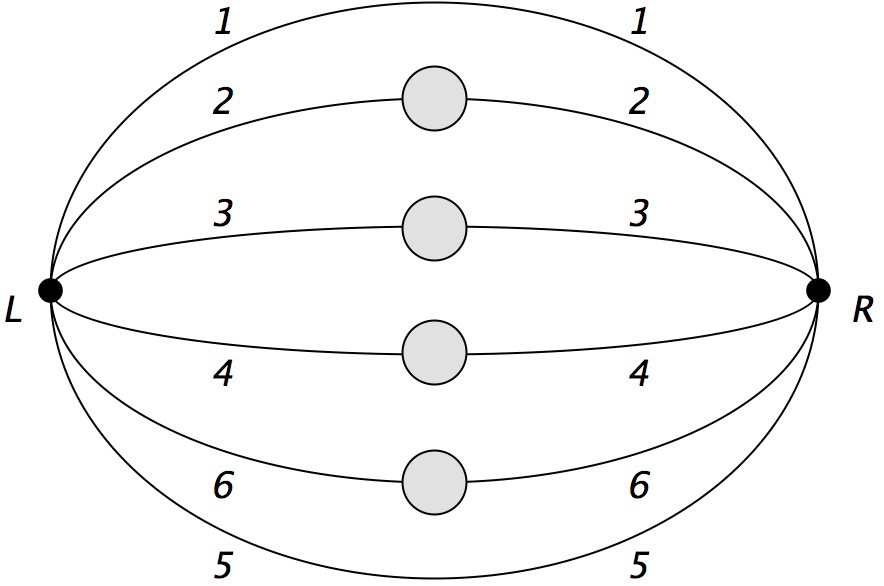}
    \caption{Requiring the 2-cycle $(\lan 1_L, 1_R\ran, \lan 5_L, 5_R\ran)$ to be maximal means it must take the above traditionally-melonic form. A similar result holds for $(\lan 1_L, 1_R\ran, \lan 2_L, 2_R\ran)$.}
    \label{doubleprism-1155}
\end{figure}

Next we consider free-energy diagrams containing the 2-cycle  $(\lan 1_L, 1_R\ran, \lan 5_L, 5_R\ran)$. Requiring all fat graphs to be planar gives rise to a free energy diagram of the form shown in Figure \ref{doubleprism-1155}. An identical result holds for free-energy diagrams containing the 2-cycle $(\lan 1_L,1_R\ran, \lan 2_L, 2_R\ran)$.

\subsubsection*{Melonic moves}
Putting all these results together, we find that this model is melonic in the conventional sense. All diagrams can be generated by replacing propagators by the elementary melon shown below in Figure \ref{elementary-melon}. There is no elementary snail, unlike the case of the wheel.
\begin{figure}
    \centering
    \includegraphics[height=1.5in]{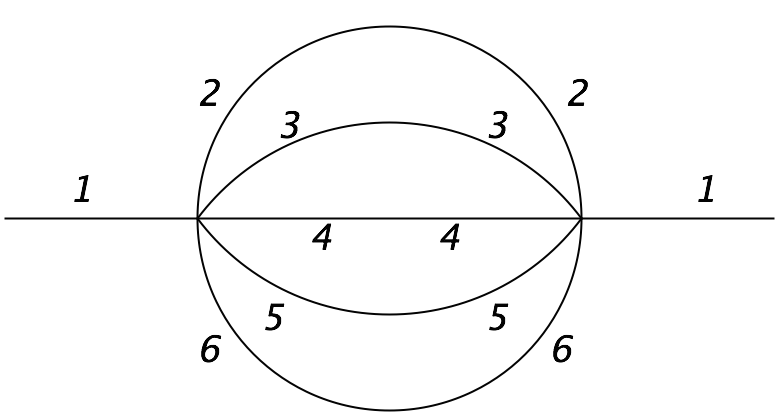}
    \caption{The theory based on the octahedron is traditionally melonic, with the above elementary melon (and its colour permutations).}
    \label{elementary-melon}
\end{figure}

\subsection{Prism interaction}
\begin{figure}[h!]
    \centering
    \begin{subfigure}[t]{0.5\textwidth}
    \centering
    \includegraphics[height=1.5in]{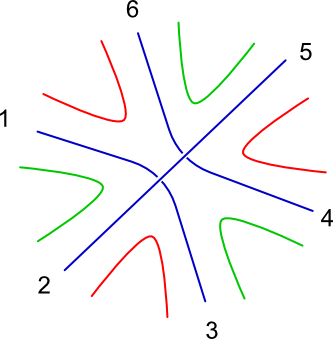}
    \caption{The prism interaction is shown as a triple-line fat vertex. }
    \end{subfigure}
    
\vspace{1cm}

    \begin{subfigure}[b]{0.3 \textwidth}\centering
    \includegraphics[height=1.5in]{labelled-vertex-red-green.png}
    \caption{The red-green fat vertex for the prism.}
    \end{subfigure}
     \begin{subfigure}[b]{0.3 \textwidth}\centering
    \includegraphics[height=1.5in]{labelled-vertex-red-blue-prism.png}
    \caption{The red-blue fat vertex for the prism.}
    \end{subfigure}
     \begin{subfigure}[b]{0.3 \textwidth}\centering
    \includegraphics[height=1.5in]{labelled-vertex-green-blue-prism.png}
        \caption{The blue-green fat vertex for the prism.}

    \end{subfigure}
    \caption{The prism interaction and its three two-colour fat-vertices are shown above. Note that, for a given choice of two-colours, we must arrange the fields in a particular cyclic order in order to maintain manifest planarity}
    \label{prismfatV}
\end{figure}

Let us now consider the prism interaction. The prism interaction and its two-colour fat vertices, are shown in Figure \ref{prismfatV}. In \cite{Azeyanagi:2017mre,GKPPT}, it was shown that the leading order diagrams arising from the prism interaction can be explicitly summed by using an auxiliary field to convert it into a quartic tetrahedron interaction.

However, if we do not introduce this auxiliary field, and simply draw Feynman diagrams using the sextic prism vertex, we find that there are maximal diagrams which are not melonic in the sextic sense. In other words, the set of maximal Feynman diagrams in the prismatic theory includes diagrams that would not be maximal in a conventional melonic theory, such as a theory based on the octahedron or a theory based on the $r=5$ maximally-single-trace interaction studied in \cite{Ferrari:2017jgw, KPP}. An example of such a diagram is shown in Figure \ref{nonmelonicMaximal}.

Because the prism interaction gives rise to a large $N$ limit that is not conventionally melonic in this sense, it is interesting to see how the method of analysis given above can be modified to characterize all maximal diagrams.

\subsubsection*{Diagrams containing a 1-cycle}
\begin{figure}
    \centering
    \includegraphics[height=1.4in]{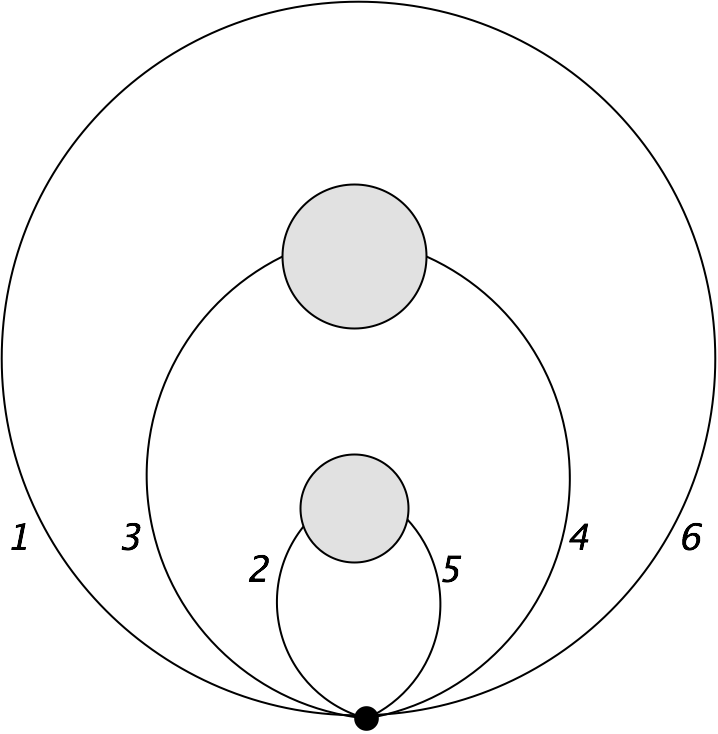}
    \caption{Any maximal Feynman diagram containing a 1-cycle passing through one prism vertex must be of this form, or related to this by permutation of colours.}
    \label{1-cycle-prism}
\end{figure}

One can check that the only maximal free energy diagram containing a 1-cycle is $\langle 1, 6 \rangle$, or its colour permutations. It takes the form shown in Figure \ref{1-cycle-prism}.

\subsubsection*{Diagrams containing a 2-cycle}

Consider two prism interaction vertices, one denoted by $L$ and the other $R$. As before, we specify a 2-cycle, by the following contractions: $(\lan X_L, Y_R\ran, \lan Z_L, W_R\ran)$. 

We can choose $X_L=1_L$ using the automorphism of the first interaction vertex and the colour permutation symmetry. The automorphism symmetry of the second vertex and the residual colour permutation symmetry group can be used to choose $Y_R=1_R$ or $2_R$. If $Y_R=1_R$ the residual colour permutation symmetry can be used to set $Z_L=2_L,~4_L,$ or $6_L$ Thus, we have the following inequivalent 2-cycles,
\begin{enumerate}
\item $(\lan1_L, 1_R\ran, \lan Z_L, W_R\ran)$
\item $(\lan1_L, 2_R\ran, \lan Z_L', W_R'\ran)$
\end{enumerate}
where $Z=2,~4,~6$. 

After removing those $2-cycles$ which give rise to fat-graphs with an odd-number of twists, we find the allowed 2-cycles are:
\begin{enumerate}
\item $(\lan1_L, 1_R\ran, \lan 2_L, 2_R\ran)$
\item $(\lan1_L, 1_R\ran, \lan 4_L, 4_R\ran)$
\item $(\lan1_L, 1_R\ran, \lan 6_L, 6_R\ran)$
\item $(\lan1_L, 2_R\ran, \lan 2_L, 1_R\ran)$
\item $(\lan1_L, 2_R\ran, \lan 5_L, 6_R\ran)$
\end{enumerate}

Let us look at the structure of the constraints imposed by requiring fat graphs planarity for free-energy diagrams containing these cycles. 

First, one can check that free energy diagrams containing the $2$-cycles $\lan 1_L, 2_R\ran, \lan 5_L, 6_R\ran$ and $\lan 1_L, 2_R\ran, \lan 2_L, 1_R\ran$  always give rise to a non-planar fat graph, and are hence non-maximal.

\begin{figure}[h!]
    \centering
     \begin{subfigure}[b]{0.3 \textwidth}
    \includegraphics[height=1.5in]{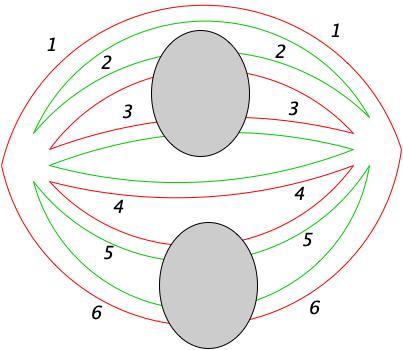}
    \end{subfigure}
     \begin{subfigure}[b]{0.3 \textwidth}
    \includegraphics[height=1.5in]{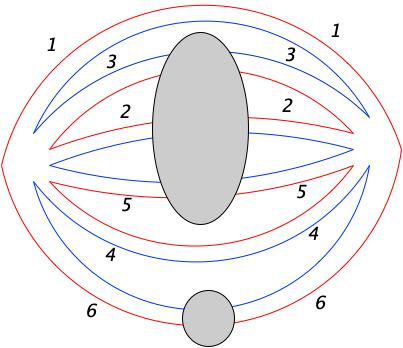}
    \end{subfigure}
     \begin{subfigure}[b]{0.3 \textwidth}
    \includegraphics[height=1.5in]{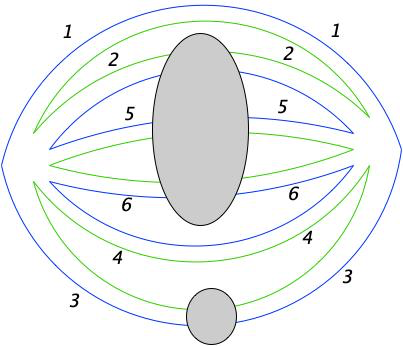}
    \end{subfigure}
    \vspace{1cm}

    \begin{subfigure}[c]{.9 \textwidth}
    \centering
    \includegraphics[height=1.5in]{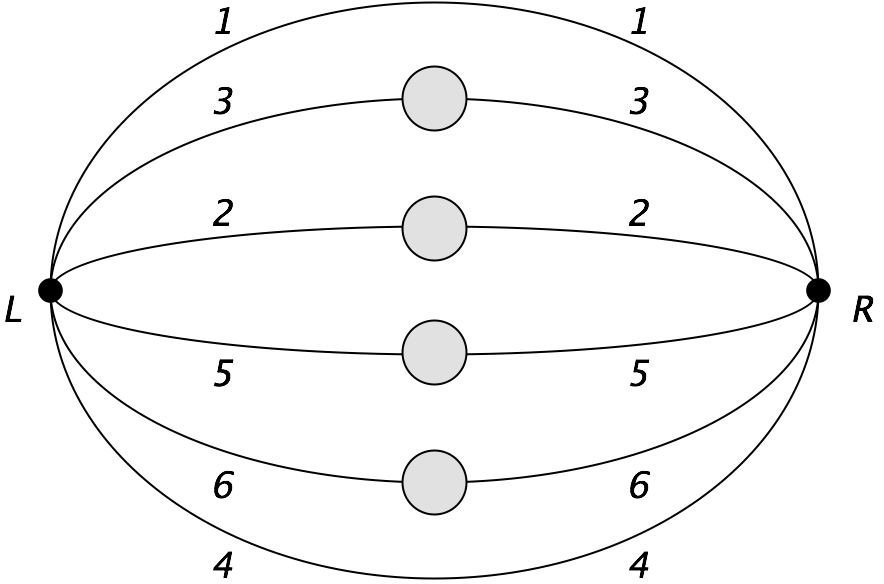}
    \end{subfigure}
    \caption{The fat graphs for the 2-cycle $(\langle 1_L, 1_R \rangle, \langle 4_L, 4_R \rangle )$ involving two prism interaction vertices are shown above. We see that this gives rise to a conventional melonic structure shown below.}
    \label{1144_prism_fatgraphs}
\end{figure}

Next consider the 2-cycle $(\lan 1_L, 1_R\ran, \lan 4_L, 4_R\ran)$. Planarity of the fat graphs restricts any free energy diagram containing this cycle to be of the form shown in Figure \ref{1144_prism_fatgraphs}. 

We then consider the case $\lan 1_L, 1_R\ran, \lan 6_L, 6_R\ran$. Free energy diagrams containing this 2-cycle (not pictured) are either a conventional melonic diagram, or a double-snail.

\noindent

\begin{figure}[h!]
    \centering
     \begin{subfigure}[b]{0.3 \textwidth}
    \includegraphics[height=1.5in]{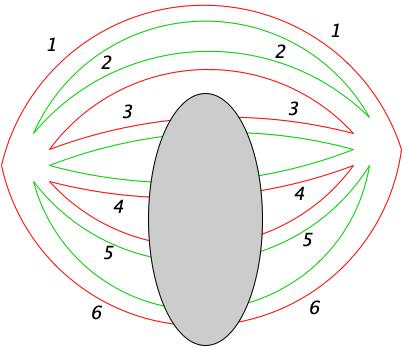}
    \end{subfigure}
     \begin{subfigure}[b]{0.3 \textwidth}
    \includegraphics[height=1.5in]{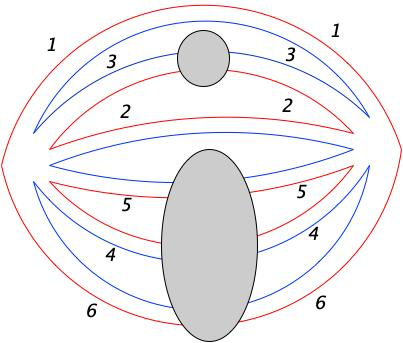}
    \end{subfigure}
     \begin{subfigure}[b]{0.3 \textwidth}
    \includegraphics[height=1.5in]{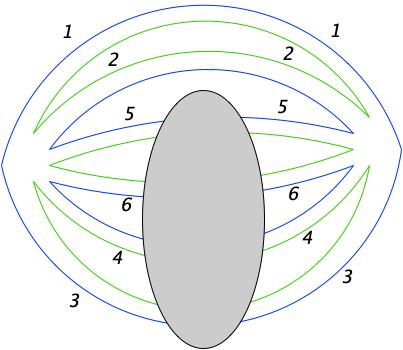}
    \end{subfigure}
    \vspace{1cm}

    \begin{subfigure}[c]{.9 \textwidth}
    \centering
    \includegraphics[height=1.5in]{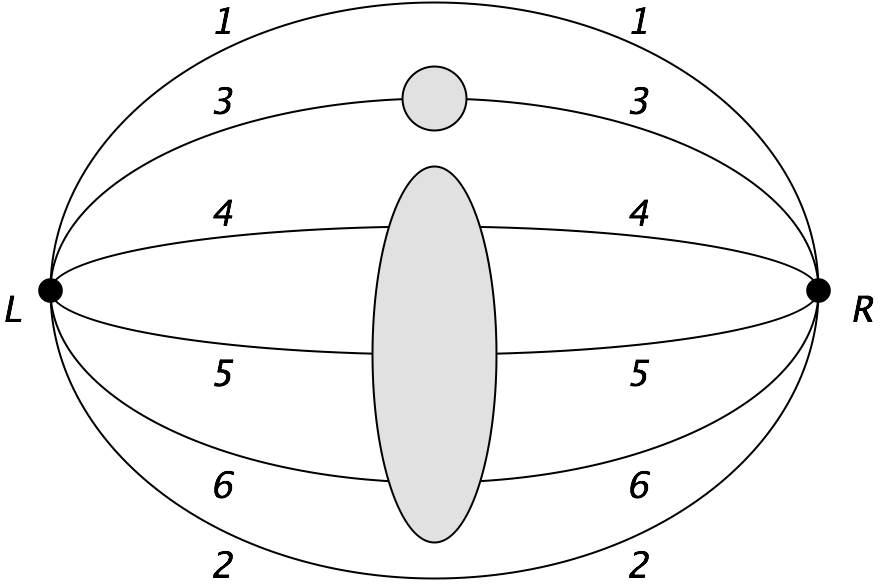}
    \end{subfigure}
    \caption{The three fat graphs for the 2-cycle $(\langle 1_L, 1_R \rangle, \langle 2_L, 2_R \rangle )$ involving two prism interaction vertices are shown above. From the blue-red fat-graph, we see that the subgraph gets split into two parts. However, there are not enough constraints to separate the subgraph into 4 components, with two external edges each. Hence the theory is not melonic in a conventional sense.}
    \label{1122_prism_fatgraphs}
\end{figure}

We finally consider free energy diagrams containing the 2-cycle $(\lan 1_L, 1_R\ran, \lan 2_L, 2_R\ran)$. We observe that the requirement of planar fat subgraphs, does not split the subgraph into 4 disconnected components, as it did for the other cases. Instead, as shown in Fig \ref{1122_prism_fatgraphs}, we are left with one subgraph with two external edges and one subgraph with six external edges.  

\subsubsection*{New melonic move}

\begin{figure}[h!]
    \centering
     \begin{subfigure}[b]{0.45 \textwidth}
     \centering
    \includegraphics[height=1.5in]{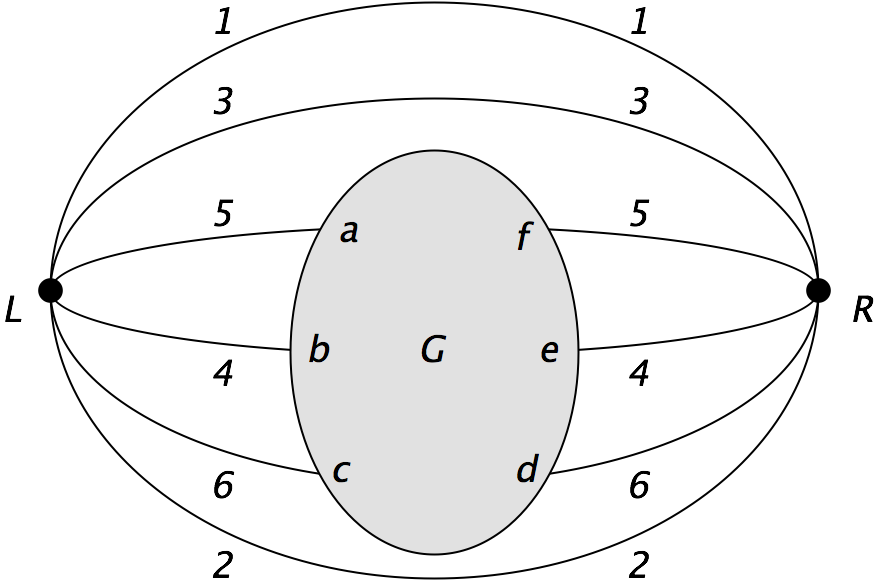}
    \end{subfigure}
     \begin{subfigure}[b]{0.45 \textwidth}
     \centering
    \includegraphics[height=1.7in]{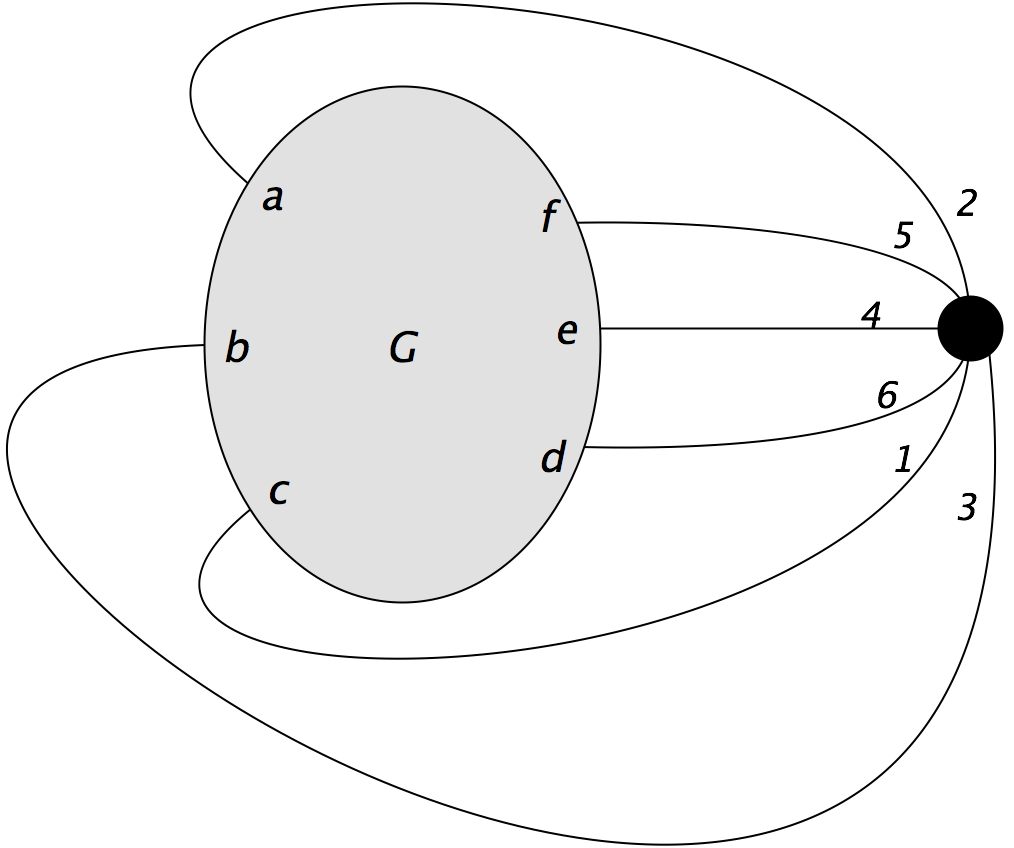}
    \end{subfigure}
    \vspace{1cm}
    \caption{The graph on the left, which originates from the Figure \ref{1122_prism_fatgraphs}, is maximal if and only if the graph on the right is maximal. One can see this by tracing each of the index contractions for each of the three $O(N)$ symmetry groups. The graph on the right is a free energy graph and must be one of the forms enumerated in the previous subsection. The above relation also gives rise to a new elementary move: of replacing one vertex (right) by two vertices (left).}
    \label{prism-cut-and-sew}
\end{figure}

In order to obtain a recursive enumeration of diagrams in this case, we need to adapt the cutting and sewing rules given earlier to the subgraph containing 6 external lines. We show how this is done in Figure \ref{prism-cut-and-sew}. By carefully following the index contractions, one can check that the diagram on the left in Figure 36 is maximal if and only if the diagram on the right is maximal.

The diagram on the right in Figure \ref{prism-cut-and-sew} (which contains one fewer interaction vertex than the original diagram) is a free energy graph that contains at least one vertex, so it must take one of the forms enumerated in the previous section. By cutting out one prism interaction vertex from any of these forms, we can determine all the possibilities for subgraph $G$ in Figure \ref{prism-cut-and-sew}. We have thus formally obtained a recursive procedure for generating all free-energy graphs.

\begin{figure}[h!]
    \centering
    \includegraphics[height=2in]{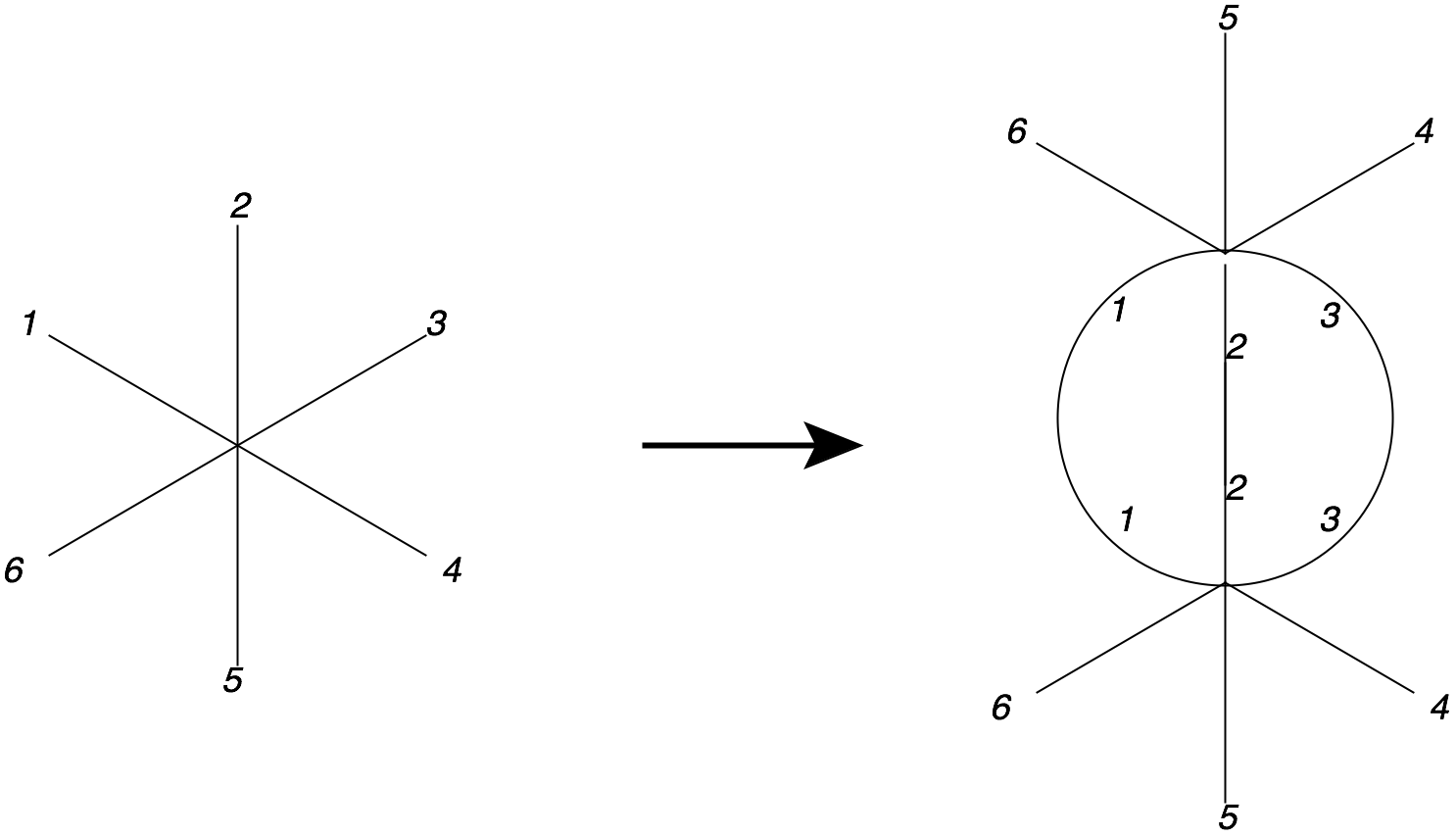}
        \vspace{1cm}
    \caption{A new melonic move, vertex expansion, is present in the prismatic model.}
    \label{post-melonic}
\end{figure}

In practice, it is helpful to translate this enumeration into the language of melonic-moves. We note that, in addition to the usual melonic move of replacing a propagator by an elementary snail or melon, Figure \ref{prism-cut-and-sew} requires us to introduce an additional move. From the possibility that the subgraph $G$ could be obtained from cutting out one interaction vertex from the non-melonic 2-cycle of Figure \ref{1122_prism_fatgraphs} itself, we obtain the new ``vertex-expansion'' move of replacing an interaction vertex by two-interaction vertices contracted in a particular way, as shown in Figure \ref{post-melonic}.\footnote{We would like to thank Adrian Tanasa and Victor Nador for discussions on this point} This vertex expansion move can be thought of as the ``inverse'' of the cutting and sewing rule of Figure \ref{prism-cut-and-sew}. All maximal diagrams can be produced by application of this melonic move, along with the melonic moves of replacing a propagator by an elementary melon or elementary snail.

One can also check that under this move, $v\rightarrow v+1$ and $f_{tot}\rightarrow f_{tot}+r$, so the maximality condition \eqref{ftot} is preserved, so all diagrams produced by this move are maximal. 

It is easy to check that the diagrams produced by this melonic move are equivalent to those produced by the auxiliary field in \cite{Azeyanagi:2017mre,GKPPT}.

\subsection{Theory with both a prism and wheel}
Let us also consider a theory with both $r=3$ maximally-single-trace interactions, the prism and wheel. The allowed 1-cycles remain those from the prism and wheel theories respectively, but the 2-cycles now also include the possibility of a closed path passing through one wheel vertex and one prism vertex. We analyze this case now. 

Let us assume the $L$ vertex is a prism and the $R$ vertex is a wheel. As shown in the appendix, the inequivalent 2-cycles are:
\begin{enumerate}
	\item $(\langle 1_L, 1_R \rangle, \langle 2_L, W_R \rangle )$, 
	\item $(\langle 1_L, 1_R \rangle, \langle 4_L, W_R' \rangle )$, 
 	\item $(\langle 1_L, 1_R \rangle, \langle 6_L, 2_R \rangle )$, 
 	\item $(\langle 1_L, 1_R \rangle, \langle 6_L, 3_R \rangle )$, 
 	\item $(\langle 1_L, 1_R \rangle, \langle 6_L, 6_R \rangle )$.
\end{enumerate}

All of the 2-cycles of the form 1, 2 and 4 contain an odd number of twists in one of the two-colour fat graphs, and are non-maximal.

Case 3 and Case 5 allow for a double snail. There are no other possibilities. In particular there is no new elementary melon containing both a wheel and a prism vertex, and the elementary moves of the prism theory and the wheel theory generate all Feynman diagrams.

\section{Comments on field-theories based on these interactions}
\label{SD-section}
In this section we discuss specific realizations of these theories. Let us focus our attention on obtaining IR fixed points with physics similar to the SYK model. A list of theories involving a single, real rank-$r$ tensor field that can be solved via the analysis given above includes:
\begin{enumerate}
  \item A quantum-mechanical theory of rank-$4$ Majorana fermions based on the octahedron interaction. 
  \item A $d<3$ dimensional theory of rank-$4$ real bosons dominated by the octahedron interaction.
  \item A $d<3$ dimensional theory of rank-$3$ real bosons dominated by the wheel interaction.
  \item A $d<3$ dimensional theory of rank-$3$ real bosons dominated by the prism interaction.
  \item A $d<3$ dimensional theory of rank-$3$ real bosons dominated by both the wheel and the prism interaction.
\end{enumerate}

All the bosonic theories face difficulties, such as a complex spectrum, in higher dimensions, with the possible exception of the last case that has not yet been carefully studied.\cite{Murugan:2017eto, Azeyanagi:2017mre, GKPPT}

We have argued that $1d$ theory of real, rank-$4$, fermionic tensors based on the octahedron is dominated by melonic diagrams. Hence we expect its large $N$ saddle point solution will proceed exactly along the lines of \cite{KPP}, and in particular we expect essentially the same spectrum as the $q=6$ SYK model. It might be interesting to study the theory more carefully, including numerical studies to compare its behaviour at finite $N$ to the rank-$5$ melonic tensor model studied in \cite{Ferrari:2017jgw, KPP} or other models  \cite{Krishnan:2017ztz,  Krishnan:2017txw, Krishnan:2018hhu, Klebanov:2018nfp, Krishnan:2018jsp, Pakrouski:2018jcc}, but we do not do this here. 

The bosonic version of this theory, which can be defined for $d<3$, also dominated by melonic diagrams. Its large $N$ saddle point solution will proceed exactly along the lines of the $q=6$ bosonic theories discussed in \cite{Giombi:2017dtl}. We illustrate this explicitly in subsection \ref{double-prism-section} below. 

For the large $N$ solution of the theory of rank-$3$ real bosons with a wheel interaction, the only difference from the traditional melonic theory is the presence of the elementary snail. This elementary snail is a tadpole, so it does not appear to affect the results of the Schwinger-Dyson equation for the exact propagator. We, therefore, again expect the large $N$ solution to again proceed along the lines of the $q=6$ bosonic tensor models discussed in \cite{Giombi:2017dtl}. However, one slightly novel feature of the wheel interaction is that it allows us to define a theory of complex bosons with $U(N)^3$ symmetry group, as its interaction graph is bi-partite. 

The large $N$ solution to the theory of rank-$3$ bosons with a prism interaction was discussed in \cite{Azeyanagi:2017mre,GKPPT}. We now have the possibility of solving for the large $N$ limit of a theory of rank-$3$ bosons with both the wheel and prism interactions. We leave this for future work.

\subsection{Real sextic bosonic theories with melonic dominance}
\label{double-prism-section}
Let us first consider the rank-$4$ bosonic theory with the octahedron interaction. The Lagrangian for this theory is:
\begin{equation}
    L = \int d^d x ~\frac{1}{2} \partial_\mu \phi\partial^\mu \phi+g
    \phi^{a_1b_1c_1d_1} 
    \phi^{a_1b_2c_2d_2}
    \phi^{a_2b_2c_1d_3}
    \phi^{a_2b_3c_3d_1}
    \phi^{a_3b_3c_2d_3}
    \phi^{a_3b_1c_3d_2}.
\end{equation}
The 't Hooft coupling for this theory is $\lambda=gN^4$.

To write down the gap equation, we need to carefully count all the melonic Wick contractions that take the form of the elementary melon in Figure \ref{elementary-melon}. Let us denote this number by $n_{\text{melon}}$.
Here $n_{\text{melon}}=6$. The gap equation in the strong coupling limit takes the form: \begin{equation}
    G^{-1}(x) = -\lambda^2 n_\text{melon} G(x)^6. 
\end{equation}

\begin{figure}
    \centering
    \includegraphics[height=1in]{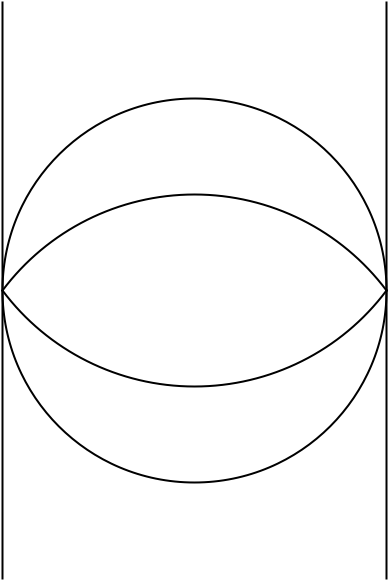}
    \caption{The integration kernel for the four-point function}
    \label{kernel}
\end{figure}

Let us now write the integration kernel:
\begin{equation}
    K(x_1,x_2,x_3,x_4) = n_{\text{kernel}} \lambda^2 G(x_{13}) G(x_{24}) G(x_{34})^4,
\end{equation}
where $n_{\text{kernel}}$ is the number of melonic Wick contractions of the form given in Figure \ref{kernel}. Clearly, for any melonic Wick contraction that takes the form of the elementary melon, one simply has to choose an internal line to cut, in order to obtain a melonic Wick contraction for the kernel, so $n_{\text{kernel}}=(q-1)n_{\text{melon}}$, where $q=6$ in our case. If we now absorb $n_{\text{melon}}$ into $\tilde{\lambda}^2=\lambda^2 n_\text{melon}$, our Schwinger-Dyson equations become:
\begin{equation}
    G^{-1}(x) = -\tilde{\lambda}^2 G(x)^6. 
\end{equation}
and
\begin{equation}
    K(x_1,x_2,x_3,x_4) = (q-1) \tilde{\lambda}^2 G(x_{13}) G(x_{24}) G(x_{34})^4,
\end{equation}
which are identical to those solved in \cite{Giombi:2017dtl}. 

An identical argument shows that the theory with only a wheel is also given by the solution in \cite{Giombi:2017dtl} for $q=6$, assuming here that the elementary snail, which is a tadpole, in the gap equation can be made to vanish, say via dimensional regularization.

\section{Discussion}
\label{conclusion}
Any tensor model with maximally single-trace interactions admits a the natural large-$N$ 't Hooft limit \cite{Ferrari:2017jgw}. In this paper, we classified all real sextic subchromatic tensor models with maximally-single-trace interactions and found only three interactions with $r<5$: the wheel (or $K_{3,3}$) interaction, the prism, and the octahedron. We showed that the theory based on the $r=4$ octahedron is dominated by melonic diagrams. We also showed that the theory based on the $r=3$ wheel (or $K_{3,3}$) interaction is dominated by melonic diagrams, with the addition of an elementary snail that should vanish in most situations.  Finally, we showed that the prism is dominated by a superset of melonic diagrams that also include diagrams generated by an additional melonic (or post-melonic) move -- vertex expansion. In all cases, these diagrams can be explicitly enumerated and summed. 

As a by-product of our arguments, we found that the diagrams which may contribute in the large $N$ limit to  general melonic theory involving maximally single trace interaction vertices may be generated by three classes of melonic moves -- replacing a propagator by an elementary snail, replacing a propagator by an elementary melon, and vertex expansion. The prismatic model is an example of a model where all melonic moves are present.

We have essentially shown that all rank-$3$ sextic tensor models are solvable in the large $N$ limit by our analysis of maximal diagrams arising from the wheel interaction. For completeness, we should also explain how to handle the non-MST interactions of \cite{GKPPT}. It is easy to see that all the non-MST interactions can be reduced to quartic pillow and double-trace interactions by the introduction of an auxiliary field, as was done for the prism in \cite{GKPPT}. Hence, any sextic rank-$3$ tensor model without a wheel interaction is effectively a quartic-tensor model and is therefore solvable in the large $N$ limit by now-standard techniques. With the analysis in this paper, one can also include diagrams arising from the wheel. We postpone a detailed study of the most general $r=3$ sextic theory and its fixed points to future work.

\begin{figure}
    \centering
    \includegraphics[height=1.5in]{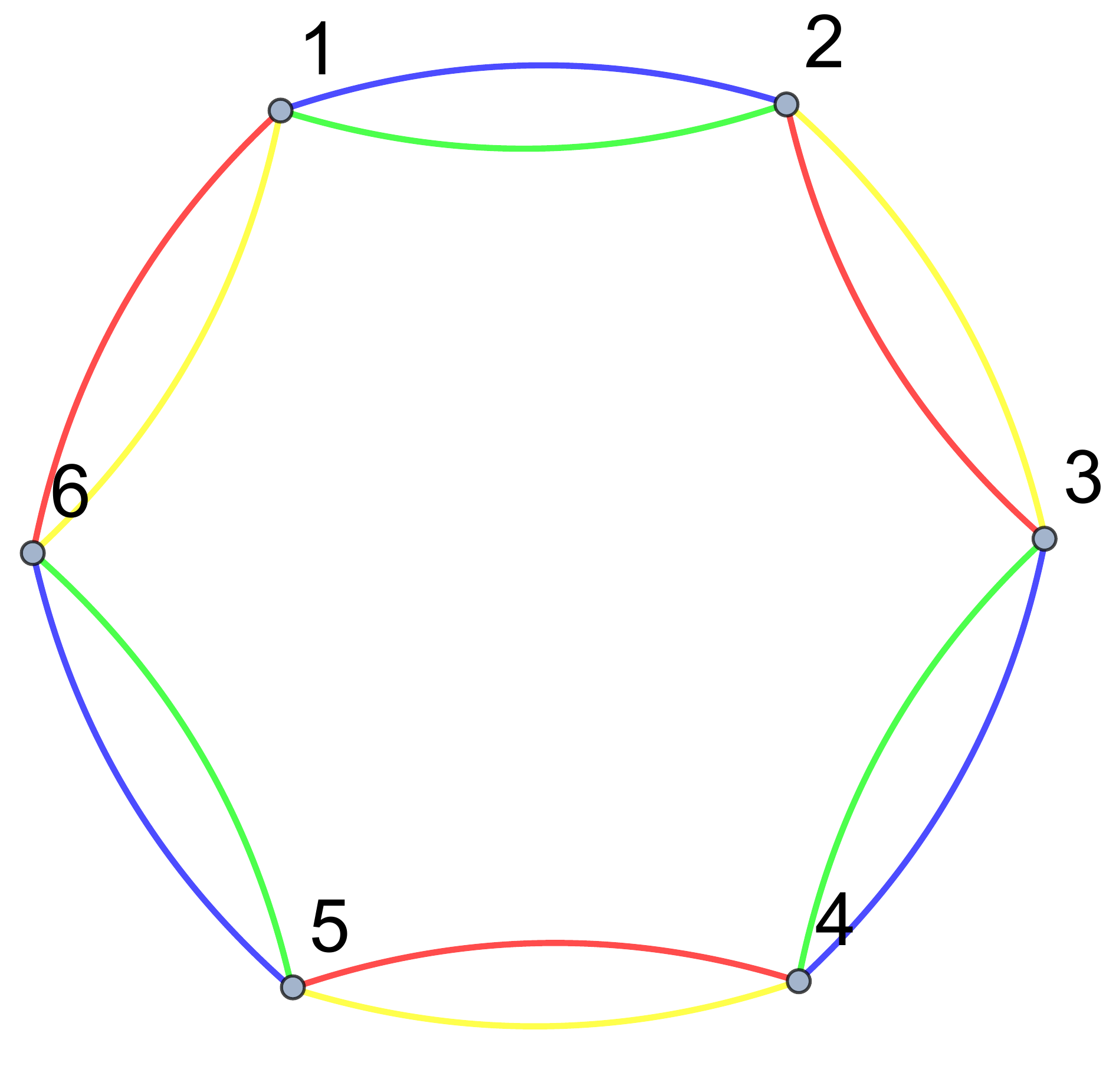}
    \caption{This is a non-MST rank-$4$ interaction whose large $N$ limit gives rise to all planar diagrams. It is equivalent to the rank-$2$ MST interaction.}
    \label{nonMSTrank4}
\end{figure}

One might ask whether all rank-$4$ sextic tensor models are solvable. This is evidently not the case -- for example, the non-MST interaction shown in Figure \ref{nonMSTrank4}, is clearly equivalent to a rank-$2$ MST interaction, and gives rise to all planar diagrams. Such an interaction would exist for any theory based on tensors of even rank. 

It would also be straightforward to extend the analysis of this paper to higher-$q$ in attempts to find new solvable large $N$ limits, perhaps similar to the prismatic limit.

\section*{Acknowledgement}
We thank Adrian Tanasa and Victor Nador for discussions. We also thank Igor Klebanov for comments on a draft of this manuscript.
RS would like to thank Bordeaux U., LPT, Orsay and IIT, Kanpur for hospitality during the course of this work. SP would like to thank IFT and ICTS, TIFR for hospitality.
\noindent
RS is partially supported by the Spanish Research Agency (Agencia Estatal 
de Investigaci\'{o}n) through the grants IFT Centro de Excelencia
Severo Ochoa SEV-2016-0597, FPA2015-65480-P and PGC2018-095976-B-C21.
SP acknowledges the support of a DST INSPIRE faculty award, and DST-SERB grants: MTR/2018/0010077 and ECR/2017/001023. 
\appendix

\section{Appendix: Finding inequivalent 2-cycles}
Here we illustrate a simple method for enumerating all inequivalent 2-cycles. We discuss only the case of a 2-cycle passing through two different wheel vertices and a 2-cycle passing through a wheel and a prism vertex. A very similar analysis applies for the other cases.

Recall that a 2-cycle is a path defined in the single-line Feynman diagram. In the single-line Feynman diagram, each vertex represents an interaction vertex, and each edge represents a Wick contraction between two fields. Consider a 2-cycle passing through 2 interaction vertices we denote as $L$ and $R$. The 2-cycle is defined by an edge contracting a field from the vertex $L$ to a field from the vertex $R$, and another edge contracting a field from the vertex $R$ to a field from the vertex $L$. Hence a 2-cycle is specified by two Wick contractions: $(\langle X_L, Y_R \rangle, \langle Z_L, W_R \rangle )$. Here $X$, $Y$, $Z$ and $W$ range from $1$ to $6$.

We consider two different 2-cycles to be equivalent if they are related to each other by either colour permutation symmetry operation or an automorphism symmetry operation. We represent the symmetry group as a group of permutations acting on the twelve labelled field-vertices $$\{1_L,2_L, \ldots 6_L, 1_R, 2_R, \ldots, 6_R\}.$$ 

Colour permutations simultaneously act on both vertices, while automorphisms can act on each vertex independently. For example, let $L$ and $R$ be two wheel interaction vertices. The 2-cycle $(\langle 1_L, 1_R \rangle, \langle 2_L, 2_R \rangle)$ is equivalent to $(\langle 5_L, 5_R \rangle, \langle 4_L, 4_R \rangle)$ by the action of the colour permutation symmetry generator $\sigma_{rg}$ defined below. The 2-cycle $(\langle 1_L, 1_R \rangle, \langle 2_L, 2_R \rangle)$ is also equivalent to $(\langle 6_L, 1_R \rangle, \langle 5_L, 2_R \rangle)$, by an automorphism symmetry acting on the vertex $L$.

Below, we enumerate all 2-cycles that are not equivalent using the symmetry operations above. Let us emphasize that the 2-cycles enumerated here are closed walks in the single-line Feynman diagram, and need not correspond to loops passing through two vertices in the multi-line Feynman diagram. 

\subsection*{Two Wheel Vertices}
The colour permutation generators act as 
\begin{equation}
    \sigma_{rg}=(1_L,5_L)(2_L,4_L)(1_R,5_R)(2_R,4_R),
\end{equation}
and
\begin{equation}\sigma_{gb}=(1_L,5_L)(4_L,6_L)(1_R,5_R)(4_R,6_R).
\end{equation} 
The automorphism symmetry group is generated by the permutations: $$(1_L,6_L)(2_L,5_L)(3_L,4_L),$$ $$(1_L,2_L)(4_L,5_L)(3_L,6_L),$$ $$(1_R,6_R)(2_R,5_R)(3_R,4_R)$$ and $$(1_R,2_R)(4_R,5_R)(3_R,6_R).$$ 

The combined symmetry group of colour permutations and automorphisms (which we are representing as a subgroup of $S_{12}$) has 216 elements.

To determine the inequivalent choices for $X_L$, we note that the orbit of $1_L$ under the combined symmetry group is $\{1_L, 2_L, 3_L, 4_L, 5_L, 6_L\}$. Hence,  any choice of $X_L$ can be related to $X_L=1_L$ without loss of generality. 

We are then left with a residual symmetry group: the stabilizer of $1_L$, which contains 36 elements. We find the orbit of $1_R$ in this residual symmetry group is $\{1_R, 2_R, 3_R, 4_R, 5_R, 6_R\}$. Hence we can take $Y_R=1_R$. We next consider the residual symmetry group that stabilizes both $1_L$ and $1_R$; this group contains $6$ elements. We find its orbits include $\{2_L, 4_L, 6_L\}$ and $\{3_L, 5_L\}$. Hence we can take $Z_L=2_L$ or $3_L$. 

The inequivalent 2-cycles so far are thus:
\begin{enumerate}
\item $(\langle 1_L, 1_R \rangle, \langle 2_L, W_R \rangle )$, 
\item $(\langle 1_L, 1_R \rangle, \langle 3_L, W_R' \rangle )$.
\end{enumerate} 

To determine the inequivalent choices for $W_R$, we again consider the residual symmetry group that stabilizes $1_L,~1_R,$ and $2_L$; this residual symmetry group contains two elements. The orbits of this symmetry group are $\{2_R\}$, $\{3_R, 5_R\}$, and $\{4_R,6_R\}$. Hence the inequivalent choices for $W_R$ are $2_R$, $3_R$ and $4_R$. 

To consider the inequivalent choices for $W'_R$, we consider the residual symmetry group that stabilizes $1_L,~1_R,$ and $3_L$; this group contains 3 elements. Its orbits are: $\{3_R\}$, $\{5_R\}$, and $\{2_R, 4_R,6_R\}$. The inequivalent choices for $W'_R$ are thus $2_R, 3_R$ and $5_R$.

\subsection*{Prism and wheel}

Let us consider a 2-cycle which intersects one prism interaction vertex and one wheel interaction vertex. Let as assume the left ($L$) vertex is a prism and the right ($R$) vertex is a wheel.

The combined colour permutation and automorphism symmetry group contains $72$ elements. The orbit of $1_L$ under this group is $\{ 1_L, \ldots , 6_L \}$, so we can take $X_L=1_L$. 

The residual symmetry group that stabilizes $1_L$ has $12$ elements. The orbit of $1_R$ under this residual symmetry group is $\{ 1_R, \ldots , 6_R \}$, so we can choose $Y_R=1_R$.

The residual symmetry group that stabilizes both $1_L$ and $1_R$ has $2$ elements. Its orbits include $\{2_L, 3_L\}$ and $\{ 4_L, 5_L\}$. Hence we can take $Z_L=2_L,~4_L,$ or $6_L$. If we choose $Z_L=6_L$, then there is still a residual symmetry group and we can take $W_R=2_R,~3_R$ or $6_R$. 

The inequivalent 2-cycles are thus:
\begin{enumerate}
	\item $(\langle 1_L, 1_R \rangle, \langle 2_L, W_R \rangle )$, 
	\item $(\langle 1_L, 1_R \rangle, \langle 4_L, W_R' \rangle )$, 
 	\item $(\langle 1_L, 1_R \rangle, \langle 6_L, 2_R \rangle )$, 
 	\item $(\langle 1_L, 1_R \rangle, \langle 6_L, 3_R \rangle )$, 
 	\item $(\langle 1_L, 1_R \rangle, \langle 6_L, 6_R \rangle )$.
\end{enumerate}

\bibliographystyle{ssg}
\bibliography{tensor}

\end{document}